\documentclass[twocolumn]{aastex631}

\def\as{\farcs}

\appdef \turnpage {%
  \AddToHookNext{shipout/after}{%
    \global\pdfpageattr\expandafter{\the\pdfpageattr/Rotate 90}%
    \AddToHookNext{shipout/after}{%
      \global\pdfpageattr\expandafter{\the\pdfpageattr/Rotate 0}%
    }%
  }%
}
\usepackage{CJKutf8}
\usepackage[caption=false]{subfig}
\usepackage{graphicx}
\usepackage{float}
\usepackage{amsmath}
\usepackage{longtable}
\usepackage{enumerate}
\usepackage{rotating}
\usepackage{comment}
\usepackage{array}

\begin{document}

\title{The ALMA Legacy survey of Class 0/I disks in Corona australis, Aquila, chaMaeleon, oPhiuchus north, Ophiuchus, Serpens (CAMPOS). I. Evolution of Protostellar disk radii}

\correspondingauthor{Cheng-Han Hsieh}
\email{cheng-han.hsieh@yale.edu}

\author[0000-0003-2803-6358]{Cheng-Han Hsieh \begin{CJK}{UTF8}{bsmi}(謝承翰)\end{CJK}}
\affiliation{Department of Astronomy, Yale University, New Haven, CT 06511, USA}

\author[0000-0001-5653-7817]{H{\'e}ctor G. Arce}
\affiliation{Department of Astronomy, Yale University, New Haven, CT 06511, USA}

\author[0000-0002-7026-8163]{Mar{\'i}a Jos{\'e} Maureira}
\affiliation{Max Planck Institute for Extraterrestrial Physics, Gießenbachstraße 1, 85748, Garching bei München, Germany}

\author[0000-0002-3972-1978]{Jaime E. Pineda }
\affiliation{Max Planck Institute for Extraterrestrial Physics, Gießenbachstraße 1, 85748, Garching bei München, Germany}

\author[0000-0003-3172-6763]{Dominique Segura-Cox}
\affiliation{Department of Astronomy, The University of Texas at Austin, 2515 Speedway, Stop C1400
Austin, Texas 78712-1205, USA}

\author[0000-0002-5065-9175]{Diego Mardones}
\affiliation{Departamento de Astronomía, Universidad de Chile, Camino El Observatorio 1515, Las Condes, Chile}

\author[0000-0003-0749-9505]{Michael M. Dunham}
\affiliation{Physics department, State University of New York at Fredonia, 280 Central Ave. Fredonia, NY 14063, USA}

\author[0000-0002-1877-8705]{Aiswarya Arun}
\affiliation{Departamento de Astronomía, Universidad de Chile, Camino El Observatorio 1515, Las Condes, Chile}

\submitjournal{ApJ}
\accepted{2024.7.8}



\begin{abstract}

We surveyed nearly all the embedded protostars in seven nearby clouds (Corona Australis, Aquila, Chamaeleon I \& II, Ophiuchus North, Ophiuchus, Serpens) with the Atacama Large Millimeter/submillimeter Array at 1.3mm observations with a resolution of 0\as1. This survey detected 184 protostellar disks, 90 of which were observed at a resolution of 14--18\,au, making it one of the most comprehensive high-resolution disk samples across various protostellar evolutionary stages to date. Our key findings include the detection of new annular substructures in two Class I and two flat-spectrum sources, while 21 embedded protostars exhibit distinct asymmetries or substructures in their disks. We find that protostellar disks have a substantially large variability in their radii across all {evolutionary classes}. In particular, the fraction of large disks with sizes above {60\,au decreases as the protostar evolves from Class 0 to Class I.} Compiling the literature data, we discovered an increasing trend of the gas disk radii to dust disk radii ratio ($R_{\rm gas,Kep}/R_{\rm mm}$) with increasing bolometric temperature (${\rm T}_{\rm bol}$).  Our results indicate that the dust and gas disk radii decouple during the early Class I stage. However, in the Class 0 stage, the dust and gas disk sizes are similar, which allows a direct comparison between models and observational data at the earliest stages of protostellar evolution. We show that the distribution of radii in the 52 Class 0 disks in our sample is in high tension with various disk formation models, indicating that protostellar disk formation remains an unsolved question.

\end{abstract}



\section{Introduction} \label{sec:intro}







Circumstellar disks play an important role in star and planet formation. Disks mediate the transfer of mass and angular momentum facilitating the accretion of material onto protostars. Their physical structure and chemical conditions also determine the resulting planetary systems. Circumstellar disks are formed from the gravitational collapse of molecular cloud cores. However, the details of their formation are complicated by the presence of magnetic fields, turbulence, and rotation.  


Theoretical studies suggest  
magnetic fields play an important role in disk formation \citep{2014ApJ...793..130L,2016MNRAS.463.4246M}. Highly pinched magnetic fields thread the protostellar disks and remove the angular momentum in a process known as magnetic braking \citep{2003ApJ...599..363A}. In the ideal magnetohydrodynamics (MHD) approximation, the magnetic braking effect is very efficient and this leads to the suppression of disk formation \citep{2008ApJ...681.1356M,2009A&A...506L..29H}. This problem is softened by the realization of the importance of non-ideal MHD effects (Ohmic diffusion, Hall effect, and ambipolar diffusion) caused by the low conductivity of the gas during the disk formation process (e.g., \citealt{2009ApJ...698..922M,2014MNRAS.438.2278M,2015ApJ...801..117T,2016A&A...587A..32M,2011ApJ...733...54K,2011ApJ...738..180L,2016MNRAS.457.1037W,2015ApJ...810L..26T,2017PASJ...69...95T,2020MNRAS.495.3795W,2023ASPC..534..317T,2019MNRAS.489.1719W,2021MNRAS.507.2354W,2020SSRv..216...43Z}). Additional explanations have been proposed for reducing the magnetic braking effects, for example, misalignment between the disk rotation axis and the magnetic field \citep{2010MNRAS.409L..39C} and turbulence \citep{2012ApJ...747...21S,2013A&A...554A..17J,2013MNRAS.432.3320S,2017ApJ...839...69M}. 

One way to progress in our understanding of disk formation is to find observables that test different disk formation models.
Gas disk radii can provide useful insights because disk sizes are a good tracer of their angular momentum. Many non-ideal MHD models offer distinct predictions for the initial distribution of protostellar gas disk sizes, for example, a bimodal distribution in disk sizes due to the Hall effect \citep{2015ApJ...810L..26T}, and the small variation of disk size centered at 18\,au predicted by ambipolar diffusion {models} (e.g., \citealt{2016ApJ...830L...8H,2021ApJ...922...36L}). 
While the advance of Atacama Large Millimeter/submillimeter Array (ALMA) with an order of magnitude increase in sensitivity and angular resolution compared to earlier radio interferometers has led the way for the characterization of disk properties, the molecular line observation towards the youngest Class 0 protostellar disks remains limited to a small sample in nearby clouds  \citep{2013A&A...560A.103M,2014ApJ...796..131O,2014ApJ...791L..38S,2014A&A...563L...3C,2014A&A...566A..74L,2017ApJ...849...56A,2017ApJ...834..178Y,2018A&A...616A..56A,2020ApJ...894...23H}. This is due to the large observational time required to resolve the small rotating disk.
Recently, the ALMA large program Early Planet Formation in Embedded Disks (eDisk) observed 12 Class 0 and 7 Class I protostellar disks in great detail, significantly increasing the number of young protostellar disks with line observations \citep{2023ApJ...951....8O,2023ApJ...951....9L,2023ApJ...951...10V,2023arXiv230615443K,2023arXiv230708952S}.
Even so, the total number of observed young protostellar disks with molecular lines falls significantly short of a population study. 

In contrast, the sample size of dust continuum observation towards protostellar disks has improved significantly over the last decade. The Protostellar Submillimeter Array Campaign (PROSAC) surveyed 20 Class 0/I disks at 2\as0 ($\sim 250-650$\,au) resolution at 0.85\,mm and 1.3\,mm wavelength using the Submillimeter Array (SMA) \citep{2009A&A...507..861J}. Subsequent efforts, such as the IRAM-PdBI survey, observed 21 Class 0 sources at 0\as3 ($\sim 120$\,au) resolution \citep{2010A&A...512A..40M,2019A&A...621A..76M} at 1.3\,mm wavelength. The Combined Array for Research in Millimeter-wave Astronomy (CARMA) was used to observe 9 more Class 0 sources in the Serpens molecular cloud at 1\as0 ($\sim 415$\,au) resolution at 1.3\,mm wavelength \citep{2011ApJS..195...21E}. Additionally, observations with the Very Large Array (VLA) were used to study 18 Class 0/I disks in the Perseus molecular cloud at $\sim$0\as05 (12\,au) resolution at 8\,mm wavelength \citep{2016ApJ...817L..14S,2018ApJ...866..161S}. These early surveys mostly focused on selected sources, and either had a relatively low angular resolution ($>100$\,au resolution) or a limited sample size.


The advent of ALMA marked a revolutionary leap forward in the field. The recent VLA/ALMA Nascent Disk and Multiplicity (VANDAM) survey of the Orion molecular clouds stands out, detecting a total of 379 protostellar disks out of 421 surveyed at an excellent angular resolution of $\sim$0\as07 (40\,au) at 0.87\,mm wavelength \citep{2020ApJ...890..130T}. To complete the sample of protostellar disks in nearby molecular clouds within 450\,pc at 0\as1 angular resolution, we initiated the ALMA Legacy survey of Class 0/I disks in Corona australis, Aquila, chaMaeleon, oPhiuchus north, Ophiuchus, and Serpens (CAMPOS). We surveyed nearly all the Class 0/I protostellar disks in seven molecular clouds: Chamaeleon I at a distance of 179 pc \citep{2018A&A...610A..64V}, Chamaeleon II at 181 pc \citep{2018A&A...610A..64V}, Corona Australis at a distance of $149$\,pc \citep{2020A&A...634A..98G}, Ophiuchus and Ophiuchus North at a distance of 144\,pc \citep{2019ApJ...879..125Z}. We also include the Serpens and Aquila molecular clouds at a distance of $436$\,pc \citep{2018ApJ...869L..33O} which is similar to Orion, allowing us to compare with the Orion Class 0/I disk survey \citep{2020ApJ...890..130T}. 

To compare our CAMPOS dust disk observation with different protostellar disk formation models, we compiled the literature data and discovered an increasing trend of the gas disk radii to dust disk radii ratio with the protostar's evolutionary stage. In particular, for Class 0 sources, the dust disk size is similar to the gas disk size, allowing a direct comparison between models with observational data. We found the size distribution of 52 Class 0 disks is in high tension with both the pure hydrodynamical models and the Hall effect models, and it is also inconsistent with the ambipolar diffusion models.



This paper is organized as follows: In \autoref{sec:observation}, we give an overview of the CAMPOS observation. In \autoref{sec:data}, we describe the detailed self-calibration tests, data reduction, imaging, and source detection for the project. In \autoref{sec:Results}, we present the main result of this paper: source detection statistics, distribution of dust disk radii, the discovery of new disk substructures, streamer candidates, and edge-on protostellar disks. We discuss our results in \autoref{sec:discussion}, and present our conclusion in \autoref{sec:conclusion}.

\section{Observation}
\label{sec:observation}

\subsection{Sample Selection}
\label{sec:sample_selection}
Our protostellar sample is derived from the \citet{2015ApJS..220...11D} catalog, which originated from two Spitzer Space Telescope Legacy Projects: the ``From Molecular Cores to Planet-forming Disks" (c2d) survey and its successor, the ``Spitzer Gould Belt (GB) Legacy Survey". An overview of the c2d survey was given by \citet{2003PASP..115..965E} with major results for major star-forming clouds highlighted in \citet{2009ApJS..181..321E}. The combined survey targeted nearly all the molecular clouds within 500\,pc of the Sun. 

We selected the Class 0, Class I, flat-spectrum sources, and early Class II protostars with bolometric temperatures ($T_{\rm bol} \le 1900\,$K) in seven molecular clouds: Serpens, Aquila, {Corona Australis,} Ophiuchus, Ophiuchus North, Chamaeleon I, and Chamaeleon II. All sources are associated with dense cores detected via submillimeter or millimeter continuum emission (wavelength $>$ 350\,$\mu$m) as noted in Table 2 of \citet{2015ApJS..220...11D}. By employing sources with established dense cores, our protostar sample is robust, avoiding reliance on assumptions regarding infrared protostar colors.


However, while our stringent criteria in dust continuum selection assure sample reliability, there's a trade-off with incompleteness. Our sample lacks representation of protostars nestled within low-mass cores, leading to a bias against the lowest-mass (luminosity) protostars. The recent complete results from the Herschel and James Clerk Maxwell Telescope Submillimetre Common-User Bolometer Array (SCUBA)-2 Legacy GB surveys allow the detection of dense cores in all GB clouds down to 0.1\,$M_\odot$ (e.g., \citet{2015MNRAS.450.1094P,2015A&A...584A..91K}). \citet{2023ApJS..266...32P} published an updated protostar catalog for the Serpens and Aquila molecular clouds, incorporating both Herschel and SCUBA data. A complete list of protostars encompassing other clouds covered by our CAMPOS survey remains pending. 


Despite the trade-off for reliability at the expense of completeness, our sizable sample ensures a fairly uniform sampling of sources within the ranges of $25\,K < T_{\rm bol} < 1900\,K$ and $0.1\,L_\odot < L_{\rm bol} < 10\,L_\odot$. {Given this range of $T_{\rm bol}$, our sample is severely incomplete for Class II sources. Hence, we report the properties of the disks around early Class II sources in this paper but exclude them from the analysis of the overall sample.}

\subsection{ALMA 1.3\,mm Continuum Observations}
\label{sec:sample_selection}

\begin{table*}[htp!]
\begin{center}
\caption{ALMA Observations summary}%
\begin{tabular}{ c c c c c c c c c}
\label{table:observation}
\\
\hline \hline
 Region\tablenotemark{a}  & Observed & Duration\tablenotemark{b} & Baseline & Number of & MRS\tablenotemark{c} & PWV\tablenotemark{d} & Calibrators\\
 & Date& (minutes) & Range (m) & Antennas& (arcsec)& (mm) & Bandpass, Flux, Phase\\
 \hline 
Aquila & 2021 Jul 6 & 32.0 & $29-3638$ & 45 & 2.3 & 0.54 & J1924-2914, J1924-2914, J1834-0301\\
ChamI (A)& 2021 Jul 20 & 22.4 & $15-3697$ & 39 & 1.5 & 1.68 & J1617-5848, J1617-5848, J1058-8003\\
ChamI (A)& 2021 Jul 22 & 16.1 & $43-3697$ & 22 & 1.2 & 0.58 & J0635-7516, J0635-7516, J1058-8003\\
ChamI (B)& 2021 Jul 20 & 24.7 & $15-3697$ & 42 & 1.5 & 1.56 & J1617-5848, J1617-5848, J1058-8003\\
ChamI (B)& 2021 Oct 24 & 23.8 & $47-8548$ & 34 & 0.9 & 0.18 & J1107-4449, J1107-4449, J1058-8003\\
ChamII & 2021 Jul 20 & 22.9 & $15-3697$ & 44 & 1.5 & 1.63 & J1617-5848, J1617-5848, J1058-8003\\
CrAus & 2021 Jul 10 & 23.6 & $29-3638$ & 46 & 1.9 & 0.66 & J1924-2914, J1924-2914, J1925-3401\\
Oph (A) & 2021 Jul 6 & 48.1 & $29-3638$ & 45 & 2.3 & 0.50 & J1427-4206, J1427-4206, J1633-2557 \\
Oph (B) & 2021 Jul 7 & 24.6 & $29-3638$ & 45 & 2.1 & 0.50 & J1517-2422, J1517-2422, J1650-2943 \\
OphN-01 & 2021 Jul 7 & 18.7 & $29-3638$ & 45 & 2.1 & 0.51 & J1550+0527, J1550+0527, J1643-0402 \\
OphN-02 & 2021 Jul 7 & 21.9 & $29-3638$ & 45 & 2.1 & 0.57 & J1517-2422, J1517-2422, J1653-1551 \\
OphN-03 & 2021 Jul 6 & 22.0 & $29-3638$ & 45 & 2.3 & 0.50 & J1517-2422, J1517-2422, J1653-1551 \\
Serp (A) & 2021 Jul 6 & 34.4 & $29-3638$ & 45 & 2.3 & 0.51 & J1924-2914, J1924-2914, J1851+0035 \\
Serp (B) & 2021 Jul 6 & 32.7 & $29-3638$ & 45 & 2.3 & 0.57 & J1924-2914, J1924-2914, J1830+0619 \\
\hline
\hline
\end{tabular}
\end{center}
\tablenotetext{a}{Cloud labels which represent the following: CorAus: Corona Australis; ChamI: Chamaeleon I; ChamII: Chamaeleon II; Oph: Ophiuchus; OphN: Ophiuchus North; Serp: Serpens. {If two executions were carried out for the same cloud, then these were labeled with A, and B in parenthesis.}}
\tablenotetext{b}{Total duration of the observation per field includes time used for  calibration.}
\tablenotetext{c}{Maximum recoverable scale.}
\tablenotetext{d}{Percipitable water vapor.}
\end{table*}

Observations for the CAMPOS project were taken between 2021 July-October (Project ID: 2019.1.01792). The project consists of 2 parts: the molecular line component and the high-resolution continuum component. For the molecular line component, a single pointing using the 12\,m array (with a resolution of 1\as0) was used to observe the morphology and kinematics of the molecular outflows and envelopes associated with the protostars in our sample. We will present the result of the line observations in future papers. 

In this paper, we focus on the high-resolution continuum component. A single pointing with the ALMA 12\,m array in configuration C43-7 (which provided a resolution of 0\as1) was used for each target. Band 6 dual polarization mode was used for the ALMA observation. The continuum was sampled in four spectral windows (SPW) centered at 224, 226, 240, and 242 GHz, {each in Time Division Mode (TDM) mode and providing 1.875\,GHz of bandwidth for a total bandwidth of 7.5\,GHz for our continuum observations.} Our observations were divided into 14 scheduling blocks, with four scheduling blocks for the Chamaeleon I molecular clouds, three scheduling blocks for  Ophiuchus North, two scheduling blocks each for the Ophiuchus and Serpens molecular clouds, and one scheduling block each for the Aquila, Chamaeleon II, and Corona Australis molecular cloud. The date of each observation, the duration (including the time spent on calibration), the baseline range, the number of antennas, the maximum recoverable scale (MRS), the precipitable water vapor (PWV), and the Bandpass, Flux, and Phase calibrators used are summarized in \autoref{table:observation}. The total time on each source was approximately 0.6 minutes. All data were calibrated using the Common Astronomy Software Application (CASA) version 6.1.1.15 \citep{2007ASPC..376..127M}.  

\section{Data}
\label{sec:data}


\subsection{Self-calibration and Imaging processing}
\label{sec:Selfcal}
We used a modified version of \verb+auto_selfcal+\footnote{This self-calibration pipeline developed by J.~Tobin can be found at: \url{https://github.com/jjtobin/auto_selfcal}.}
to conduct the self-calibration of the entire sample. The most significant modification was to change the mask used for cleaning from being created automatically by the CASA \verb+tclean+ (\verb+automasking+) task to a manually-inputted mask created before the self-calibration process. In \autoref{Appendix:Selfcal_test} we provide more details on the procedure and the different tests we performed. We first ran the self-calibration pipeline with \verb+automasking+ to obtain the first round of self-calibrated measurement sets. We then applied \verb+tclean+ to image these measurement sets with \textit{uniform}, \textit{briggs0.5}, and \textit{natural} weighted models with the \textit{mtmfs} deconvolver. We then identified local peaks in these 3 maps and used them as a guide to create self-calibration masks centered on sources with clear detection. With the self-calibration masks ready, we {now run the modified self-calibration pipeline with manually-inputted self-calibration masks on the \textit{uniform}, \textit{briggs0.5}, and \textit{natural} weighted models with the \textit{mtmfs} deconvolver. }

{Once the self-calibration was complete, we then proceeded with a  two-stage imaging process. The purpose of the first step is to produce images from which we can  draw the final clean masks on all the identified sources.} We applied \verb+tclean+ on the final self-calibrated measurement sets, and cleaned down to $2\sigma$ using the same mask used for self-calibration. We then drew new clean masks on the naturally weighted maps, including all the sources that were detected with a signal-to-noise (S/N) greater than or equal to 4 in all three maps (\textit{uniform}, \textit{briggs0.5}, and \textit{natural} weighted maps), and detected with S/N $\ge 5$ in at least one of the three maps. We also drew clean masks on possible candidate sources on the position of the known protostars or visually identified local peaks with sizes greater than the beam near other protostars.

{We conducted the second stage of the imaging process using the larger final clean masks based on the naturally weighted maps.  In this final step, we ran \verb+tclean+ on the  three final self-calibrated measurement sets
with the \textit{multi-scale} deconvolver  in order to better recover the large-scale emission (1--2\arcsec).} We cleaned down to $2\sigma$ using three different weighting parameters (\textit{natural}, \textit{briggs0.5}, and \textit{uniform}) to produce three different images with different sensitivity and angular resolution. For each weighting scheme, we used the  self-calibrated visibilities with the same weighting scheme to achieve the best signal-to-noise ratio. In \autoref{Appendix:Self-cal_fields}, we summarize all the image properties of each field before and after the self-calibration.


\subsection{Source Detection process}
\label{sec:source_search}
We applied {SciPy's \verb+ndimage.maximum_filter+} to the final cleaned images to identify the sources. The maximum filter replaces each pixel value of an image with the maximum {pixel value within the filter scale. We set the filter scale to be the same size as the beam size to filter out structures smaller than the beam size.} We then identified all the local maxima in the maps with S/N $\ge 4$ in all 3 maps (\textit{uniform}, \textit{briggs0.5}, and \textit{natural} weighted maps), or that have S/N $\ge 5$ in at least one of the maps. Lastly, we also added one more source that has S/N $\ge 4$ in two of the 3 maps. While many continuum surveys use S/N $\ge 3$ as a detection threshold (eg. \citet{2020ApJ...890..130T}), we adopted a more conservative detection threshold S/N $\ge 5$. For ALMA Band 6 observations, the 0.2 primary beam cutoff (i.e., the radius at which the map reaches a sensitivity that is 20\% of that of the phase center) results in a field of view of 37\as0. For a beam size of 0\as1, there are $1.1\times 10^5$ beams in each map. Adopting $3\sigma,\,4\sigma,\,5\sigma$, based on statistics we would expect around 290, 6, and 0.06 artificial sources in each map respectively. All the sources detected in our survey except one are detected at least above $5\sigma$. 

\section{Results}
\label{sec:Results}

Our ALMA continuum images revealed the detailed disk structures on scales between 0\as1 and 1\as1. This corresponds to a physical scale of 15--{164}\,au for sources in {Corona} Australis, 14--158\,au for sources in Ophiuchus and Ophiuchus North, and 18--{199}\,au for sources in Chamaeleon. As for the Serpens and Aquila molecular clouds, these clouds are located at a distance of 420\,pc and 440\,pc respectively \citep{2018ApJ...869L..33O}. Thus,  for the targeted sources in these clouds  we probe scales  of 42--462\,au, and 44--484\,au similar to those probed by the Orion and Perseus Class 0/I disk surveys \citep{2018ApJ...867...43T,2020ApJ...890..130T}. 
In this section, we present all the high-resolution continuum images and sources detected in our survey. 


\begin{table}[]
\centering
\caption{Fields and detection of sources per cloud}%
\begin{tabular}{ c c c c }
\hline \hline
 Cloud\tablenotemark{a}  & Targeted & Detected\tablenotemark{b} & Sources Detected\tablenotemark{c} \\
 \hline 
Aql & 18 & 18 & 32\\
CorAus & 11& 10 & 19\\
Cham I & 13& 11 & 13\\
Cham II & 3& 3 & 3\\
Oph & 41& 38 & 52\\
OphN & 3& 3 & 3\\
Serp & 36& 35 & 62\\
\hline
Total & 125 & 118 & 184\\
\hline
\hline
\end{tabular}
\tablenotetext{a}{Cloud labels as in \autoref{table:observation}.}
\tablenotetext{b}{Number of fields with detection.}
\tablenotetext{c}{Total number of sources detected.}
\label{table:dection_stat}
\end{table}

\subsection{Source Detection Statistics}
\label{sec:source_search_results}

\begin{table*}[htp]
\centering
\caption{New sources and sources first detected at millimeter wavelength in CAMPOS survey}%
\begin{tabular}{ l c l c }
\hline \hline
 Source Name  & Cloud   & Note & Ref.\\
 \hline 
 New source\\
 \hline
 SerpS10-mm & Aquila   & Not detected before. &  \\
 ChamI-9 mm & ChamI   & Not detected before. &  \\
 Ser-emb 30C & Serpens  & Not detected before. &  \\
 \hline 
 New companion\\
 \hline
SerpS-MM2 b & Aquila   & 2MASS J18293891-0151063 source is resolved into 2 sources. & 1 \\
SerpS-MM6 b & Aquila   & MAMBO source, SerpS-MM6, is resolved into 2 sources. & 2 \\
eHOPS-aql-96 b & Aquila  & eHOPS-aql-96 is resolved into 2 sources. &3 \\

IRAS 18278-0158 b & Aquila  & eHOPS-aql-135 is resolved into 2 sources. & 3 \\
eHOPS-aql-139 b & Aquila   & eHOPS-aql-139 is resolved into 2 sources. & 3\\
eHOPS-oph-20c & Oph & WL 20E resolved into 2 sources. & 4, 5\\
Ser-emb 10B & Serpens & Ser-emb 10 or eHOPS-aql-15 resolved into 2 sources. & 6\\
Ser-emb 8B & Serpens  & Previously identified as a condensation. & 7\\
Ser-emb 8C & Serpens  & Previously identified as a condensation. & 7\\
Ser-emb 6C & Serpens  & SMM1b is resolved into 2 sources. & 8, 9\\
Ser-emb 12A & Serpens  & Serpens SMM 10 IR is resolved into 3 sources. & 10\\
Ser-emb 12C & Serpens & Serpens SMM 10 IR is resolved into 3 sources. & 10\\
Ser-emb 15B & Serpens  & Serp-emb 15 is resolved into 2 sources. & 9 \\
eHOPS-aql-86B & Serpens & Serpens SMM 3 is resolved into 2 sources. & 11\\
Ser-emb 24A & Serpens & Ser-emb 24 is resolved into 2 sources. & 12\\
\hline
First Subm.\tablenotemark{a}\\
\hline
SSTgbs J1829381-015100 & Aquila   & Detected by Spizter at 3.6-8 $\mu m$. Core observed by Herschel in Far infrared. & 13 \\ 
SSTgbs J1830469-015651 & Aquila & Detected by Spizter at 3.6-8 $\mu m$. Chandra X-ray source 260.  & 14, 15\\
SSTgbs J1829386-015100 & Aquila  & Detected by Spizter at 3.6-8\,$\mu m$. &  14 \\  
SSTgbs J1829419-015011 & Aquila  & Detected by Spizter at 3.6-8\,$\mu m$. &  14 \\ 
MIRES G028.6593+03.8185 & Aquila  & Detected by Spizter at 3.6-8\,$\mu m$. MYStIX Candidate Protostar. & 16, 17 \\
V* HO Cha a & ChamI  & V* HO Cha or ISO\,126 binary was detected in the VLT K band. & 18, 19 \\
Ser-emb 30A & Serpens & Chandra X-ray source 49. & 20 \\
\hline
\hline
\end{tabular}
\tablenotetext{}{\textbf{References for Note:} (1) \cite{2003yCat.2246....0C} (2) \cite{2011A&A...535A..77M} (3) \cite{2023ApJS..266...32P} (4) \cite{2013ApJS..205....5H} (5) \cite{2008ApJS..179..249D} (6) \cite{2009ApJ...692..973E} (7) \cite{2021A&A...655A..65T} (8) \cite{2019A&A...632A.101T} (9) \cite{2019ApJ...871..149F} (10) \cite{2022ApJ...937...29F} (11) \cite{2014ApJ...797...76L} (12) \cite{2009ApJ...692..973E} (13) \cite{2015A&A...584A..91K}  (14) \cite{2018AJ....155..241W} (15) \cite{2013ApJ...779..113M} (16) \cite{2016ApJ...833..193R} (17) \cite{2018AJ....155...99W}  (18) \cite{2017ARep...61...80S}  (19) \cite{2016MNRAS.458.2476S} (20) \cite{2007A&A...463..275G}  }
\tablenotetext{a}{First detection at the millimeter wavelength presented in this paper.}
\label{table:new_source}
\end{table*}

Out of the 125 embedded protostellar systems initially targeted, 
our ALMA CAMPOS survey detected protostars in 118 fields. Within these fields, we detected a total of 184 embedded protostars, including many multiple systems. 
Among these sources, 53  were associated with Class 0 systems, 71 with Class I systems, 33 with flat-spectrum systems, and 27 with early Class II systems.
The evolutionary class and estimate of ${\rm T}_{\rm bol}$ for all sources in the  Aquila and Serpens molecular clouds were obtained from the latest eHOPS survey \citep{2023ApJS..266...32P}. For the remaining clouds, we utilized unpublished eHOPS results (Pokhrel, private communication). In instances where CAMPOS sources lacked corresponding entries in the eHOPS survey, we relied on the ${\rm T}_{\rm bol}$ and classification from \citet{2015ApJS..220...11D}. 

The evolutionary stage classification and estimate of ${\rm T}_{\rm bol}$  for each source was obtained by constructing Spectral Energy Distribution (SED) plots using photometry and spectra  from 1 to 850$\,\mu m$ \citep{2023ApJS..266...32P}. This involved data from various catalogues and telescopes such as the Two Micron All Sky Survey (2MASS) point source catalog \citep{2006AJ....131.1163S}, observations by the Spitzer Space Telescope \citep{2004ApJS..154...25R,2004ApJS..154...10F}, the Herschel Space Observatory \citep{2010A&A...518L...1P}, and the James Clerk Maxwell Telescope \citep{2018ApJS..238....8K}. The SEDs were constructed using telescopes with diverse apertures, ranging from the Infrared Array Camera (IRAC) for Spitzer with a FWHM of about  2\as0 to Herschel-SPIRE 500 $\mu$m detector with an FWHM of 37\as4. In situations where the eHOPS data could not resolve close binaries or multiples, we adopted the same ${\rm T}_{\rm bol}$ and evolutionary class measured for sources within the ALMA field. For detailed detection statistics in each molecular cloud, please refer to \autoref{table:dection_stat}.

We cross-referenced all 184 protostars identified in our ALMA survey with established literature catalogs, listing the detected sources and alternative names used in the literature in \autoref{Appendix:Source_summary}, while presenting their images in \autoref{Appendix:Image_gallery}. Remarkably, our survey unveiled {18 new protostellar sources}, and includes 7 known protostellar sources for which we report the first detection at millimeter wavelengths. These sources are summarized in \autoref{table:new_source}. Among these, {nine} were part of close binary systems, namely IRAS 18278-0158 b, eHOPS-aql-139 b, V* HO Cha a, ChamI-9 mm, Ser-emb 15B, {SerpS-MM6 b, SSTgbs J1830469-015651}, eHOPS-aql-86B, and Ser-emb 24A. Additionally, five protostars were associated with triple systems (eHOPS-oph-20c, Ser-emb 12A, Ser-emb 12C, Ser-emb 30C, Ser-emb 30A). {Eleven} were linked with complex multiple systems harboring at least 4 protostars each (SSTgbs J1829381-015100, SerpS-MM2 b, SSTgbs J1829386-015100, SSTgbs J1829419-015011, MIRES G028.6593+03.8185, SerpS10-mm, IRAS 18278-0158 b, eHOPS-aql-139 b, Ser-emb 8B, Ser-emb 8C, Ser-emb 6C). A comprehensive analysis of protostar multiplicity will be presented in forthcoming publications. 


Within our CAMPOS survey, there are no source detections in 5 of the 125 fields observed. The Class I protostar, ChamI-03 (J110658.0-772248), and Class II source, CrAus-06 (J190153.7-370033) were not detected. The flat-spectrum systems, Oph-24 (J162721.8-242727), Serp-32 (J183000.3+010944), and ChamI-04 (J110716.1-772306) were not detected in our survey. However, the Spitzer Space Telescope \citep{2015ApJS..220...11D} clearly detected all these sources, suggesting their millimeter dust continuum emission is below our sensitivity limits. Additionally, two sources, the Class 0 protostar Oph-04-0 (J162614.6-242507) and the Class I protostar Oph-37-0 (J163152.0-245726), exhibited faint dust continuum emission (at a $\le 3\sigma$ level) in our ALMA observations at the position of the sources identified by the Spitzer Space Telescope. {Due to the low S/N, these two sources are not included in the analysis. } 

We also identified 4 potential candidates. These sources exhibited a signal-to-noise ratio of $\ge 3\sigma$ in at least one of the three maps of the same field imaged with different weighting schemes (i.e., natural, Briggs 0.5, and uniform), or represented a local peak connected to a known protostar. {These four sources are the Class 0 candidate  CrAus-04-1, the Class I candidate  ChamI-02-1, and the Class II candidates  Serp-14-1 and Serp-26-3. To assign an evolutionary class to these sources we assumed they are coeval to their nearest source and assign them the same class as their neighbor.}
While ChamI-02-1 and Serp-14-1 were detected at $11 \sigma$ and $5 \sigma$ respectively, additional higher resolution observations are essential to confirm these candidates' nature, as some may represent extended emission from the surrounding dusty envelope. Detailed images showing (and a table listing) all the candidate sources and faint emission sources are provided in \autoref{Appendix:D}. 



\subsection{Distribution of Protostellar Dust Disk Radii}
\label{sec:Disk_R}

\subsubsection{Measuring disk radii}

\begin{figure*}[tbh!]
\includegraphics[width=1.0\textwidth]{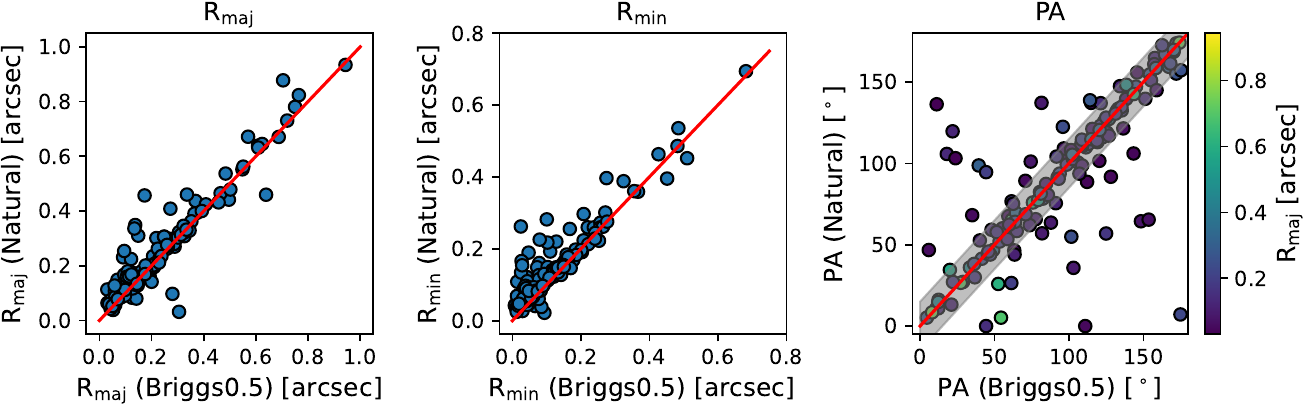}
    \caption{Comparison of disk major axis (Rmaj), minor axis (Rmin), and position angle (PA) for naturally weighted and Briggs 0.5 weighted maps. Unresolved sources are not included. The red line represents identical measurements from the two maps. The gray shaded region represents a $\pm 10^\circ$ spread in the position angle measurements.} 
\label{fig:Rcompare}
\end{figure*}

\begin{figure}[tbh!]
\includegraphics[width=\columnwidth]{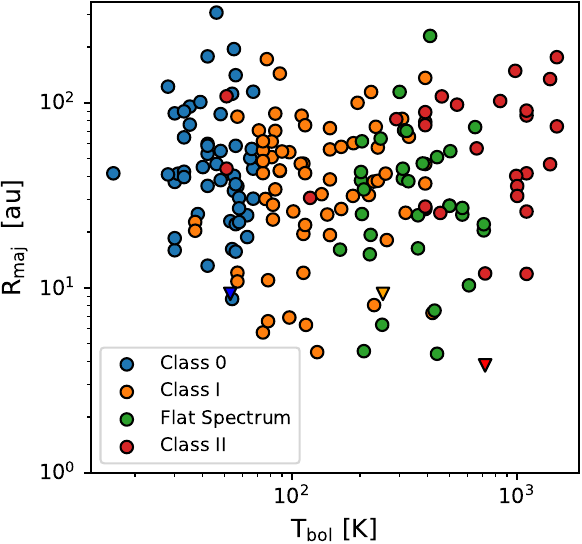}
    \caption{Dust disk major axis (Rmaj) versus bolometric temperature (${\rm T}_{\rm bol}$) of the source. The blue, orange, green, and red represent Class 0, Class I, flat-spectrum, and Class II sources respectively.  We exclude the circumbinary disks in our samples: SMM 2, IRAS16293-2422 A, and the VLA1623A. The triangles represent the upper limit radius of unresolved disks, which corresponds to half the beam size. } 
\label{fig:R_Tbol}
\end{figure} 

\begin{table*}[]
\centering
\caption{Dust Disk Radii Distribution Properties}%
\begin{tabular}{ l c c c c c }
\hline \hline
 Sample  & Mean ${\rm R}_{\rm disk}$  & $\sigma \left( log_{10}\left( \frac{R_{disk}}{1 au} \right) \right)$& Median ${\rm R}_{\rm disk}$ & References\\
 & (au) &  & (au) & \\
 \hline
 All Sources \\
 \hline 
Class 0 & $  59.3 \pm 7.4 $ & $ 0.33_{-0.17}^{+0.02 }$ & $ 41.4_{\;26.4}^{\;67.8}$ & 1\\
Class I & $ 46.0 \pm 4.0 $ & $ 0.36_{-0.16}^{+0.02 }$ & $ 39.9_{\;22.9}^{\;61.4}$ & 1\\
Flat & $ 41.6 \pm 7.2 $ & $ 0.38_{-0.06}^{+0.03 }$ & $ 27.8_{\;19.3}^{\;50.8}$& 1 \\
Class II & $ 66.9 \pm 8.4 $ & $ 0.37_{-0.32}^{+0.05 }$ & $ 56.7_{\;31.0}^{\;94.2}$ & 1\\
\hline
Subsample by cloud and Class\tablenotemark{a}\\
\hline
Ophiuchus (Class 0)& $ 35.4 \pm 8.0 $ & $ 0.26_{-0.01}^{+0.01 }$ & $ 31.2_{17.9}^{42.0}$ & 1\\
Ophiuchus (Class I)& $ 26.4 \pm 6.4 $ & $ 0.41_{-0.06}^{+0.03 }$ & $ 12.0_{\;7.2}^{\;33.4 }$ & 1\\
Ophiuchus (Flat)& $ 38.4 \pm 12.9 $ & $ 0.41_{-0.07}^{+0.04 }$ & $ 27.4_{\;13.9}^{\;40.2 }$ & 1\\
Ophiuchus (Class II)& $ 89.6 \pm 15.6 $ & $ 0.33_{-0.12}^{+0.05 }$ & $ 83.4_{\;49.8}^{\;126.2 }$ & 1\\
Serpens (Class 0)& $ 77.3 \pm 11.7 $ & $ 0.25_{-0.06}^{+0.03 }$ & $ 54.9_{\;40.6}^{\;88.3 }$ & 1\\
Serpens (Class I)& $ 48.6 \pm 7.4 $ & $ 0.25_{-0.19}^{+0.05 }$ & $ 44.2_{\;25.7}^{\;55.3 }$ & 1\\
Serpens (Flat)& $ 51.0 \pm 5.7 $ & $ 0.15_{-0.04}^{+0.01 }$ & $ 54.7_{\;40.1}^{\;62.9 }$ & 1\\
Serpens (Class II)& $ 72.0 \pm 11.5 $ & $ 0.25_{-0.04}^{+0.02 }$ & $ 82.3_{\;39.9}^{\;100.3}$ & 1\\
Aquila (Class 0)& $ 49.9 \pm 17.6 $ & $ 0.37_{-0.07}^{+0.08 }$ & $ 24.6_{\;20.8}^{\;65.8 }$ & 1\\
Aquila (Class I)& $ 61.8 \pm 7.3 $ & $ 0.23_{-0.07}^{+0.03 }$ & $ 57.3_{\;36.6}^{\;78.8 }$ & 1\\
Cor Australia (Class 0)& $ 36.8 \pm 9.3 $ & $ 0.3_{-0.02}^{+0.02 }$ & $ 25.1_{\;17.9}^{\;43.3 }$ & 1\\
Chamaeloeon (Class I)& $ 51.4 \pm 7.5 $ & $ 0.16_{-0.02}^{+0.02 }$ & $ 50.9_{\;34.1}^{\;61.5}$ & 1\\
Chamaeloeon (Class II)& $ 39.3 \pm 11.9 $ & $ 0.44_{-0.02}^{+0.1 }$ & $ 36.2_{\;15.4}^{\;54.1}$ & 1\\
\hline
Orion (Class 0)& $ 44.9_{-3.4}^{+5.8} $ & $ 0.38_{-0.004}^{+0.002}$ & $ 48.1_{\;24.5}^{\;79.6 }$ & 2\\
Orion (Class I)& $ 37.0_{-3.0}^{+4.9} $ & $ 0.42_{-0.005}^{+0.001}$ & $ 38.1_{\;17.5}^{\;64.0 }$ & 2\\
Orion (Flat)& $ 28.5_{-2.3}^{+3.7} $ & $ 0.38_{-0.007}^{+0.001}$ & $ 30.9_{\;13.0}^{\;51.3 }$ & 2\\
\hline
\hline
\end{tabular}
\tablenotetext{}{\textbf{References:} (1) This Work. (2) \citet{2020ApJ...890..130T}. Note: The sub- and superscripts on the median values correspond to the ﬁrst and third quartiles of the distributions and are absolute values, not relative to the median. In addition, circumbinary disks are excluded from the sample.}
\tablenotetext{a}{We only report values with a sample size greater than 5.}
\label{table:disk_radii}
\end{table*}

\begin{figure*}[tbh!]
\includegraphics[width=1.0\textwidth]{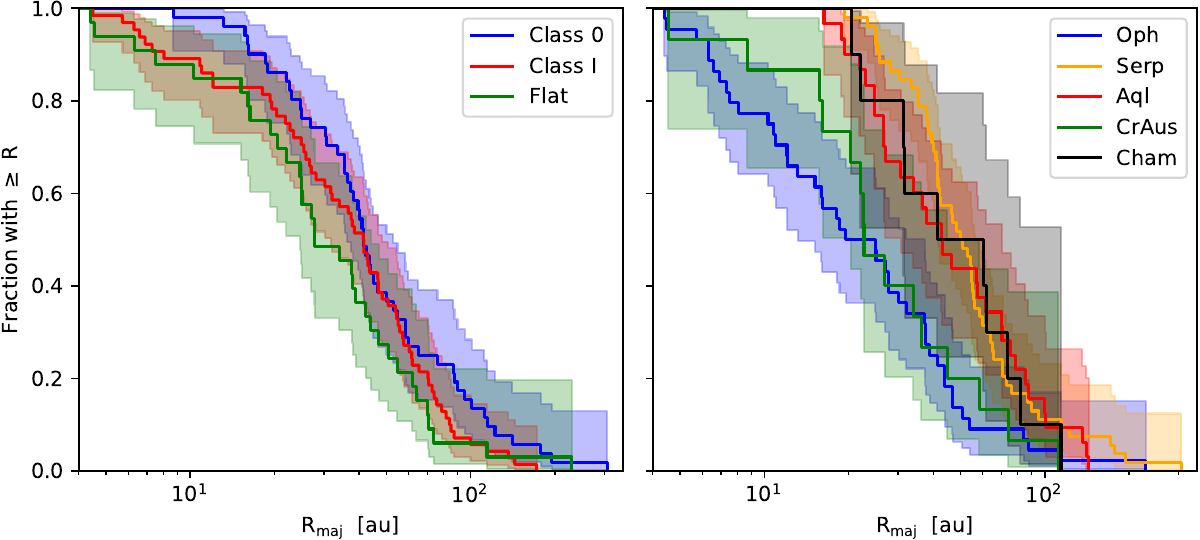}
    \caption{Cumulative distributions of the dust disk radii for the CAMPOS survey sample. The cumulative distributions reveal the fraction of {Class 0 protostellar disks are larger than Class I and flat-spectrum protostars} (left panel) and a large variation of disk radii between different molecular clouds (right panel). The blue, {red}, and {green} shaded regions mark the Class 0, Class I, and flat-spectrum sources, respectively. The shaded region marks the $1\sigma$ uncertainty. The right panel shows the cumulative distributions of dust disk radii for different clouds in the CAMPOS sample. 
    The names shown in the legend correspond to the cloud names as in \autoref{table:observation}.
    For both panels, circumbinary disks (SMM2, IRAS16293-2422 A, and VLA1623A) are excluded. 
    } 
\label{fig:R_cumulative}
\end{figure*}


We characterized the protostellar disk radii using the CASA {\it imfit} task to fit a 2D Gaussian to the continuum data. The simple Gaussian fitting enables us to quantify the size of the emission and its corresponding orientation. It is important to note that the 2D Gaussian fitting might not reflect the exact radius of the disks, potentially due to contamination with envelope material \citep{2024arXiv240112142T}. This concern is more significant for the most embedded sources and those observed with lower resolution due to their greater distances (see \S \ref{sec:Rdust_Comparison_with_theory} for a detailed discussion). However, this approach has been used in previous large surveys  (e.g, Orion: \citet{2020ApJ...890..130T}, Perseus: \citet{2016ApJ...818...73T}). Thus for easier comparison between younger protostellar disks, we adopted the same definition and methods employed in the Orion Class 0/I disk survey by \citep{2020ApJ...890..130T}. We used the $2\sigma$ size of the deconvolved major axis as disk radius. We multiplied the FWHM of the Gaussian deconvolved radius by a factor of 2/2.355 to obtain the disk radius (${\rm R}_{\rm 95}$), which includes $\sim 95\%$ of the total flux density within the fitted Gaussian\footnote{The FWHM of a Gaussian roughly equals 2.355$\sigma$.}.  


We used CASA {\it imfit} to fit an elliptical Gaussian on the three sets of images obtained with different weighting schemes. Despite the different S/N ratios and angular resolutions of these images, measurements from the three maps are consistent for most sources. Measurements from the uniform and Briggs 0.5 weighted maps are nearly identical, while we see more differences between the naturally weighted and Briggs 0.5 weighted maps, as illustrated in the comparison plots for the disk major axis (${\rm R}{\rm maj}$), minor axis (${\rm R}{\rm min}$), and position angle (PA) in \autoref{fig:Rcompare}.

In \autoref{fig:Rcompare}, the red line represents identical measurements from the two maps in each plot. For some sources with sizes close to the beam size (0\as1), we find that the naturally weighted maps tend to yield larger deconvolved major and minor axes due to the larger angular beam size. Unresolved point sources with CASA-fitted deconvolved major and minor radius sizes of 0 are excluded from this plot. Regarding the position angle, the majority of the sources fall within the gray-shaded region, representing a $\pm 10^\circ$ spread in the position angle measurements. Most of the sources with significant differences in position angle measurements have small disk sizes, comparable to the beam size.

We constructed the final disk size table by first adopting the measurements from the uniform weighted maps which have the smallest beam size. For sources 
in which the Gaussian fitting procedure failed (due to low S/N or being unresolved),
we adopted the measurements from the Briggs 0.5 weighted maps and, in cases where those were not usable, we used the results from the naturally weighted maps as a final resort.  


For sources that exhibit emission profiles that deviate significantly from a Gaussian shape, potentially due to disk substructures, edge-on configurations, or have poor CASA {\it imfit} results due to low S/N, we used the 5$\sigma$ contour in the Briggs 0.5 weighted maps to measure the radius. The sources for which we used this procedure  are ISO Oph 2a, DoAr 20, ISO-Oph 17, DoAr 25, Elias 2-24, Oph-emb 22, Elias 2-27, DoAr 29, Oph-emb-20, CFHTWIR-Oph 79, EDJ 1013,  IRAS 16459-1411, IRS 2, SMM 2, ISO-ChaI 101, ISO-ChaI 207, eHOPS-oph-20a, Ser-emb 2, Ser-emb 13, Ser-emb 6D, Ser-emb 12C, Ser-emb 15B, eHOPS-aql-86B, and MIRES G028.6593+03.8185. The measured dust disk sizes for all the sources detected are listed in \autoref{Appendix:Source_summary}.



\subsubsection{Evolution of disk radii}

Our results indicate that protostellar disks have substantially large variability in disk radii across all evolutionary classes. In \autoref{fig:R_Tbol}, we plot the disk's major axis with respect to the bolometric temperature, which is commonly used as an indicator of the protostellar evolutionary stage \citep{1993ApJ...413L..47M,1995ApJ...445..377C}, with certain caveats \citep{2013AN....334...53F}. The blue, orange, green, and red symbols represent Class 0, Class I, flat-spectrum, and Class II sources respectively. Triangular markers denote the upper limit of unresolved point sources, set at half the beam size. A notable observation is that early Class II disks tend to be larger and exhibit less variability in disk size. This phenomenon arises because our sample is only complete up to $T_{\rm bol} \sim$ 1900\,K and is incomplete for Class II. We thus omit Class II sources from further  analysis. 

\begin{figure*}[tbh!]
    \includegraphics[width=.99\textwidth]{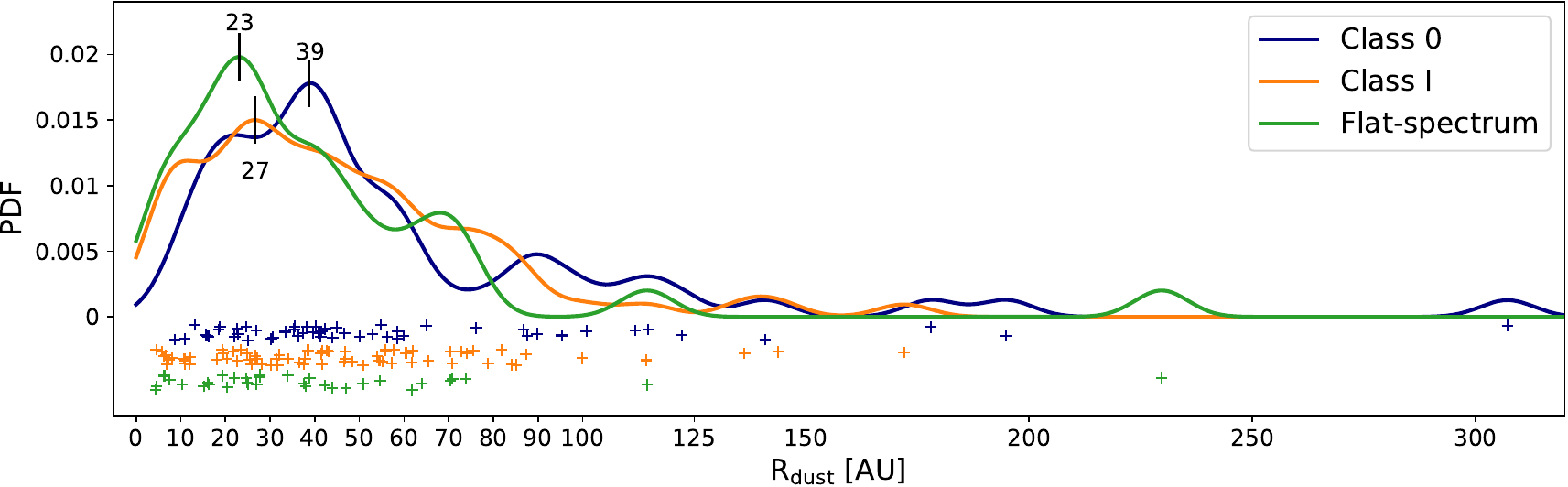}
    \caption{Dust disk radius distribution for the protostellar disks detected in the CAMPOS survey at different evolutionary stages. Note that circumbinary disks are excluded and half of the beam size is adopted for the radii of the unresolved disks. The radius distribution is computed with a Gaussian kernel density estimation. Crosses mark  measurements for each source. We also label the radius of the peak of the distribution for Class 0, Class I, and flat-spectrum sources. Overall, the peak disk dust radius decreases  from 39\,au to 23\,au (or 41\,\%) between Class 0 and the flat-spectrum phase.} 
\label{fig:Radius_KDE}
\end{figure*}

{We find Class 0 protostellar disks are larger than Class I and flat-spectrum protostars (\autoref{fig:R_cumulative}).} In \autoref{table:disk_radii}, our survey shows that the median radii for Class 0, Class I, and flat-spectrum sources are always smaller than the mean radii, indicating the presence of large disks in the population. Between Class 0 and Class I, the mean radius decrease from  59.3$\pm 7.4$\,au to 46.0$\pm 4.0$\,au, while the median radius stays roughly constant only decreasing from 41.4\,au to 39.9\,au. This suggests a significant decrease in the fraction of large disks with sizes above the median radii between Class 0 and Class I. To highlight the evolution of disk radii distribution between different evolutionray classes, we plotted the protostellar disk radius cumulative distribution in \autoref{fig:R_cumulative} and the corresponding probability density function in \autoref{fig:Radius_KDE}. In both figures, we excluded the circumbinary disks (SMM 2, IRAS16293-2422 A and VLA1623A) and Class II disks. {The plots in \autoref{fig:R_cumulative} were created by using the Python package \textit{lifeline} \citep{cameron_davidson_pilon_2019_2652543}. The left-censored fitting function of the Kaplan-Meier estimator for survival analysis was employed to account for the upper limits of unresolved sources.} In \autoref{fig:R_cumulative} (left panel), the blue, {red, and green} shaded regions represent the Class 0, Class I, and flat-spectrum sources respectively. While the three distributions overlap, a clear separation between the {Class 0 and Class I} can be seen for {all disk radii except 40--60\,au} (see \autoref{fig:R_cumulative}). The data {also} shows a decreasing trend in the fraction of large disks with sizes above {60\,au as the protostar evolves from Class 0 to Class I}. 


We computed the kernel density distribution of disk radii with a Gaussian Kernel shown in \autoref{fig:Radius_KDE} to compare the disk radii distribution between each evolutionray class. {The sizes of the unresolved point sources are adopted to be half of the beam size.}
The crosses mark the disk size measurements for each source. We also labeled the most probable radii for each class. The probability density function for each class in \autoref{fig:Radius_KDE}  show  a Gaussian-like distribution with a long tail towards large disk radii. To quantify the  distributions, we fit them with an exponentially modified Gaussian distribution (EMG), which is composed of the sum of independent normal and exponential functions, given by:
\begin{equation}
        f(x ; \mu, \sigma, \lambda)=\frac{\lambda}{2} e^{\frac{\lambda}{2}\left(2 \mu+\lambda \sigma^2-2 x\right)} {\rm erfc}\left(\frac{\mu+\lambda \sigma^2-x}{\sqrt{2} \sigma}\right),
\end{equation}
where $\mu$, $\sigma^2$ represent the Gaussian mean and variance, respectively, and $\lambda$ represents the exponential rate. We find that for Class 0: $\mu = 16.4 \pm 0.2$, $\sigma = 12.5 \pm 0.3$, and $\lambda = 0.030 \pm 0.001$; Class I: $\mu = 13.6 \pm 0.3$, $\sigma = 19.8 \pm 0.3$, and $\lambda = 0.037 \pm 0.001$; and flat-spectrum sources: $\mu = 9.2 \pm 0.2$, $\sigma = 11.7 \pm 0.3$, and $\lambda = 0.036 \pm 0.002$. Overall, the mean disk dust radius of the Gaussian component ($\mu$) and the most probable radius (marked peaks in \autoref{fig:Radius_KDE}) decrease significantly with evolutionary class. Between Class 0 and the flat-spectrum phase, the most probable radius decreases from 39\,au to 23\,au or by 41\,\%, and the Gaussian component ($\mu$) also decreases by 44\,\%. The exponential rate parameter ($\lambda$) increases as the protostar evolves, which indicates a shortening of the positive tail at  larger radius. The general increasing trend of $\lambda$ with evolutionary stage is consistent with the drop in fraction of the large disk with size above {60}\,au over time.

Lastly, we conducted the Anderson-Darling test to assess the statistical differences among the disk radii distributions of each evolutionary class.  As indicated in \autoref{table:disk_radii_stat}, we found that the likelihood of Class 0 and Class I sources, Class 0 and flat-spectrum sources, Class I and flat-spectrum sources of being drawn from the same distribution are 0.23, 0.06, and $>$0.25, respectively.  Although there is no statistical evidence suggesting Class I and flat-spectrum sources are drawn from different distributions, there is a weak statistical significance (p-value: 0.06) suggesting Class 0 and flat-spectrum sources are drawn from distinct distributions. The high p-value is consistent with \autoref{fig:R_Tbol} showing the large scatter in disk radii across all classes. 

\begin{table}[]
\centering
\caption{Dust Disk Radii Sample Comparisons (Anderson-Darling Tests)}%
\begin{tabular}{ l c c  }
\hline \hline
 Sample  & Disk Radius  & Number of  \\
 & Probability& Sources\\
 \hline
 All Sources \\
 \hline 
Class 0  vs. Class I & 0.23 & (52, 70) \\
Class 0 vs. Flat & 0.06 & (52, 33) \\
Class I vs. Flat & $>$0.25 & (70, 33) \\
\hline
Ophiuchus only& \\
\hline
Class 0  vs. Class I& 0.09 & (8, 20) \\
Class 0 vs. Flat& $>$0.25 & (8, 16) \\
Class I vs. Flat &$>$0.25 & (20, 16) \\
\hline
Serpens only & \\
\hline
Class 0  vs. Class I& 0.03 & (27, 20) \\
Class 0 vs. Flat& $>$0.25 & (27, 7) \\
Class I vs. Flat & $>$0.25 & (20, 7) \\
\hline
Aquila only & \\
\hline
Class 0  vs. Class I& 0.03 & (7, 21) \\
\hline
Class 0 only\tablenotemark{a}&\\
\hline
Aql vs. Serp& 0.004 & (7, 27) \\
Aql vs. CrAus & $>$0.25 & (7, 10) \\
Aql vs. Oph \& OphN & $>$0.25 & (7, 8) \\ 
CrAus vs. Serp & 0.002 & (10, 27) \\
CrAus vs. Oph \& OphN & $>$0.25 & (10, 8) \\
Serp vs. Oph \& OphN & 0.01 & (27, 8) \\
\hline
Class I only&\\
\hline
Aql vs. Serp& 0.13 & (21, 20) \\
Aql vs. Cham I\,\&\,II & $>$0.25 & (21, 6) \\
Aql vs. Oph \& OphN & 0.001 & (21, 20) \\
Serp vs. Cham I\,\&\,II & $>$0.25 & (20, 6) \\
Serp vs. Oph \& OphN & 0.002 & (20, 20) \\
Cham I\,\&\,II vs. Oph \& OphN & 0.02 & (6, 20) \\
\hline
flat-spectrum only&\\
\hline
Serp vs. Oph \& OphN & 0.02 & (7, 16) \\
\hline
\hline
\end{tabular}
\tablenotetext{a}{Abbreviation of cloud names are the same as in \autoref{table:observation} 
}
\label{table:disk_radii_stat}
\end{table}

\begin{figure}[tbh!]
\includegraphics[width=\columnwidth]{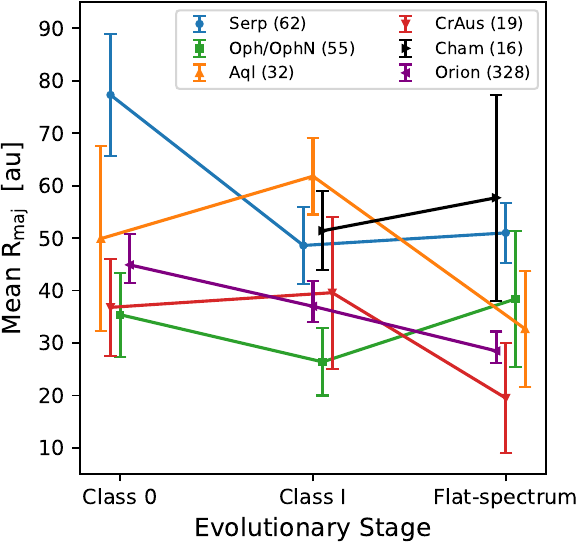}
    \caption{Evolution of mean disk major axis (Rmaj) at the different protostellar evolutionary stages in different clouds. The blue, green, orange, red, black, and purple represent sources from the Serpens, Ophiuchus, Aquila, Corona Australis, Chamaeleon, and Orion respectively. {The Orion data is from the VANDAM survey, which has a similar angular resolution (0\as1) to our data \citep{2020ApJ...890..130T}.} The number behind the labels represents the sample size. Circumbinary disks (SMM 2, IRAS16293-2422 A, and VLA1623A) are excluded. We include the deconvolved radii calculated for the marginally resolved disks. For sources without a deconvolved radius (point source), we adopted half of the beam size as their disk size. }
\label{fig:R_Class}
\end{figure}


\subsubsection{Disk radii comparison across molecular clouds}

Our data shows that even within the same evolutionray class, protostellar disks show substantially large variability in disk radii. One possible source of this large variability is  variations between sources of the same class in different clouds (See \autoref{fig:R_cumulative} right panel).
The blue, orange, red, green, and black-shaded regions represent the cumulative distribution of dust disk radii for Ophiuchus \& Ophiuchus North (hereafter Ophiuchus), Serpens, Aquila, Cor Australis, and Chamaeleon  I \& II (hereafter Chamaeleon), respectively. The mean and median dust disk radii and the log dispersion for each cloud and evolutionary stage are shown in \autoref{table:disk_radii}. Since our survey is only complete up to ${\rm T}_{\rm bol} \le 1900\,$K, we excluded the Class II sources in \autoref{fig:R_cumulative}. We find that Ophiuchus has the smallest disks followed by {Corona} Australis. In contrast, disks in Serpens, Aquila, and Chamaeleon molecular clouds are generally larger. 


We conducted the Anderson-Darling test between clouds that have more than 5 disks and found 2 distinct radii distributions at different evolutionary stages. For Class 0 disks, disk sizes in Aquila, Ophiuchus, and Corona Australis are consistent with being drawn from the same distribution. Class 0 disks in Serpens are generally larger and seem to be drawn from a different population than the other four clouds. 
For Class I disks in the Serpens, Aquila, and Chamaeleon molecular clouds, the sizes are consistent with being drawn from the same distribution. Class I disks in Ophiuchus in contrast are much smaller and appear to be drawn from a different distribution. For flat-spectrum sources, disks in Serpens and Ophiuchus molecular clouds are consistent with being drawn from different distributions. The p-values, and sample sizes are tabulated in \autoref{table:disk_radii_stat}.





Significant variation in disk radii for sources at a similar evolutionary stage, in different clouds (\autoref{table:disk_radii_stat}),  instead of a universal trend, highlights the influence of local environments on disk radii evolution. The mean radii distribution for each cloud evolves differently as illustrated in \autoref{fig:R_Class}. The error bars represent the $1\sigma$ dispersion of the disk radii distribution. The number following the name of the cloud shown in the legend indicates the sample size for that cloud. For comparison, we also included the sources from the Orion Class 0/I disk survey \citep{2020ApJ...890..130T}. From our CAMPOS survey, we find that the Serpens and Ophiuchus molecular cloud, which represents more than 60\% of our entire CAMPOS samples, show statistically significant evidence suggesting the decrease of dust disk {mean radius} between Class 0 and Class I. In contrast, for the Aquila molecular cloud, there is tentative evidence that the dust disk {mean radius} increases between Class 0 and Class I sources. Between the Class I and the flat-spectrum phase, the {mean radius} stays constant for the Serpens molecular cloud, but for the Ophiuchus molecular cloud, the disk {mean radius} increase to sizes similar to the Class 0 disks. For other clouds, the sample sizes are too small to be statistically significant. 





\subsection{Streamer Candidates}
\label{sec:streamer}

\begin{figure*}[tbh!]
    \includegraphics[width=.99\textwidth]{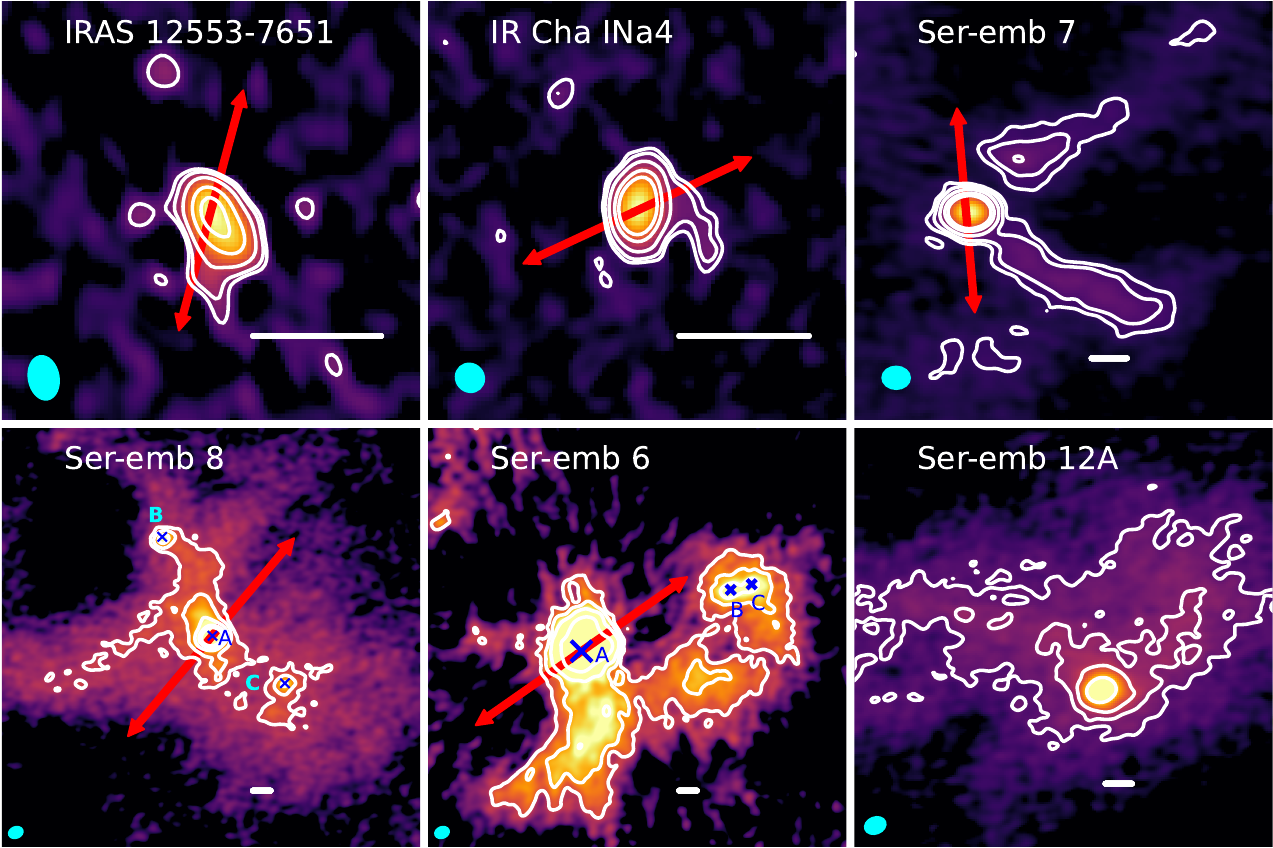}
    \caption{Possible streamers detected in our ALMA Band 6 CAMPOS dust continuum survey. The cyan-filled ellipse represents the synthesized beam size. The white line marks a scale of 100\,au. The red arrows show the direction of the protostellar outflow traced by our $1''$  resolution CO data. When multiple protostars are present in the field of view, {the positions of the protostars are marked by blue crosses with their names labeled. The white contours are at 3, 5, 10, 20, 30$\sigma$ with $\sigma=0.13, 0.09, 0.14, 0.26$\,mJy for IRAS\,12553-7651, IR Cha INa4, Ser-emb 7, Ser-emb 8, respectively. For Ser-emb 6 and Ser-emb 12, the contours are at 5, 10, 20, 60$\sigma$ with $\sigma=0.62, 0.10$\,mJy, respectively.}  Ser-emb-12A does not show a clear CO outflow at $1''$ ($\sim$420\,au) resolution. 
    } 
\label{fig:Streamer}
\end{figure*} 

Past observation of embedded protostellar envelopes (e.g., L1157, L1165, TMC-1A, Per-bolo 58 and CB230) have revealed complex velocity structures \citep{2012ApJ...748...16T,2015ApJ...812...27A,2017ApJ...849...89M,2023ApJ...944..222S}. Detailed Position-velocity (PV) diagram modeling of these envelopes shows that their velocity structures are inconsistent with an axisymmetric collapse or solid-body rotation. Some of these envelope kinematics can be well described by an infalling filament model, suggesting that the accretion in envelopes is non-axisymmetric \citep{2012ApJ...748...16T}. 

Many of the protostellar envelopes that show evidence of non-axisymmetric infall reveal flow-like structures that appear to be funneling cloud material from large scales to the inner envelope or disk. These ``accretion flows" or ``streamers" have been discovered around some Class 0 systems (Lupus3-MMS: \citet{2022ApJ...925...32T}, Per-emb-2: \citet{2020NatAs...4.1158P}, IRAS16293 binary: \citet{2022A&A...658A..53M}, VLA1623A: \citet{Marvis_Master_thesis,2020ApJ...894...23H}), as well as Class I systems like Per-emb-50 \citep{2022A&A...667A..12V}, and the SVS13A binary system \citep{2023A&A...669A.137H}). Additionally, similar features have been observed in more evolved Class I/II systems such as HL Tau \citep{2022A&A...658A.104G,2019ApJ...880...69Y}, and DR Tau \citep{2022A&A...658A.104G,2023ApJ...943..107H}.

The discovery of these accretion flows or streamers provided evidence of asymmetric accretion from the envelope. 
Streamers link the small ($\sim 10-100$\,au) disk scales  with the larger ($\sim 10,000$\,au core scales  \citep{2020NatAs...4.1158P,2022A&A...667A..12V}. The collision between the streamer material and the disk can form accretion shocks, traced by the enhancement of SO,   which can cause the outer disk to deviate from Keperian rotation \citep{2020ApJ...894...23H}. The connection between streamers and disks has not been fully explored.


Our CAMPOS continuum survey reveal extended structures around six possible ``streamer candidates" shown in \autoref{fig:Streamer}. Out of the six candidates, three of them are Class 0 systems (Ser-emb 6, Ser-emb 7, Ser-emb 8) and 3 are Class I systems (Ser-emb 12A, IRAS 12553-7651, IR Cha INa4). 
Note that these 6 sources are only candidates and line observations are needed to analyze the kinematics of these structures to determine whether these trace material moving towards the sources or outflow cavity walls.  

The emission that extends beyond the disk structure for two of the 6 sources that we labeled as being candidate streamers (IRAS 12553-7651 and IR Cha INa4) is considerably more compact than the extended structure in the other four sources. In the case of IRAS 12553-7651, a minor dust tail south of the protostellar disk aligns closely with the protostellar outflow axis, which has a position angle around $-15^\circ$. The alignment with the protostellar outflow axis suggests the dust tail might be tracing the walls of the outflow cavity rather than infalling gas. Conversely, the dusty tail in IR Cha INa4 is perpendicular to the outflow direction. However, due to the modest signal-to-noise ratio for sources in the Chamaeleon molecular cloud,
caused by fewer antennas and elevated water vapor levels relative to observations of sources in other molecular clouds,
we are not certain whether the tail is real or caused by residual artifacts from the deconvolution process. 



Ser-emb 6 system is a complex Class 0 system with 5 embedded protostars. Three of the five protostars ({marked by blue crosses}) are located within the embedded filament structure shown in \autoref{fig:Streamer}. The brightest source in the field, Ser-emb 6A, drives a clear CO outflow {marked by the red arrow} with a position angle of $\sim140^{\circ}$. 
We compared our CO outflow data and the CARMA observations by \citet{2019ApJ...871..149F} and found that the extended continuum breaks into 2 parts separated by the Ser-emb 6A protostellar outflow. The northern part includes two sources {outside the field shown in \autoref{fig:Streamer}, while the southern part is the region shown in the figure}.

Ser-emb 7, a Class 0 source with a bolometric temperature of 57 K \citep{2023ApJS..266...32P}, is surrounded by complex filamentary structures. While the ``$<$" shaped dust structure is perpendicular to the outflow direction, the protostellar outflow traced by CO is very wide and shows an ``X" shape that overlaps with the dusty structure. Thus, the dust structure might be tracing the walls of the protostellar outflow cavity rather than infalling streamers. Molecular line data is needed to determine the nature of this dust filamentary structure. In addition, the ``$<$" shaped structure is part of a much larger tilted ``K" shaped structure that extends about 4\as0 ($\sim 1800$\,au) towards the north, as revealed by CARMA observations   \citep{2019ApJ...871..149F}.


Ser-emb 8A also known as S68N \citep{1994ApJ...424..222M} is a Class 0 protostar with a bolometric temperature of 33\,K \citep{2023ApJS..266...32P}. Three protostars, including  Ser-emb 8A (brightest source), are marked by {blue} crosses in \autoref{fig:Streamer}. Another protostar, Class 0 protostar Ser-emb 8(N) \citep{1994ApJ...424..222M}, not shown in the field of view is located north of Ser-emb 8A. From the large-scale CARMA observation, an extended filament can be seen connecting all 4 protostars \citep{2019ApJ...871..149F}. Ser-emb 8 drives a molecular outflow in the southwest-northeast direction marked in red, which is perpendicular to the filament direction \citep{2014ApJS..213...13H}. 

\subsection{Annular Substructures}
\label{sec:substructures}

\begin{table}[]
\centering
\caption{Disk Substructure Table}%
\begin{tabular}{ c c c c }
\hline \hline
  Source   & $T_{\rm bol}$ & Class & Detected  \\
  Name  & (K) & & by\\
 \hline
 Circumbinary Disks\\
 \hline
VLA 1623A & 30 & 0 & (1)\\
IRAS 16293-2422 A & 31 & 0 & (2)\\
SMM 2 & 72 & I\tablenotemark{a} & This work\\
\hline
Annular Substructures\\
\hline
CFHTWIR-Oph 79 & 112 & I & eDisk\tablenotemark{b}\\
Oph-emb 22 & 224 & I & ODISEA\tablenotemark{c}\\
IRS 2 & 235 & I & This work\\
ISO-Oph 17 & 290 & II & ODISEA\\
Oph-emb-20 & 310 & Flat\tablenotemark{a} & This work \\
Elias 2-27 & 410 & Flat\tablenotemark{a} & DSHARP\tablenotemark{d}\\
ISO-ChaI 101 & 650 & Flat & This work\\
DoAr 29 &  840  & II & ODISEA\\
Elias 2-24 & 980 & II & ODISEA\\
ISO Oph 2a & 1100 & II& ODISEA \\
DoAr 20 & 1100 & II & DSHARP \\
IRAS 16459-1411 & 1400 & II & DSHARP \\
EDJ 1013 & 1500 & II & ODISEA\\
DoAr 25 & 1500 & II & ODISEA \\
\hline
Asymmetric disk \\
\hline
IRAS 16293-2422 B & 31 & 0 & (2)\\
Ser-emb 6A  & 42 & 0 & (3)
\\
eHOPS-aql-86A & 46 & 0 & (3) \\
Ass Cha T 1-15 & 1400& II & This work\\
\hline
\hline
\end{tabular}
\tablenotetext{a}{Classification updated using the latest e-HOPS survey  (see \citealt{2023ApJS..266...32P}, and Pokhrel, private communication.}
\tablenotetext{b}{Also known as GY263, observed by the eDisk survey at 5\,au resolution \citep{2023ApJ...958...20N}.}
\tablenotetext{c}{Observed by the ODISEA survey at 5\,au resolution \citep{2021MNRAS.501.2934C}.}  
\tablenotetext{d}{Observed by the DSHARP survey at 5\,au resolution \citep{2018ApJ...869L..41A}.}
\tablenotetext{}{\textbf{References:} (1) \citet{2018ApJ...861...91H} (2) \citet{2020ApJ...904..185O,2022ApJ...941L..23M} (3) \citet{2009ApJ...692..973E}.
 }
\label{table:disk_substructure}
\end{table}

\begin{figure*}[tbh!]
    \includegraphics[width=.99\textwidth]{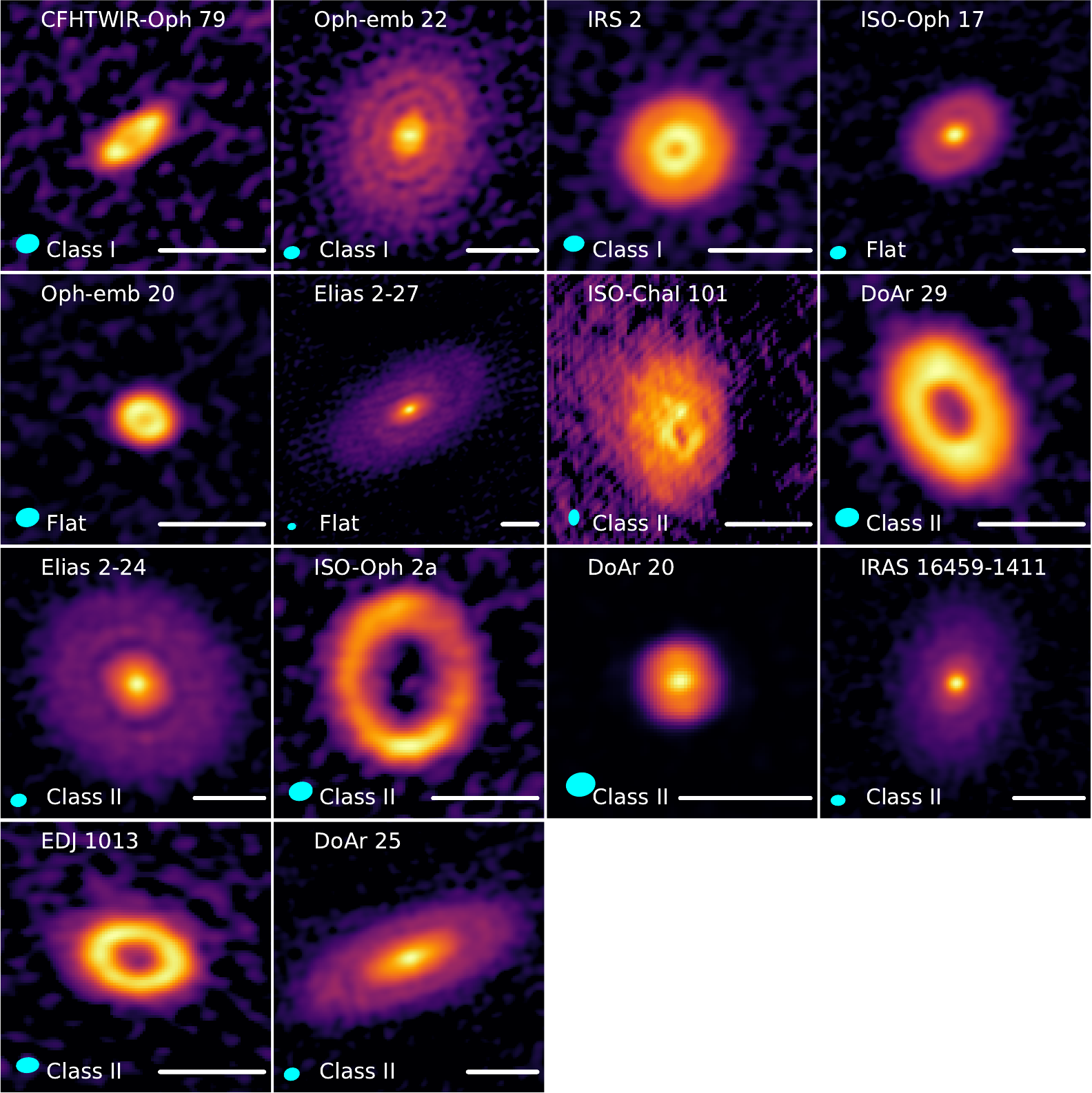}
    \caption{Sources with a clear detection of disk annular substructures or spirals in our ALMA Band 6 CAMPOS dust continuum survey. The cyan-filled ellipse represents the synthesized beam size. The white scale bar marks a scale of 100\,au. The sources are arranged in an increasing order of evolutionary sequence based on their bolometric temperature ($T_{\rm bol}$) from left to right and from top to bottom. {Their corresponding $T_{\rm bol}$ values are shown in \autoref{table:disk_substructure}. }  
    } 
\label{fig:disk_substructure}
\end{figure*}

One of the major problems in planet formation is understanding the formation and growth of planetesimals (e.g. \citet{2014prpl.conf..547J}). ALMA has been a game changer in this field with resolved observations of continuum emission directly tracing how millimeter/centimeter-sized particles are distributed in protoplanetary disks. In particular, many evolved Class II and Class III protoplanetary disks are observed with unprecedented sensitivity and resolution revealing single or multiple dust rings/gaps and spirals \citep{2016ApJ...820L..40A,2018ApJ...869L..41A,2016PhRvL.117y1101I,2016ApJ...829L..35T,2017A&A...600A..72F,2018ApJ...869...17L,2018ApJ...866L...6C,2019AJ....158...15P,2019ApJ...871....5T,2021MNRAS.501.2934C}. Interestingly, these substructures are also detected in young Class I disks ($\le 1$~Myr) (e.g. HL Tau: \citet{2015ApJ...808L...3A}, WL 17: \citet{2017ApJ...840L..12S}, Elias 42: \citet{2018MNRAS.475.5296D}, GY 91: \citet{2018ApJ...857...18S}, IRS 63: \citet{2020Natur.586..228S}). Studies have shown that gaps in these young disks ($\le 1$~Myr) are consistent with the existence of fully formed planets  \citep{2015ApJ...808L...3A,2017ApJ...851L..23C}. In addition, simulations have shown that planets with mini-Neptune-mass \citep{2019AJ....158...15P} or Earth mass \citep{2017ApJ...843..127D} can produce detectable disk substructures with the minimum gap opening mass depending on the viscosity and the scale height of the disk \citep{2006Icar..181..587C,2013ApJ...769...41D}. If yet-undetected planets are responsible for the gaps and rings observed in Class I disks, then rapid planet formation (within $1\,$Myr) is needed to reach the minimum gap opening mass (e.g., \citet{2018ApJ...857...18S,2020Natur.586..228S}). In addition, the dust masses of embedded disks are a factor of 3 to 5 times larger than Class II disks indicating that the core of the giant planets may need to begin forming during the protostellar phase \citep{2020A&A...640A..19T}. However, rapid planet formation within $\sim 1$\,Myr remains a challenge for the core accretion model \citep{1980PThPh..64..544M,1996Icar..124...62P} due to the long time scales required to form planetary cores with a mass larger than the critical mass needed for runaway gas accretion at radii $\ge 10\,$au \citep{2011ApJ...738...35K}.



While embedded protoplanets can create disk substructures, they are not the only way to form them. 
Many different mechanisms have been proposed to explain the disk substructures (See \citet{2023ASPC..534..423B} for the review), such as the sintering of dust aggregates near the snow line \citep{2015ApJ...806L...7Z}, reconnection of toroidal magnetic fields in the disk-wind system \citep{2018MNRAS.477.1239S,2019MNRAS.484..107S,2019A&A...625A.108R}, secular gravitational instability (e.g. \citealt{2011ApJ...731...99Y,2014ApJ...794...55T,2019ApJ...881...53T,2020ApJ...900..182T}), eccentric instability \citep{2021ApJ...910...79L}, irradiation instability \citep{2021ApJ...923..123W} and viscous ring-instabilities \citep{2018A&A...609A..50D}. However, the dominant mechanism behind substructure formation remains unclear, and it is debated whether the observed substructure morphology results from the combined effects of multiple processes.


While the origins of disk substructures remain debated, investigating their presence in young Class 0/I systems holds particular significance. These substructures represent localized pressure bumps, which slow or trap drifting solids and can facilitate planetesimal formation via gravitational or streaming instabilities \citep{2002ApJ...580..494Y,2005ApJ...620..459Y}. If such substructures are common in young Class 0/I disks, then planetesimals or planetary systems can be created much more efficiently \citep{2018ApJ...869L..41A}. 
Currently, the search for disk substructures in Class 0/I disks remains limited (e.g., \citealt{2020Natur.586..228S,2020ApJ...902..141S,2021MNRAS.508.2583Z,2023ApJ...951....8O}). Out of the 300 protostellar disks observed at a $\sim$40\,au spatial resolution, the Orion survey only found 7 Class 0/I protostellar disks associated with substructures \citep{2020ApJ...890..130T,2020ApJ...902..141S}. However, the detected rings and cavities have sizes larger than 50\,au, possibly representing close-separation binary formation at early times instead of planet formation \citep{2020ApJ...902..141S}. In a different approach, the eDisk collaboration conducted observations of Class 0/I disks at $\sim$7\,au (0\as04) resolution, but the sample size was limited to 19 sources \citep{2023ApJ...951....8O}. Out of the 19 sources, only 3 sources (L1489IRS, IRAS04169+2702, and OphIRS 63), show visually identified ring-like structures possibly suggesting that disk substructures are rare in the Class 0 and early Class I stage \citep{2023ApJ...951....8O}. 


Our CAMPOS survey is designed to balance between angular resolution and sample size. Among the 184 disks in our survey, 74 have been observed with a resolution of 15 au, and 16 with a resolution of 18 au, making it the most extensive high-resolution ($\le$ 18 au) and uniform search for disk substructures around young Class 0/I protostars. Our findings, including all sources with clear disk substructure detections, are presented in \autoref{fig:disk_substructure}, \autoref{fig:circumbinary_disk}, and \autoref{fig:asymmetric_disk}. Out of the 21 disks discovered with substructures or asymmetric flux distribution, 5 of them are newly discovered. Cross-matching them with the \citet{2015ApJS..220...11D} and the eHOPS catalog (Pokhrel, private communication) reveals that two of them are Class I sources, two are flat-spectrum, and one is a Class II source. A comprehensive summary of the discovered disk substructures is provided in \autoref{table:disk_substructure}. Detailed modeling of these newly discovered disk substructures will be presented in future papers.


\subsection{Circumbinary disks}
\label{sec:circumbinary_disk}

\begin{figure*}[tbh!]
    \includegraphics[width=.99\textwidth]{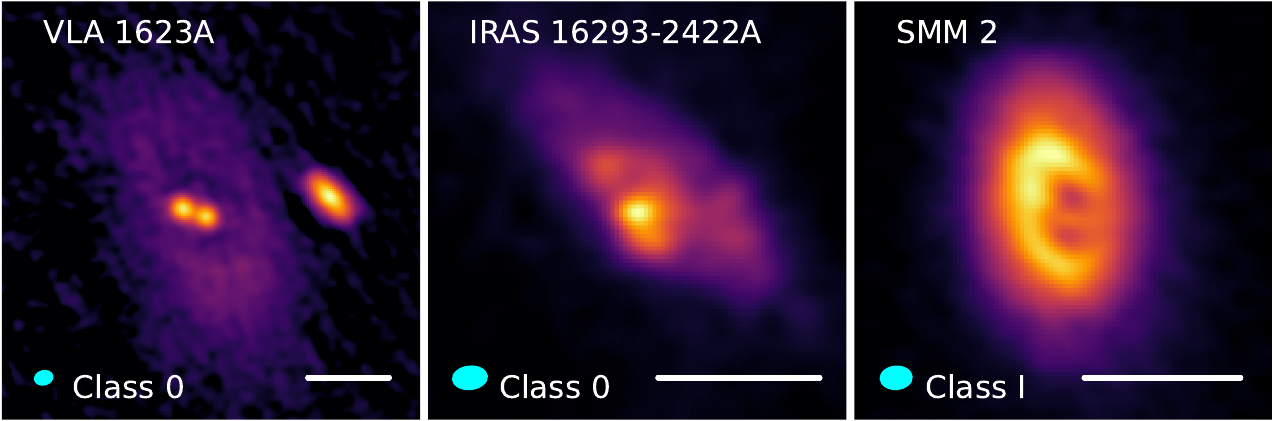}
    \caption{Circumbinary disks detected in the CAMPOS survey. The cyan-filled ellipse represents the synthesized beam size. The white line marks a scale of 100\,au.} 
\label{fig:circumbinary_disk}
\end{figure*}

In the CAMPOS survey we detected 3 circumbinary disks, VLA1623 A, IRAS 16293-2422A, and SMM 2 as shown in \autoref{fig:circumbinary_disk}. These circumbinary disks are associated with Class 0 or extremely young Class I systems. The absence of circumbinary disks around more evolved flat-spectrum sources or Class II sources suggests that the lifetime of circumbinary disks is shorter than that of circumstellar disks. In what follows, we will provide comments on the 3 young circumbinary disks detected in the survey. 

VLA 1623A is a well-known Class 0 multiple system and has been considered to be the prototype of Class 0 objects \citep{1993ApJ...406..122A,2000ApJ...529..477L,2011MNRAS.415.2812W,2013A&A...560A.103M,2015A&A...581A..91S,2018A&A...617A.120M,2020ApJ...894...23H,2022MNRAS.515..543C,2022ApJ...927...54O,2024MNRAS.528.7383C}. Recent ALMA observation by \citet{2021ApJ...912...34H} revealed misaligned twin protostellar outflows from VLA 1623A. The  axes of these two outflows are inclined by 70$^\circ$ to each other, suggesting a misalignment of a similar magnitude  between the two protostellar disks within the circumbinary system \citep{2021ApJ...912...34H}. 

In contrast, our new 0\as1 ($\sim$14\,au) resolution CAMPOS data challenges the misaligned protostellar disk interpretation proposed by \citet{2021ApJ...912...34H}. As depicted in \autoref{fig:circumbinary_disk}, the resolved circumstellar disks around VLA 1623Aa and VLA 1623Ab in the center of the circumbinary disks are aligned. We used CASA imfit to constraint their position angles to be $40.1 \pm 8.8^\circ$ and $48.0 \pm 9.7^\circ$. As for the inclination angle, assuming thin circular disks, the major-to-minor axis ratio gives inclination angles of $56.5 \pm 8.1^\circ$ and $59.0 \pm 9.0^\circ$ respectively. The larger circumbinary disk has an inclination angle of $55^\circ$ \citep{2013A&A...560A.103M}. From our data, we conclude that the individual circumstellar disks around VLA1623Aa and VLA1623Ab are co-planer with each other as well as the circumbinary disk. This alignment is a natural consequence of angular momentum conservation, with the circumbinary disk serving as the mass reservoir for the two smaller circumstellar disks. 


Given  the disks in the binary system are in fact aligned, a more natural explanation for the observed outflow misalignment   is that one of the  protostellar outflows originated from the neighboring protostar VLA 1623B \citep{2020ApJ...894...23H}. Another possible scenario is that the outflows arise form the binary components, but their axes   are not aligned with the  rotation axes of the circumstellar disks. Future higher angular resolution observations using  optically thin tracers are needed to determine the origin of the misaligned protostellar outflows.


SMM\,2 is an early Class I source with a bolometric temperature of 72\,K. With our high-resolution ALMA continuum data, we discovered the presence of two substantial holes within the SMM\,2 disk shown in \autoref{fig:circumbinary_disk}.  Higher angular resolution dust continuum data indicates that at the center of the disk lies two distinct sources connected by a dusty stream (Maureira, private communication). The SMM2 disk is therefore a protostellar  circumbinary disk.


IRAS16293-2422 A1 and IRAS16293-2422 A2 are Class 0 protostars member of a binary system that are separated by a plane-of-sky distance of 54\,au \citep{2012A&A...544L...7P,2018A&A...614A..20D,2022ApJ...941L..23M}. This system is a well-known prototypical hot corino  \citep{2016A&A...595A.117J,2016A&A...590L...6C,2018ApJ...854...96O,2019MNRAS.490...50D}. 
Three-dimensional modeling of dust and gas shows that the extended 4 $M_\odot$ envelope around IRAS16293-2422 A is connected to the nearby protostar, IRAS16293-2422 B, by a dust filament \citep{2018A&A...612A..72J}. Within the extended IRAS16293-2422 A circumbinary envelope, \citet{2022ApJ...941L..23M} discovered multiple 10\,au dust hot spots. These emission peaks do not show strong variation in the spectral index, indicating that they are due to high-temperature spots instead of optical depth variations. Detailed analysis and comparison with the emission from complex organic molecules show these substructures are due to mechanical heating from shocks \citep{2022ApJ...941L..23M}.

\subsection{Asymmetries in disks}
\label{sec:asymmetric_disk}

\begin{figure*}[tbh!]
    \includegraphics[width=.99\textwidth]{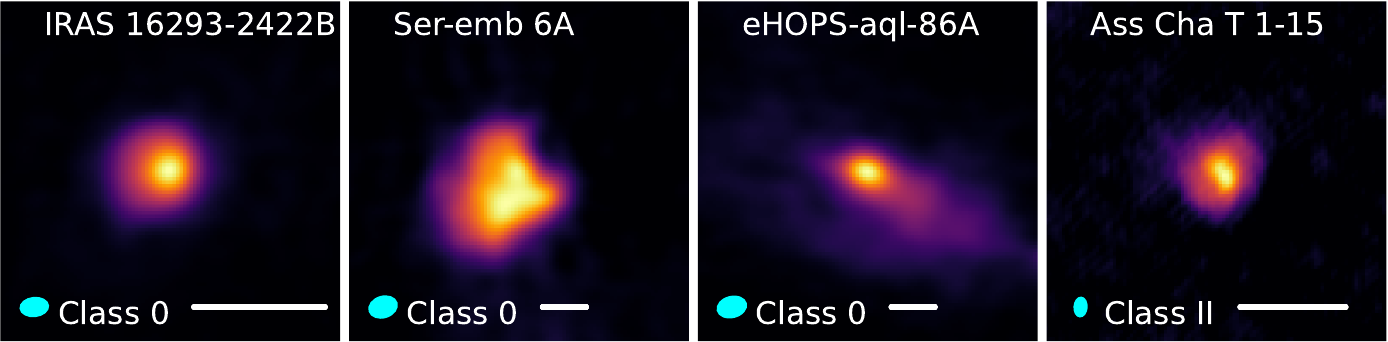}
    \caption{Asymmetric disks or embedded disk systems detected in the CAMPOS survey. The cyan-filled ellipse represents the synthesized beam size. The white line marks a scale of 100\,au.} 
\label{fig:asymmetric_disk}
\end{figure*}

The CAMPOS survey detected 4 highly nebulous sources with asymmetric flux distribution, Ass Cha T 1-15, IRAS 16293-2422B, Ser-emb 6A, and eHOPS-aql-86A shown in \autoref{fig:asymmetric_disk}. All sources except Ass Cha T 1-15 are young Class 0 protostars. Ass Cha T 1-15 also known as 2MASS J11074366-7739411 is a 1.01\,$M_\odot$ Class II M0 pre-main sequence star \citep{2013A&A...560A.100O}. The dust continuum emission of {Ass Cha T 1-15} in \autoref{fig:asymmetric_disk} has a size expected for a disk ($\sim$100\,au). It remains unclear why the disk appears asymmetric with the peak of the dust continuum shifted to one side.   

The other 3 sources all show extended dust continuum emission resembling embedded envelopes. Ser-emb 6A also known as SMM 1a \citep{1993A&A...275..195C,2017ApJ...847...92H,2019ApJ...871..149F} is the brightest Class 0 source in the Serpens Main molecular cloud.  SMM 1a powers a one-sided high-velocity molecular jet ($\sim 80$\,km\,s$^{-1}$) inside an ionized outflow cavity \citep{2016ApJ...823L..27H}. Previous ALMA observations identified this system as a Class 0 protobinary system \citep{2019ApJ...871..149F}, however from our high-resolution ALMA observation, we found this region has 5 embedded protostars. In particular, the nebulous dust emission around the intermediate mass Class 0 source, SMM 1a, was previously identified as an unusually massive disk ($\sim 1\,M_\odot$) \citep{1999ApJ...513..350H,2009ApJ...707..103E}. The source shows significant brightness variation in submillimeter wavelengths. Archival data of the James Clerk Maxwell Telescope (JCMT) Gould Belt Survey and the JCMT Transient Survey data shows that the brightness of SMM 1a increased by $\sim 2\%$ yr$^{-1}$ from 2012 to 2016 \citep{2017ApJ...849..107M}. 
Its brightness further increased by $\sim 5\%$ yr$^{-1}$ during the 18 months of the JCMT Transient Survey from 2017 December to 2018 June \citep{2018ApJ...854...31J}. The huge brightness variations and the amorphous shape of the continuum emission as seen in our high-resolution map suggest the source is deeply embedded within an infalling envelope.     

eHOPS-aql-86A also known as Ser-SMM3 is a single source Class 0 protostar first identified by \citep{1993A&A...275..195C}.  Early observation by \citet{1997ApJ...486L..59H} found molecular hydrogen knots that arise from the bow shocks of Ser-SMM3 jets. Ser-SMM3 is a well-known source for astrochemistry studies \citep{2013A&A...558A..88D,2021A&A...656A.146M,2022ApJ...934..153W,2023A&A...672A.122K}. The source is surrounded by an extended 800\,au envelope similar to IRAS 16293-2422A, and the flux peak is off-centered within the envelope. Future observation at longer (optically thin)  wavelengths and dust modeling similar to \citet{2022ApJ...941L..23M} is needed to search for possible substructures within the extended envelope. 

\subsection{Edge-on Disks}
\label{sec:edge_on_disk}

\begin{figure*}[tbh!]
    \includegraphics[width=.99\textwidth]{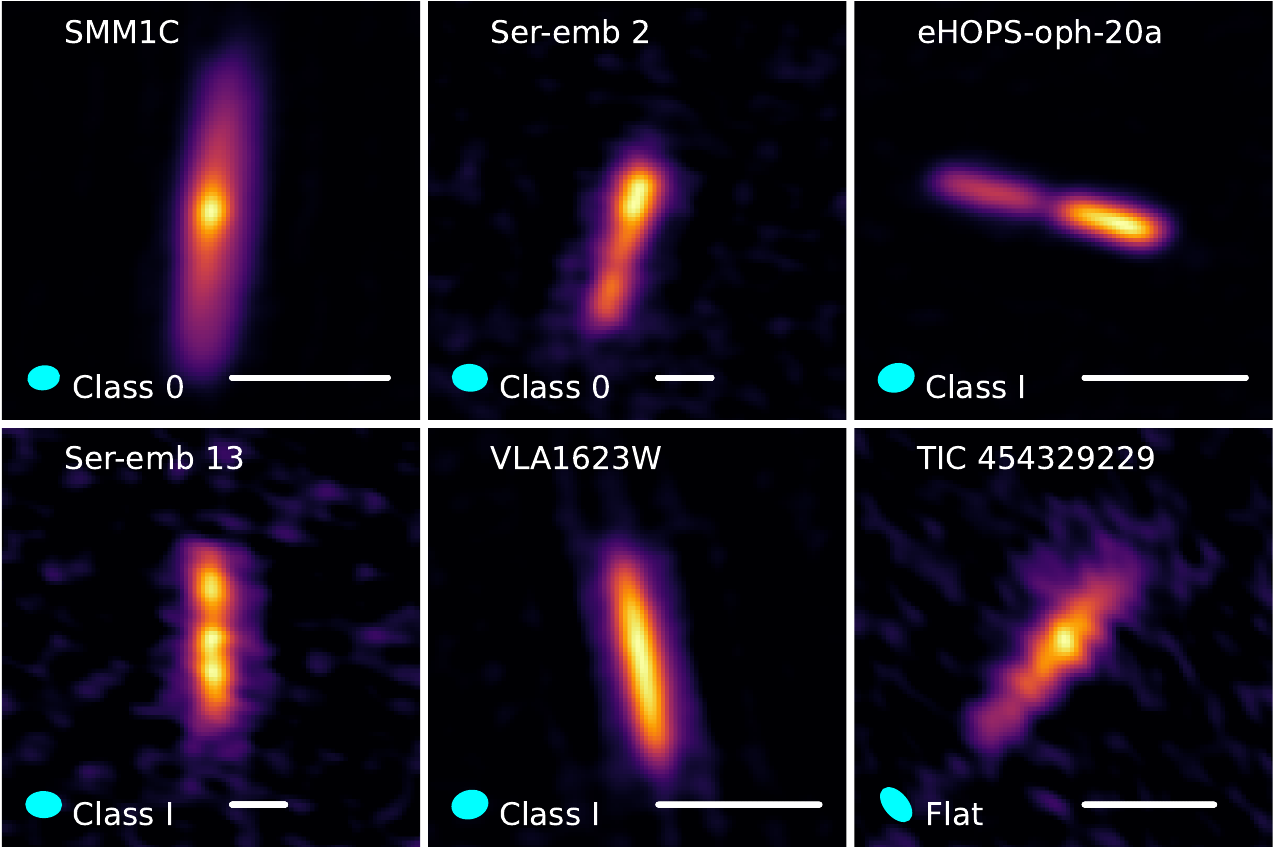}
    \caption{Edge-on disks with inclination angle greater than $75^\circ$ detected in the CAMPOS survey. The cyan-filled ellipse represents the synthesized beam size. The white line marks a scale of 100\,au. Three  disks (eHOPS-oph-20a, Ser-emb 13, and Ser-emb 2) show possible signs of substructures as evidenced by the asymmetric intensity profiles along their major axis. } 
\label{fig:Edge_on_disk}
\end{figure*} 

\begin{table}[]
\centering
\caption{Edge-on disks 
detection summary}%
\begin{tabular}{ c c c c c }
\hline \hline
 Source Name  & inc [$^\circ$] & & {Source Name} & inc [$^\circ$]\\
 \cline{1-2} \cline{4-5}
TIC 454329229	&77.9	$\pm$4.8& &SMM1C	& 77.4	$\pm$0.3	\\
VLA1623W	& 80.5	$\pm$0.7& &  eHOPS-oph-20a	& 85.0	$\pm$0.4	\\
Ser-emb 13	& 85.1	$\pm$0.4& &Ser-emb 2	&75.1 $\pm$2.5\\
\hline
\hline
\end{tabular}
\tablenotetext{}{\textbf{Note:} Column title inc represents the disk inclination angle. $90^\circ$ is edge-on and $0^\circ$ is face-on. Inclination angle derived from disk major and minor axis. The inclination angle error is derived by assuming 10\,\% error in the disk major and minor axis measurements.}
\label{table:edge-on_disk}
\end{table} 

Edge-on circumstellar disks are particularly interesting for optical and infrared surveys, due to their distinctive orientation. The disk occults the central star, enhancing contrast for scattering light imaging \citep{2014IAUS..299...99S}. This configuration allows ice signatures to be observed in absorption against the continuum produced by the warmer central disk region. This makes the edge-on disks a perfect laboratory for constraining the abundance and distribution of ice in circumstellar disks \citep{2017ApJ...834..115T,2023arXiv230502355S}. 


At longer millimeter or sub-millimeter wavelengths, edge-on disks can be used to measure the thickness of disks, trace their vertical structure, and determine the degree of dust settling in the disk mid-plane \citep{2020A&A...642A.164V,2023arXiv230502338S}. Understanding dust settling sets constraints on the vertical shear instability (VSI) of disks \citep{2022A&A...668A.105D}. 
VSI arises from the vertically sheared angular velocity profile coupled with rapid cooling of the gas. It produces upward and downward vertical streams of gas that slowly oscillate, which stirs up the dust particles in the mid-plane \citep{2022A&A...658A.156L,2022A&A...668A.105D}. Edge-on disks are the perfect targets to test for the vertical mixing of dust particles and compare them with  predictions of VSI models. Furthermore, these disks serve as optimal candidates for exploring rotational features in protostellar jets, a key element in discerning jet launching models \citep{2018ApJ...856...14L}.


Despite their significant research appeal, edge-on disks are scarce. Demographic studies of edge-on disks identified as such through their SEDs have claimed that edge-on disks are underrepresented in low-mass star-forming regions observed by Spitzer \citep{2014IAUS..299...99S,2023ApJ...945..130A}. In our CAMPOS survey of 184 disks in seven nearby molecular clouds, we discovered six edge-on disks with an inclination angle greater than $75^\circ$. The 3.3\% detection rate is much lower than the 25.9\% prediction based on simple geometric arguments. The edge-on-disk occurrence rate is also underrepresented in our resolved millimeter continuum disk survey. The lack of highly inclined disks is due to the bias of deriving the inclination angle. The inclination angle is derived from the ratio of major and minor axes of the disk, assuming a circular flat disk. Due to the disk thickness, deviations from the flat disk assumption lead to underestimations of the derived inclination angles, particularly in edge-on disks. Consequently, the derived inclination angles represent lower limits of the true inclination angles.

We list the detected edge-on disks in \autoref{table:edge-on_disk}. Here, $90^\circ$ signifies an edge-on view and $0^\circ$ denotes face-on orientation.
Among the six edge-on disks, eHOPS-oph-20a, Ser-emb 13, and Ser-emb 2, show signs of substructure that cannot be identified as either ring or spiral. eHOPS-oph-20a shows an intensity dip at the position of the protostar with one side significantly brighter than the other. This raises the possibility that eHOPS-oph-20a comprises two closely orbiting disks, potentially separated by a mere 20\,au in the plane of the sky. The true nature awaits clarification through future line observations.

Similarly, Ser-emb 2 also shows asymmetric brightness with one side much brighter than the other. Interestingly, Ser-emb 13 shows three bright spots evenly spaced along the disk's major axis. The study by \citet{2021ApJ...912..164D} shows that for an inclined thin ring with a vertical dust scale height, the major axis will appear  brighter than the minor axis, due to optical depth effects. It is possible that the three bright spots of Ser-emb 13 can be explained by an inclined ring scenario, where  the central bright spot corresponds to the central protostar. Higher resolution observations and detailed modeling are needed to determine the nature of these sources.

\section{Discussion}
\label{sec:discussion}

\subsection{Evolution of Dust disk radii: Observational Trends}
\label{Evolution_of_R}
\begin{figure*}[!htp]
\centering
\subfloat{%
  \includegraphics[width=0.48\textwidth]{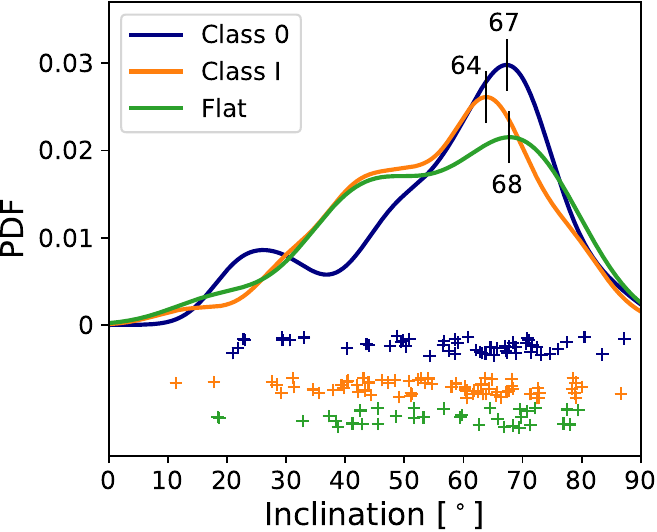}%
}\quad
\subfloat{%
  \includegraphics[width=0.48\textwidth]{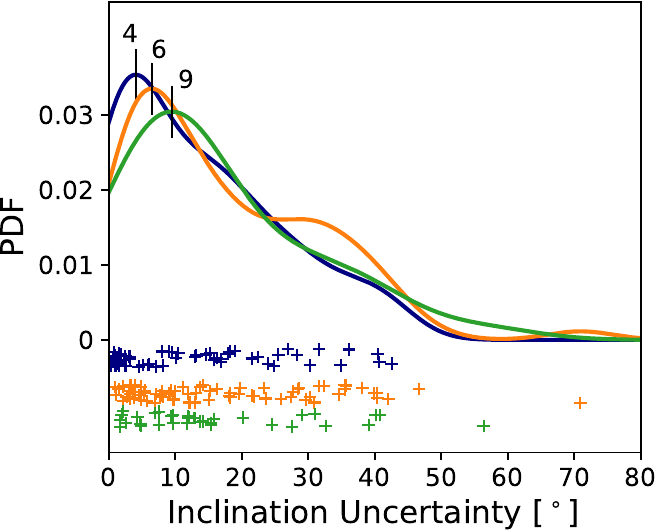}
}
\caption{\emph{Left}: Inclination angle distribution of the protostellar disks detected in the CAMPOS survey. The inclination angles are derived from the deconvolved disks major and minor axes, and assuming all are flat circular disks. The inclination distribution is computed using a kernel density estimation with a Gaussian Kernel. Note that 0$^\circ$ is face-on, and 90$^\circ$ is edge-on. The blue, orange, and green lines represent Class 0, Class I, and flat-spectrum sources, respectively. The distribution drops significantly for edge-on sources due to the intrinsic disk scale height and the breakdown of the flat disk assumption. The crosses mark the inclination angle for each source. We also labeled the peak of the distribution. \emph{Right}: The inclination angle uncertainty distribution. Considering the uncertainty (right panel), the inclination angle distribution between the Class 0, Class I, and flat-spectrum sources (left panel) are consistent with each other. } 
\label{fig:Inc_distribution}
\end{figure*}

Instead of a universal trend, our CAMPOS survey found significant variation in disk radii among the same evolutionary class between different clouds, which highlights the influence of local environments on disk radii evolution (See \autoref{table:disk_radii}). For a combined sample of disks in all 7 clouds, the Anderson-Darling Test shows that Class 0 and Class I disk radii distributions are consistent with being drawn from the same distribution. The large p-value is a result of dust disk radii varying significantly among the same classes for different molecular clouds. This suggests that the study of disk radii evolution should take into account cloud variation (\autoref{fig:R_cumulative}, right panel).



For each molecular cloud, we separated the disks by class and conducted Anderson-Darling Tests to compare the evolution of the disk radii distribution. The tests show that out of the 7 molecular clouds, only Serpens, Ophiuchus, and Aquila have statistically distinct Class 0 and Class I disk radii distribution with p-values of 0.03, 0.09, and 0.03 respectively. Between Class 0 and Class I, disk radii cumulative distributions show that the fraction of large disks with sizes above the {60 au} decreases with evolution (\autoref{fig:R_cumulative}). Separating the disks by both cloud and class shows that this trend is dominated by disks in the Serpens and Ophiuchus molecular clouds (\autoref{fig:R_Class}). The decrease in the average dust continuum radius between Class 0 and Class I is consistent with the VANDAM survey which observed more than 300 protostars in the Orion region with the same angular resolution (0\as1) as our CAMPOS survey \citep{2020ApJ...890..130T}. On the other hand, the average disk size in the  Aquila molecular cloud  seems to increase  between the Class 0 and Class I stages. 


For Class I and flat-spectrum sources, the evolutionary trend for disk radii is unclear. Anderson-Darling Tests of Serpens and Ophiuchus and Ophiuchus North molecular clouds show Class I and flat-spectrum disk radii are drawn from the same distribution. All other clouds do not have statistically significant samples. 

Conflicting evolutionary trends can be observed across different clouds. The disk average disk size in Aquila and Corona Australis decrease over time, but the average disk radius in Ophiuchus seem to increases over time. In contrast, the average disk radius in Serpens and Chamaeleon molecular clouds stay roughly constant. Interestingly, \citet{2020ApJ...895..126H} found the 58 evolved Class II/III (age of 2–3 Myr old) disks in the Chamaeleon I have a $R_{\rm 90}$ median radius of 43.1\,au, which is consistent with the Class I $R_{\rm 90}$ median radius of 42.0\,au  measured from our CAMPOS survey (note that $R_{\rm 90} \sim 0.825 R_{\rm 95}$). This implies no significant evolution of the average dust disk size in the Chamaeleon I molecular cloud. 


The different disk radii evolutionary trends between the clouds might be attributed to the  imperfect protostellar evolutionary stage tracers  ---bolometric temperature and spectral index--- used in the classification of protostars. Estimates of the bolometric temperature and spectral index are  inclination-dependent. For example, sources with a face-on disk or pole-on outflow will have outflow cavity walls along the line-of-sight of the observer, allowing more infrared emission to be detected the observer and thus classified as more evolved sources. This effect is more prominent in Class I and flat-spectrum sources, as these have significantly larger outflow cavities  compared to Class 0 sources \citep{2023ApJ...947...25H}. To test this interpretation, we plot the protostellar disk inclination distribution for different evolutionary stages in \autoref{fig:Inc_distribution}. Considering the uncertainty of the inclination angle measurements (right panel of \autoref{fig:Inc_distribution}), the inclination angle distribution for Class 0, Class I, and flat-spectrum sources are consistent with each other. The lack of inclination dependence between evolutionary stages implies that misclassification due to the inclination effects are minimal in our sample. 
Thus, the data suggests that instead of a universal trend, local cloud environments play an important role in shaping the disk size evolution. 

\subsection{Evolution of Dust disk size: Comparison with theory}
\label{sec:Rdust_Comparison_with_theory}

We found that disk size decrease between Class 0 and Class I for Serpens and Ophiuchus molecular cloud, but increase for sources in the Aquila molecular cloud.


For Aquila disks, the increase in size as protostars evolve from Class 0 to Class I is consistent with the prediction of non-magnetized hydrodynamic models with inside-out collapse. For a collapsing and rotating dense core, if magnetic fields are neglected, then centrifugal radius will grow as time cubed, $t^3$ \citep{1976ApJ...210..377U,1981Icar...48..353C,1984ApJ...286..529T}. In addition, the standard models of viscosity-driven disk evolution involving the $\alpha$-prescription also naturally predict the growth of disk radii \citep{1973A&A....24..337S}. In the context of non-ideal magnetohydrodynamics (MHD) simulations, the influence of non-ideal MHD effects (e.g. ambipolar diffusion) diminishes the magnetic braking effects over time and results in disk radii growth \citep{2014MNRAS.438.2278M,2015ApJ...810L..26T,2016A&A...587A..32M,2016ApJ...830L...8H}. Consequently, regardless of the presence or absence of magnetic fields, an increase in disk size is anticipated, which is consistent with the evolutionary trend shown in the Aquila molecular cloud. 

For the Serpens and Ophiuchus molecular cloud, the decrease in the average dust disk radius between Class 0 and Class I seems to contradict theoretical models.  However, it is important to point out that the theoretical models give predictions of gas disk sizes, not that of the dust disk measured in our observations. Dust disk radii are expected to be smaller than the gas disk radii due to the radial drift of dust grains. The decrease in the average dust disk radii in Serpens and Ophiuchus as the protostars evolve could be explained by one of the following scenarios: 
\begin{enumerate}
    \item Class 0 protostellar disks are contaminated by the envelope resulting in larger disk radii.
    \item Radial drift of dust resulting in smaller dust radii over time.
    \item Disk winds and protostellar outflows removing the angular momentum of the disk \citep{2016ApJ...818..152B}. 
\end{enumerate}

Our ALMA observations are sensitive only to compact structures with sizes smaller than 2\as0. Thus we expect minimal envelope contamination, as large envelope emission would be filtered out by our interferometric observations. In addition, if the dense inner envelope emission were to blend with the disk emission,  we would expect the intensity distribution to be more isotropic on the plane of the sky. If that were the case we would expect these source to have major to minor axes ratios of one, and thus would appear to have a face-on inclination. 
Yet, we found the distribution of inclination angles for Class 0 sources (where we would expect envelope contamination to be more prominent) is consistent with the envelope-depleted flat-spectrum sources as shown in \autoref{fig:Inc_distribution}. Both inclination distributions are nearly consistent with a random distribution of disk inclinations, and we did not detect an overabundance of face-on disks in our sample. This implies that most of the continuum structures detected in our survey are not contaminated by the isotropic dense inner envelope. 

While the inclination angle distribution rules out possible  contamination by an isotropic dense inner envelope as a cause for larger disks in more embedded sources, it does not rule out the possibility of overestimating disk sizes for young sources due to the  
contamination of asymmetric large-scale structures such as streamers, filaments, or outflow cavity walls. These sources with extended dust emission are shown in \autoref{fig:Streamer} and \autoref{fig:asymmetric_disk}. The boundaries of protostellar disks are sometimes connected with accreting streamers or protostellar outflow cavity walls. These sources with extended emission may be contaminated by accreting streamers (\autoref{fig:Streamer}) or compact inner envelopes (\autoref{fig:asymmetric_disk}). Even so, sources with extended emissions are infrequent in our sample. More importantly, Orion, Aquila, and Serpens molecular clouds are located at a similar distance and observed at similar angular resolution ($\sim 40\,au$), but show distinct disk radii distributions or evolutionary trends. This indicates that the protostellar disk emission dominates the dust continuum emission. 


Another possible explanation for the decrease of dust disk radii between Class 0 and Class I in the Ophiuchus and Serpens molecular clouds is dust radial drift. If radial drift is responsible for the decrease in dust disk radii, then its timescale should be smaller or comparable to the half-life of the Class 0 stage. We conducted an order of magnitude analysis and compared the two timescales to assess this scenario. The drift timescale ($\tau_{\mathrm{drift}}$) is given by 
\begin{equation}
\tau_{\mathrm{drift}}=\frac{\Delta r V_{\mathrm{k}}}{\mathrm{St} c_{\mathrm{s}}^2} \gamma^{-1}
\end{equation}, 
where $V_k$ is the Keplerian velocity {at the outer disk boundary of 40 au}, $\Delta r$ is the distance the dust moves due to  radial drift, $c_s$ is the sound speed, $St$ is the stokes number of the dust particle, and
\begin{equation}
\gamma=\left|\frac{\mathrm{d} \ln P}{\mathrm{~d} \ln r}\right|
\end{equation} is the absolute value of the power-law index of the gas pressure profile \citep{2012A&A...539A.148B}. Between Class 0 and Class I, the Ophiuchus mean protostellar disk radius decreases $\sim25\%$ from 35.4\,au to 26.4\,au between Class 0 and Class I sources ({see} \autoref{table:disk_radii}). Therefore, we assume $\Delta r\sim10\,$au. We estimate the Keplerian velocity at the outer disk boundary of 40\,au. For this, we assume a typical Class 0 mass of 0.1\,$M_\odot$ \citep{2021ApJ...909...11J}, which gives a Keplerian velocity of $V_k\sim1.5$\,km\,s$^{-1}$ at the radius of interest. 
Assuming an isothermal midplane temperature of 10\,K, the corresponding sound speed is 0.2\,km\,s$^{-1}$. Supposing an equation of state of $P=\rho c_s^2$, and a disk density profile of $\rho(r) \propto r^{-1}$, this gives  $\gamma \sim 1$.

The Stokes number ($St$) is given by the ratio of the particle stopping time ($t_{stop}$) to the turn-over time of the largest turbulent eddy ($t_L$). Assuming Epstein drag law, compact spherical particles, an isothermal gas density profile, and $t_L =\Omega_K^{-1}$ (Keplerian angular frequency), then the particles near the disk mid-plane have Stokes number of:
\begin{equation}
\mathrm{St}=\frac{a \rho_{\mathrm{s}}}{\Sigma_{\mathrm{g}}} \frac{\pi}{2}
\end{equation},
where $a$ is the particle radius, $\rho_s$ is the internal density of dust grains which is adopted to be 1.6\,g\,cm$^{-3}$ \citep{2012A&A...539A.148B}, and $\Sigma_g$ is the gas surface density. Our ALMA 1.3 mm continuum maps mostly trace millimeter dust grains, so we adopt $a\sim 10^{-3}$\,m. Assuming a typical protostellar disk {surface} density of $\Sigma_g\sim 10$\,g\,cm$^{-2}$ \citep{2019MNRAS.484.1574T}, we estimate $St\sim0.025$. This gives us a dust drift timescale ($\tau_{\rm drift}$) of $\sim 1.3\times10^{5}$ years for the Ophiuchus molecular cloud. This timescale is a factor of three longer than the half-life of the Class 0 stage (5\,$\times 10^4$ years) derived from the Spitzer Space Telescope “cores to disks” (c2d) and “Gould Belt” (GB) Legacy surveys \citep{2018A&A...618A.158K}.

While simulations show that radial drift is possible in protostellar disks \citep{2010A&A...513A..79B}, the lifetime for Class 0 is too short compared to the dust drift timescales. In addition, it takes time for the dust to grow from small $\sim0.1\mu m$ grains inherited from the ISM to roughly millimeter-sized grains that are affected by gas drag \citep{2019ApJ...878..116P}. Moreover, the high accretion rate in the Class 0 phase indicates that the gas and dust in the protostellar disk are constantly replenished  \citep{2020A&ARv..28....1L}, thus a much shorter dust radial drift timescale is needed to offset the dust accreted from outside the disk.

The mean radius of protostellar disks in the Serpens molecular cloud decreases $\sim 37\%$ (from 77.3\,au to 48.6\,au)  between Class 0 and Class I,  more than the decrease in disk size seen in the Ophiuchus molecular cloud.  Therefore, the decrease in dust disk size between Class 0 and Class I in the Ophiuchus and Serpens molecular clouds cannot be explained by radial drift alone.


Another possible explanation for the decrease of disk dust radii is the impact of MHD disk winds on the disk. Removal of angular momentum by a disk wind can keep the disk radii small (e.g., \citealt{2016ApJ...818..152B}), and possibly cause the disk radii to decrease with time. However, it is important to point out that such predictions are for gas radii. Thus, only gas disk radii can be tested against predictions from theory and simulations. In the next section, we will investigate the evolution of disk gas radii to disk dust radii ratio, and show that dust and gas radii decouple during the early Class I stage. Because of the decoupling of dust and gas radii at early Class I stage, it remains unclear whether or not disk winds are responsible for the decrease in disk radii between Class 0 and Class I for sources in the Serpens and Ophiuchus molecular cloud.  


\subsection{The decoupling of dust and gas disk in early Class I phase: Evolution of gas disk radius to dust disk radius ratio}
\label{sec:RgasRmm}

\begin{figure}[tbh!]
    \includegraphics[width=\columnwidth]{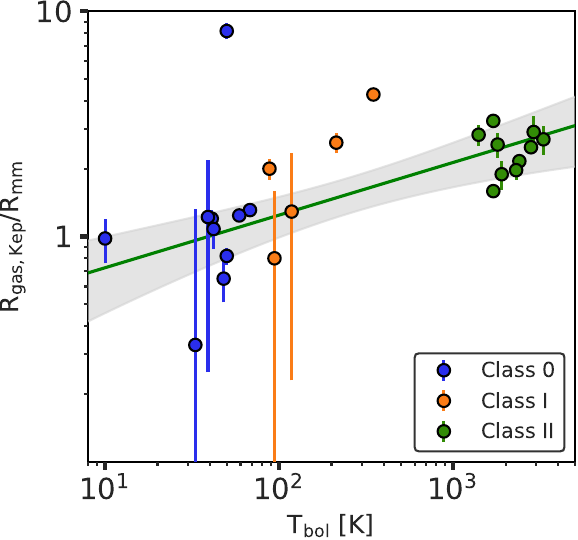}
    \caption{The evolution of the ration of gas disk radius to dust disk radius  ($R_{\rm gas,Kep}/R_{\rm mm}$) for young stellar objects as a function of evolutionary stages traced by their bolometric temperature ($T_{\rm bol}$),  using data from the literature. We separate the data into Class 0 (blue), Class I (orange), and Class II (green). {The measurements and the corresponding uncertainties are given in \autoref{table:disk_RgasRdust}.} The green line shows the best-fitted power-law Bayesian model with $y = a\times x^b$, where $a = 0.42 \pm 0.12$, and  $b = 0.23 \pm 0.01$. The gray region marks the $2\sigma$ uncertainty. Note that the Class 0 outlier, IRAS 15398-3359 (in the upper part of the plot), is excluded from the fit. {The estimated values of $R_{\rm gas,Kep}/R_{\rm mm}$  might be affected by the different methods and line tracers used to measure them. Yet, even using different methods and line tracers, $R_{\rm gas,Kep}/R_{\rm mm}\sim 1$} for Class 0 protostars. } 
\label{fig:RgasRmm_Tbol}
\end{figure}

Most measurements of disk sizes around the youngest embedded protostars come from (sub)-millimeter dust continuum observations (e.g., \citealt{2019A&A...621A..76M,2020ApJ...890..130T} or see \autoref{sec:Disk_R}). These continuum observations probe the distribution of millimeter-sized dust grains and reveal the evolution of disk solids. The dust particles might be aerodynamically decoupled from the gas and migrate inwards due to radial drift. On the other hand, molecular line observations encode important information about the disks' gaseous content evolution, especially the angular momentum transport \citep{2016ApJ...818..152B,2019FrASS...6...54P,2022MNRAS.512.2290T}). The predictions of hydrodynamic models \citep{1999ApJ...525..330Y,2005MNRAS.356.1201B,2009MNRAS.392..413S,2010ApJ...708.1585K,2018MNRAS.475.5618B}, non-ideal magnetohydrodynamical (MHD) models \citep{2006ApJ...647..374G,2008A&A...477....9H,2008ApJ...681.1356M,2009A&A...506L..29H,1984ApJ...286..529T} and the latest MHD disk-wind-driven disk accretion models \citep{2019FrASS...6...54P,2022MNRAS.512.2290T} should be compared to gas disks, which require significantly more time to observe (compared to dust continuum measurements). Besides several case studies of individual sources and a few surveys with a relatively small sample of less than 20 sources, large systematic high-resolution CO disk surveys of the youngest Class 0/I protostars remain limited (e.g., \citet{2023ApJ...951....8O}). 

Currently, the sample size of existing high-resolution CO disk surveys is too small to conduct a population study that will constrain the disk formation mechanism. In contrast, the CAMPOS survey and the Orion VANDAM survey \citep{2020ApJ...890..130T} in total cover more than 500 protostellar disks in nearby star-forming regions at 0\as1 resolution. This extensive dataset holds the key to conducting a comprehensive demographic study, shedding light on the intricate processes governing the formation and evolution of protostellar disks. It is therefore important to understand when, in disks, do millimeter-sized dust grains decouple from the gas and how reliably dust traces the gas in disks in order to determine how protostellar disks form and evolve. 


Detailed comparisons between the gas disk radii and dust disk radii have been carried out for more evolved Class II disks \citep{2022ApJ...931....6L}, however, this has never been conducted for younger Class 0/I sources. As a result, the evolutionary trend of how the gas disk radii to dust disk radii ratio evolves has not been studied. We compiled the existing literature for disks with both a measurement of the gas disk radius, defined by the Keplerian rotation radius ($R_{\rm gas,Kep}$), and the dust disk radius ($R_{\rm mm}$) measured by the millimeter continuum data in \autoref{table:disk_RgasRdust}. We list both the methods and lines used to determine the radii in \autoref{table:disk_RgasRdust}. If the disk radii uncertainty is not given in the literature, we adopted half of the beam size as the uncertainty in the disk radius measurement. We cross-matched the sources with \citet{2015ApJS..220...11D} to obtain their bolometric temperature ($T_{\rm bol}$), which is an indicator of the protostellar age \citep{1995ApJ...445..377C}.

\begin{table*}[]
\centering
\caption{Disk properties from literature}%
\begin{tabular}{ c c c c c c c}
\hline \hline
 Source & $T_{\rm bol}$    & $R_{\rm gas,Kep}$ & $R_{\rm mm}$ & ratio &Ref & Note \\
 ID & (K) & (au)& (au) &\\
 \hline 
L1489 IRS&213&600 $\pm$ 62 &230 $\pm$ 3 &2.61 $\pm$ 0.27
& 1, 2 & $R_{\rm gas}$: C$^{18}$O PV fit\tablenotemark{a}, $R_{\rm mm}$: Visibility fit \\
IRAS 04302+2247&88&620 $\pm$ 62&310 $\pm$ 10&2.00 $\pm$ 0.21& 3 &$R_{\rm gas}$: By eye, $R_{\rm mm}$: By eye\\ 
L1527 IRS&41&102 $\pm$ 6&85 $\pm$ 6&1.20 $\pm$ 0.11 & 4, 5 & $R_{\rm gas}$: C$^{18}$O PV fit, $R_{\rm mm}$: 5$\sigma$ contour
\\
CB 68&50&25.7 $\pm$ 0.5&31.3 $\pm$ 2.5&0.82 $\pm$ 0.07& 6 &$R_{\rm gas}$: C$^{18}$O PV fit, $R_{\rm mm}$: CASA imfit\\
Ced110 IRS4A&68&120 $\pm$ 5&91.7 $\pm$ 0.2&1.31 $\pm$ 0.05 & 7 & $R_{\rm gas}$: C$^{18}$O PV fit, $R_{\rm mm}$: COG\tablenotemark{b}\\
IRS5N&59&77.0 $\pm$ 2.3&62.0 $\pm$ 0.5&1.24 $\pm$ 0.04 & 8 & $R_{\rm gas}$: C$^{18}$O PV fit, $R_{\rm mm}$: CASA imfit\\
IRAS 16253-2429&42&16 $\pm$ 3&14.9 $\pm$ 0.3&1.08 $\pm$ 0.20 & 9 & $R_{\rm gas}$: UCM\tablenotemark{c}, $R_{\rm mm}$: CASA imfit\\
IRAS 15398-3359&50&31.2 $\pm$ 2.4&3.82 $\pm$ 0.09&8.17 $\pm$ 0.66 & 10 & $R_{\rm gas}$: SO PV fit, $R_{\rm mm}$: CASA imfit\\
Oph IRS63&348&265 $\pm$ 9&62 $\pm$ 2&4.27 $\pm$ 0.21 & 11 &  $R_{\rm gas}$: $^{13}$CO PV fit, $R_{\rm mm}$: GoFish\tablenotemark{d} fit\\
VLA1623A CB&10&210 $\pm$ 45\tablenotemark{e}&215 $\pm$ 15&0.98 $\pm$ 0.22 & 12, 13 & $R_{\rm gas}$: C$^{18}$O PV fit, $R_{\rm mm}$: By eye\\
HH211&33&10 $\pm$ 30 &30 $\pm$ 6&0.33 $\pm$ 1.00& 14  &$R_{\rm gas}$: SO disk\tablenotemark{f}, $R_{\rm mm}$: CASA imfit\\
HH212&48&44 $\pm$ 9\tablenotemark{g}&68 $\pm$ 4&0.65 $\pm$ 0.14 & 15 & $R_{\rm gas}$: HCO$^+$ PV fit, $R_{\rm mm}$: Intensity profile\tablenotemark{h}\\
Lupus 3 MMS&39&100 $\pm$ 53&82 $\pm$ 49&1.22 $\pm$ 0.97 & 16 & $R_{\rm gas}$: C$^{18}$O PV fit, $R_{\rm mm}$: CASA UVmodelfit \\
TMC-1A&118&90 $\pm$ 74&70 $\pm$ 4&1.29 $\pm$ 1.06 & 17 &$R_{\rm gas}$: C$^{18}$O PV fit, $R_{\rm mm}$: CASA imfit \\
L1551 IRS 5&94&64 $\pm$ 64 &80 $\pm$ 5&0.80 $\pm$ 0.80 & 18 &$R_{\rm gas}$: CS PV fit, $R_{\rm mm}$: CASA imfit \\
Sz 65&3300&178 $\pm$ 25 &66 $\pm$ 2&2.70 $\pm$ 0.39 & 19 & $R_{\rm gas}$: CO Moment 0, $R_{\rm mm}$: COG\tablenotemark{i}\\
Sz 91&2900&358 $\pm$ 64&123 $\pm$ 3&2.91 $\pm$ 0.53& 19 & $R_{\rm gas}$: CO Moment 0, $R_{\rm mm}$: COG\\
Sz 98&1900&279 $\pm$ 41&148 $\pm$ 3&1.89 $\pm$ 0.28& 19 & $R_{\rm gas}$: CO Moment 0, $R_{\rm mm}$: COG\\
Sz 100&2400&121 $\pm$ 8 &56 $\pm$ 2&2.16 $\pm$ 0.16 & 19 & $R_{\rm gas}$: CO Moment 0, $R_{\rm mm}$: COG\\
Sz 123A&2300&118 $\pm$ 10 &60 $\pm$ 2&1.97 $\pm$ 0.18& 19 & $R_{\rm gas}$: CO Moment 0, $R_{\rm mm}$: COG\\
MY Lup&2800&192 $\pm$ 7 &77 $\pm$ 3&2.49 $\pm$ 0.13 & 20 & $R_{\rm gas}$: CO Moment 0, $R_{\rm mm}$: COG\\
SR 4&1400&82 $\pm$ 7 &29 $\pm$ 2&2.83 $\pm$ 0.31 & 20 & $R_{\rm gas}$: CO Moment 0, $R_{\rm mm}$: COG\\
DoAr 25&1700&233 $\pm$ 6 &147 $\pm$ 2&1.59 $\pm$ 0.05& 20 &$R_{\rm gas}$: CO Moment 0, $R_{\rm mm}$: COG\\
DoAr 33&1800&64 $\pm$ 6 &25 $\pm$ 2&2.56 $\pm$ 0.32 & 20 &$R_{\rm gas}$: CO Moment 0, $R_{\rm mm}$: COG\\
WaOph 6&1700&297 $\pm$ 7 &91 $\pm$ 4&3.26 $\pm$ 0.16 & 20 &$R_{\rm gas}$: CO Moment 0, $R_{\rm mm}$: COG\\
\hline
\hline
\end{tabular}
\tablenotetext{}{\textbf{References:} (1) \citet{2020ApJ...893...51S} (2) \citet{2023ApJ...951...11Y} (3) \citet{2023ApJ...951....9L} (4) \citet{2023ApJ...951...10V} (5) \citet{2012Natur.492...83T} (6) \citet{2023arXiv230615443K}(7) \citet{2023arXiv230708952S} (8) \citet{2023ApJ...954...69S} (9) \citet{2023ApJ...954..101A} (10) \citet{2023ApJ...958...60T} (11) \citet{2023ApJ...951...11Y} 
(12)\citet{2013A&A...560A.103M} (13) \citet{2018ApJ...859..165S} (14) \citet{2018ApJ...863...94L} (15) \citet{2017ApJ...843...27L} (16) \citet{2017ApJ...834..178Y} (17) \citet{2015ApJ...812...27A} (18)  \citet{2014ApJ...796...70C} (19) \citet{2018ApJ...859...21A} (20) \citet{2018ApJ...869L..41A}}  
\tablenotetext{a}{Gas disk radius derived using the Position-velocity (PV) diagram method. Here Keplerian rotation ($v\propto r^{-0.5}$) is fitted to the PV diagram. The gas disk radius is determined by the radius at which the velocity profile changes from $v\propto r^{-0.5}$ to $v\propto r^{-1}$ (consistent with an infalling envelope with conservation of angular momentum).}
\tablenotetext{b}{Radius derived from the Curve-of-growth (COG) method by \citet{2016ApJ...828...46A}.} 
\tablenotetext{c}{Determined from the disk centrifugal radius from the UCM envelope model \citep{1976ApJ...210..377U,1981Icar...48..353C} with the specific angular momentum of $j\sim45\,$km\,s$^{-1}$\,au \citep{2019ApJ...871..100H}, and central stellar mass of 0.14\,$M_\odot$.} 
\tablenotetext{d}{2$\sigma$ radius from the GoFish fit code developed by \citet{2019JOSS....4.1632T}.}
\tablenotetext{e}{\citet{2013A&A...560A.103M} reported a Keplerian rotation radius of 180\,au assuming a distance of 120\,pc. \citet{2018ApJ...859..165S} also report a dust radius of 180\,au assuming the same 120\,pc distance. We adjusted the radius by using the latest Gaia distance estimate for Ophiuchus (144\,pc) \citep{2019ApJ...879..125Z}.}
\tablenotetext{f}{The radius of the rotating disk atmosphere detected in SO. SO PV modeling was shown in Figure 4 of \citet{2018ApJ...863...94L}, but a break-point between infall and Keplerian rotation is not detected. We assume a factor of 3 error in the derived gas disk radii due to the lack of break-point between infall and Keplerian rotation.} 
\tablenotetext{g}{Centrifugal Barrier is adopted to be the gas disk radius.}
\tablenotetext{h}{Half width half maximum of the intensity profile.}
\tablenotetext{i}{For all the Class II sources with $T_{\rm bol} > 1400$\,K, the sources are from Table 2 of \citet{2022ApJ...931....6L}.
}
\label{table:disk_RgasRdust}
\end{table*}

We found that the gas disk radius to dust disk radius ratio ($R_{\rm gas,Kep}/R_{\rm mm}$) increases as a protostar evolves (see \autoref{fig:RgasRmm_Tbol}). We separated the data into Class 0 (blue), Class I (orange), and Class II (green). Excluding the outlier IRAS 15398-3359, the Class 0 source with the highest $R_{\rm gas,Kep}/R_{\rm mm}$ ratio, we found the average $R_{\rm gas,Kep}/R_{\rm mm}$ for Class 0, Class I, and Class II are $1.0 \pm 0.3$, $2.2 \pm 1.2$, and $2.4 \pm 0.5$, respectively. To quantify the relationship between $R_{\rm gas,Kep}/R_{\rm mm}$ and $T_{\rm bol}$, we adopted a simple Bayesian power-law model and used the \verb+emcee+ package \citep{2013PASP..125..306F} to draw 2,000 posterior samples in 50 sampling chains to conduct the Monte Carlo Markov chain (MCMC) modeling. We set the burn-in period to be 1,000 and from the MCMC modeling we found:
\begin{equation}
    \frac{R_{\rm gas,Kep}}{R_{\rm mm}} = (0.4 \pm 0.1)\times \left( \frac{T_{\rm bol}}{\rm K}\right)^{0.23 \pm 0.01},
    \label{eq:RgasRmm_Tbol}
\end{equation}
shown as the green solid line in \autoref{fig:RgasRmm_Tbol}. The gray-shaded region represents the $2\sigma$ uncertainty.

Note that we strongly caution that the evolutionary trend shown in \autoref{fig:RgasRmm_Tbol} is not robust due to different angular resolutions, different line tracers, and a variety of methods used to obtain $R_{\rm gas,Kep}$ and $R_{\rm mm}$  (\autoref{table:disk_RgasRdust}). For example, the Keplerian disks of Lupus 3 MMS, TMC-1A, and L1551 IRS 5 are not well resolved, and this resulted in higher uncertainties in the gas disk radius measurements \citep{2017ApJ...834..178Y,2015ApJ...812...27A,2014ApJ...796...70C}. In addition, we noticed that gas-to-dust disk radii ratios derived from SO significantly deviate from the results derived from C$^{18}$O. The Class 0 sources IRAS 15398-3359 and HH211 have gas-to-dust disk radii ratios of 8.2 and 0.3 respectively. This significantly deviates from the Class 0 average of 1. This suggests that SO, a well-known shock tracer, is not a good line tracer for gas disk radius. Furthermore, the Class II gas radii traced by the integrated intensity map (Moment 0) of CO might be more extended than the Keperlian rotation modeling with C$^{18}$O. As a result, we do not know whether the trend in $R_{\rm gas,Kep}/R_{\rm mm}-T_{\rm bol}$ seen in \autoref{fig:RgasRmm_Tbol} between Class I and Class II is due to the evolution of disk radii or the systematic error from using different methods and tracers. 

In contrast to  Class II sources, most Class 0/I gas disk radii in \autoref{fig:RgasRmm_Tbol} are traced by the same method (Keplerian rotation modeling) and the same molecular line tracer (C$^{18}$O). The Class 0/I gas radius is defined as the transition radius between  Keplerian rotation and infall in the C$^{18}$O position-velocity diagram (PV). 
The current data suggests that for Class 0 protostellar disks, the C$^{18}$O Keplerian rotating gas disk is comparable in size to the millimeter dust disk. This allows a direct comparison of our millimeter dust disk observation with different hydrodynamical and non-ideal MHD disk formation models to constrain the main mechanisms behind the Class 0 disk formation. 

Class 0 sources have gas disk to dust disk radii ratio much smaller than early Class I protostellar disks with bolometric temperature between 70 and 400\,K ($R_{\rm gas,Kep}/R_{\rm mm}\sim 2.2$). This implies that the dust and gas disk radii rapidly decouple around early Class I. The evolution of $R_{\rm gas,Kep}/R_{\rm mm}$ ratio cannot be explained by the evolution of dust disk radiii alone. The $R_{\rm gas,Kep}/R_{\rm mm}$ ratio increases by a factor of 2.2 between Class 0 and the early Class I, while the average dust disk radius only decrease by $\sim$30\,\% (e.g., in the Ophiuchus molecular cloud). Nevertheless, the rapid evolution of $R_{\rm gas,Kep}/R_{\rm mm}$ and $R_{\rm mm}$ during the early Class I phase highlights an important timescale for the decoupling of gas disk radius and dust disk radius. The underlying mechanism for setting this timescale is unclear and should be explored by future studies.

\subsection{Class 0 protostellar disks are small}
\label{Size_distribution_disk}

Our CAMPOS survey found that for the Aquila, Corona Australis, Ophiuchus North, Ophiuchus, and Serpens molecular clouds, the median Class 0 dust disk radius is 41.5\,au, which is consistent with the median Class 0 dust disk radius of 48.1\,au in the Orion molecular cloud \citep{2020ApJ...890..130T}.  Together with our CAMPOS survey, a large majority of Class 0 disks in at least 8 molecular clouds (Aquila, Corona Australia, Ophiuchus North, Ophiuchus, Serpens, Orion A, Orion B, and Perseus) have been fully sampled\footnote{Based on the \citet{2015ApJS..220...11D} and \citet{2016ApJS..224....5F} catalog.} \citep{2018ApJ...867...43T,2020ApJ...890..130T}. It is now well established that Class 0 disks 
have a typical disk radius of 40--50\,au \citep{2018ApJ...867...43T}. 


While there is a consensus that the majority of Class 0 disks are small with disk radii of 40--50\,au, it remains debated whether or not the distribution is skewed with a long positive tail towards larger disks. The CALYPSO survey of 16 Class 0 protostars shows that less than 25\,\% of Class 0 disks are larger than 60\,au \citep{2019A&A...621A..76M}, and from this, the authors concluded that large Class 0 disks are rare. In contrast, the Orion protostellar disk survey shows that 46\% of Class 0 protostars have dust radius greater than 50\,au, and argues that Class 0 disks larger than 50\,au are not rare \citep{2020ApJ...890..130T}. From our CAMPOS survey, we found that out of 52 Class 0 circumstellar disks in the 7 nearby clouds, only 14 disks (27\,\%) are larger than 60\,au, this is in strong agreement with \citet{2019A&A...621A..76M}. Interestingly, 21 disks (40\,\%) are larger than 50\,au, this is also in agreement with \citet{2020ApJ...890..130T}. From our CAMPOS survey, the distribution of Class 0 dust radii shows the existence of a long tail towards  larger radii as shown in \autoref{fig:Radius_KDE}. 

\subsection{The formation of protostellar disks: Tension with hydrodynamic, Hall effect and ambipolar diffusion models}
\label{sec:Disk_formation_theory}

The study of gas disk radii offers a robust test for different disk formation models. From our results, we see that Class 0 gas disk radii can be traced by the dust disk radii ($R_{\rm gas,Kep}/R_{\rm mm}\sim 1$), allowing a direct comparison between our observational data and the model predictions (see \autoref{sec:RgasRmm}). Hydrodynamical models of infalling rotating cores with angular momentum conservation predict that rotation-supported disks can quickly grow to sizes larger than $100\,$au in only a few thousand years \citep{1999ApJ...525..330Y,2003ApJ...595..913M,2005MNRAS.356.1201B,2009MNRAS.392..413S,2010ApJ...708.1585K,2018MNRAS.475.5618B}. These hydrodynamical models are in strong tension with the much smaller $\sim$40\,au Class 0 disk radii found in our CAMPOS survey. In order to form a 50\,au disk from a uniformly rotating core with a size of 0.1\,pc, density profile ($\rho \propto r^{-2}$), and $\beta = 0.02$ \citep{1993ApJ...406..528G}, more than 75\,\% of the initial angular momentum needs to be removed \citep{2019A&A...621A..76M}. Without an additional mechanism to remove initial core angular momentum, disks formed in infalling rotating cores are expected to be a factor of 2.5 times larger than the  Class 0 disks typically observed.  

Currently, the only known mechanism that can effectively remove angular momentum in infalling rotating cores is magnetic torques \citep{2014prpl.conf..173L}. In particular, the evolution of magnetic fields is determined by non-ideal effects like ambipolar diffusion. \citet{2016ApJ...830L...8H} developed an analytical model for the protostellar disk radius. 
Assuming the magnetic braking timescale is similar to the rotation timescale, and a Keplerian rotating disk in vertical hydrostatic equilibrium, then
according to \citet{2016ApJ...830L...8H}
 the disk radius ($r_{\mathrm{d}, \mathrm{AD}}$) can be written as:
\begin{equation}
\begin{aligned}
r_{\mathrm{d}, \mathrm{AD}} \simeq & 18 \mathrm{au} \\
& \times \delta^{2 / 9}\left(\frac{\eta_{\mathrm{AD}}}{0.1 \mathrm{~s}}\right)^{2 / 9}\left(\frac{B_z}{0.1 \mathrm{G}}\right)^{-4 / 9}\left(\frac{M_{\mathrm{d}}+M_*}{0.1 M_{\odot}}\right)^{1 / 3},
\end{aligned}
\end{equation}
where $\eta_{\mathrm{AD}}$ is the ambipolar diffusion timescale, $\delta$ is a scaling factor on the order of a few, {$B_{\rm z}$ is the projected core magnetic field strength parallel to the disk rotation axis}, and $M_{\mathrm{d}}+M_*$ is the sum of the disk and the protostar mass. This model predicts {a small disk size of $\sim20$\,au and} a weak disk radius dependence on all parameters, suggesting that even in environments with different magnetic field strengths or protostars with different masses, Class 0 disks should not vary much in size. In contrast, the observed Class 0 disk size distribution has a large variation in disk radii and the {average disk radius} is a factor of 2.2 times larger than the ambipolar diffusion {models} \citep{2016ApJ...830L...8H,2021ApJ...922...36L}.

{The disk size difference between the ambipolar models and observation could be explained by a weaker magnetic field. \citet{2024A&A...682A..30L} shows that variations in the mass-to-flux ratio significantly influence disk size. The study finds that ambipolar diffusion models with a mass-to-flux ratio of 10  reproduce better the observed disk sizes. Another} possible explanation for the disk size difference is the lack of small dust grains due to the grain coagulation process \citep{2012MNRAS.422.1263H}. Small grains ($< 0.1 \mu m$) are the main charge carriers, responsible for the coupling of neutral matter to the magnetic field at densities below $10^{10}$\,cm$^{-3}$ \citep{2016ApJ...830L...8H}. The depletion of very small grains with size 10-100\,\r{A} can enhance the ambipolar diffusion by 1 to 2 orders of magnitude and trigger the early formation of larger rotationally supported disks \citep{2018MNRAS.473.4868Z}. 
{The large uncertainty in the dust size distribution and its corresponding ambipolar diffusivity coefficient is one of the major difficulties in predicting disk sizes from ambipolar diffusion models \citep{2020A&A...641A..39S,2020A&A...643A..17G,2023MNRAS.518.3326L}. Instead of assuming an unknown dust size distribution and ambipolar diffusivity coefficient to predict disk size, an alternative approach is to work backward. Starting from observation and assuming the disk radius is set by ambipolar diffusion,  \citet{2024ApJ...968...26T} derived the first estimation of the ambipolar diffusivity coefficient for a Class 0 protostellar disk.}



{Ambipolar diffusion is not the only non-ideal MHD process that can influence disk sizes during their formation. In particular, the Hall effect predicts a bimodal distribution of Class 0 disk radii \citep{2015ApJ...810L..26T}.} When the disk rotation vector and the disk magnetic field are anti-parallel, the Hall effect weakens the magnetic braking effect and allows a 20\,au disk to form. Likewise, when they are parallel, the magnetic braking effect is enhanced, resulting in a very small ($<$1\,au) disk \citep{2015ApJ...810L..26T}. Given that the Hall effect does not play a role during cloud core formation \citep{2004Ap&SS.292..317W}, the probability of having an anti-parallel versus a parallel configuration should be the same. Thus the Hall effect predicts bimodality in Class 0 disk sizes \citep{2015ApJ...810L..26T}. If the Hall effect were an important factor in disk formation, then we would expect half of the Class 0 disks to have radii of less than 1\,au, thus unresolved. In our CAMPOS results, only 1 out of 52 Class 0 sources are unresolved and we do not see a bimodal distribution of Class 0 disk radii. The disk around the Class 0 source B335, which has a disk radius much smaller than 10\,au \citep{2015ApJ...812..129Y,2019A&A...631A..64B} is an exception rather than the norm. Recent analytical and numerical studies by \citet{2020SSRv..216...43Z} and \citet{2021ApJ...922...36L} show that the bimodal distribution from the Hall effect is transient and only lasts a few thousand years. Our results {show that the Hall effect} is not the dominant factor in protostellar disk formation. {They are either inconsistent  with the bimodal distribution of Class 0 disk sizes predicted by the Hall effect or indicate that the Hall effect is transient and therefore not observable.} Furthermore, even in the anti-parallel case, {a disk radius of 20\,au is still a factor of two smaller than the observed average disk radius in Class 0 sources of 40\,au.} Thus, our results are in high tension with the Hall effect model \citep{2015ApJ...810L..26T}.  

In summary, our work reveals that protostellar disk formation is magnetically regulated. Hydrodynamical models without consideration of magnetic fields cannot explain the disk radii distribution of Class 0 protostars. Our observations are in high tension with the bimodal disk size distribution predicted by the Hall effect \citep{2015ApJ...810L..26T}. The large spread of Class 0 disk radii centered at 40\,au is also inconsistent with the small {average} disk radius (18\,au) and the small variation of disk radii predicted by models that include ambipolar diffusion (e.g., \citealt{2016ApJ...830L...8H,2021ApJ...922...36L}). Additional parameters such as turbulence {\citep{2015MNRAS.446.2776S}}, dust distribution \citep{2018MNRAS.473.4868Z}, and the inclination angle between the magnetic field and the angular momentum axis \citep{2018ApJ...868...22T,2020ApJ...898..118H,2024ApJ...961L..28L} might also play a role in shaping disk  sizes. Future studies are needed to explain the observed Class 0 disk size distribution. 




\section{Conclusion}
\label{sec:conclusion}
We have conducted an ALMA high-resolution (0\as1) continuum survey of nearly all known Class 0, Class I, and flat-spectrum protostars in seven nearby molecular: clouds, Corona Australis, Aquila, Chamaeleon I \& II, Ophiuchus, Ophiuchus North, and Serpens. We also include young Class II sources still surrounded by their natal envelopes, with ${\rm T}_{\rm bol}\sim 1900$\,K. In total, we detected 184 protostellar disks. In this paper, we present the first paper of the CAMPOS survey, which covers the self-calibration and imaging of all continuum sources. 
Our main results are summarized as follows.

\begin{enumerate}
    \item {Class 0 protostellar disks are in general larger than Class I and flat-spectrum protostars.} The mean dust radii for the Class 0, Class I, and flat-spectrum protostars are $59.3 \pm 7.4$, $46.0 \pm 4.0$, and $41.6 \pm 7.2$\, au. Using a Gaussian kernel density estimation (see \autoref{fig:Radius_KDE}), we found that the most probable radii for Class 0, Class I, and flat-spectrum protostars are 39, 27, and 23\,au. {The disk radii distributions for Class 0 and Class I are consistent with being drawn from the same sample. In contrast, the distributions for Class 0 and flat-spectrum sources are inconsistent with being drawn from the same distribution.  } 

    \item We conducted Anderson-Darling tests and found significant variation in disk radii among the same class between different clouds, which highlights the influence of local environments on disk size evolution (see \autoref{table:disk_radii_stat} and \autoref{fig:R_Class}).
        
    \item We found a statistically significant correlation (p-value $<$0.06) suggesting that the {average dust disk radius} decrease between Class 0 and Class I phase for the Serpens and Ophiuchus molecular clouds (See \autoref{table:disk_radii_stat}). We showed that the dust drift timescale is longer than the Class 0 half-life, and the 30\,\% decrease in {average dust disk radius} between Class 0 and Class I cannot be explained by radial drift alone (see \autoref{fig:Radius_KDE}). 


    \item For Class 0 sources the dust disk radii are similar in size to the gas disk radii (see conclusion point 8), allowing a direct comparison between the gas disk radii from simulations and our observed dust disk radii. Our work reveals that protostellar disk formation is magnetically regulated. Hydrodynamical models without consideration of magnetic fields cannot explain the disk size distribution of Class 0 protostars. In addition, the CAMPOS data is in high tension with the bimodal disk size distribution predicted by disk formation models that include the Hall effect \citep{2015ApJ...810L..26T}. Our data is also inconsistent with the small disk size and the small variation of disk sizes predicted by models that include the ambipolar diffusion (e.g., \citealt{2016ApJ...830L...8H,2021ApJ...922...36L}). 

    \item From our CAMPOS survey, we found 21 protostellar disks with substructures or asymmetries as shown in \autoref{fig:disk_substructure}, \autoref{fig:circumbinary_disk} and \autoref{fig:asymmetric_disk}. Out of the 21 disks with substructures, 5 of them are newly discovered. Two of them are Class I (SMM 2, IRS 2), two are flat-spectrum sources (Oph-emb-20, ISO-ChaI 101), and one is a Class II source (Ass Cha T 1-15). The discovered disks with substructures are summarized in \autoref{table:disk_substructure}. 
    
    \item We cross-matched all protostars identified in our ALMA survey with the known literature catalogs. In total, we {detected 18 new protostellar sources (\autoref{table:new_source})}. All the ALMA protostellar sources detected by our ALMA survey are listed in \autoref{Appendix:Source_summary} with images shown in  \autoref{Appendix:Image_gallery}. 
    
    \item We highlight the sources with many large-scale extended structures in our CAMPOS survey as possible ``streamer candidates", shown in \autoref{fig:Streamer}. In addition, we report the detection of edge-on sources (\autoref{fig:Edge_on_disk}), which will be interesting for follow-up study. 

    \item We compiled the literature data and discovered a strong increasing trend of the gas disk radii to dust disk radii ratio ($R_{\rm gas,Kep}/R_{\rm mm}$) as a function of bolometric temperature (${\rm T}_{\rm bol}$) as shown in \autoref{fig:RgasRmm_Tbol}. We found that Class 0 protostellar disks have $R_{\rm gas,Kep}/R_{\rm mm}\sim 1.0$, early Class I protostellar disks ($70 \le T_{\rm bol} \le 400$) have $R_{\rm gas,Kep}/R_{\rm mm}\sim 2.2$., and Class II disks have $R_{\rm gas,Kep}/R_{\rm mm}\sim 2.4$. The $R_{\rm gas,Kep}/R_{\rm mm}$ ratio increases rapidly during the early Class I stage ($T_{\rm bol}\sim$ few 100\,K). This sets an important timescale for the decoupling of dust and gas disk radii. Future studies are needed to precisely capture the rapid evolution of $R_{\rm gas,Kep}/R_{\rm mm}$ in early Class I, and reveal the mechanism behind the decoupling of gas and dust radii. However, we strongly caution that the results from the literature we used to find the evolutionary trend may suffer from large uncertainties due to the variety of methods used to obtain $R_{\rm gas,Kep}$,  line tracers, and sensitivity limits.

\end{enumerate}

\section*{Acknowledgments}

C.H.H. and H.A are supported in part by NSF grants {AST-1714710}. D.S.C. is supported by an NSF Astronomy and Astrophysics Postdoctoral Fellowship under award AST-2102405. Special thanks to e-HOPS group, Riwaj Pokhrel, and Thomas Megeath for providing the unpublished e-HOPs catalog for protostar classification. This paper makes use of the following ALMA data: ADS/JAO. \#2019.1.01792.S. ALMA is a partnership of the ESO (representing its member states), NSF(USA), and NINS (Japan), together with the NRC (Canada), NSC and ASIAA(Taiwan), and KASI(Republic of Korea), incooperation with the Republic of Chile. The Joint ALMA Observatory is operated by the ESO, AUI/NRAO, and NAOJ. The National Radio Astronomy Observatory is a facility of the National Science Foundation operated under a cooperative agreement by Associated Universities, Inc. 


%

\vspace{5mm}
\facilities{ALMA}
\software{Astropy:  \citet{2013ascl.soft04002G,2018AJ....156..123A,2022ApJ...935..167A}, CASA: \citet{2007ASPC..376..127M} {SciPy:  \citet{2020NatMe..17..261V}, Lifeline: \citet{cameron_davidson_pilon_2019_2652543}}}




\bibliography{CAMPOS_disk.bib}

\nocite{*}



\appendix

\section{Self-calibration pipeline modifying and testing}
\label{Appendix:Selfcal_test}

We used the modified version of the self-calibration pipeline developed by J.~Tobin to conduct the self-calibration for the entire sample\footnote{The self-calibration pipeline can be found on:\url{https://github.com/jjtobin/auto_selfcal}.}. The self-calibration pipeline runs multiple iterations with different values of the \textit{solints} parameter \citep{2022arXiv220705591R}. {This parameter controls the interval of time of the observations used to determine the self-cal solution. The optimal solution for the phase calibration can be achieved by finding the minimum useful time interval so as to minimize the atmospheric variation while maintaining enough S/N for the self-calibration (see \citet{2022arXiv220705591R} for more information). }


{ALMA continuum observation combines four spectral windows to form a single image. For each spectral window, the search for the minimum useful solution time interval begins by time-averaging over each observation scan (the duration spent continuously on a particular source) with \textit{solint}=\textit{inf}. The process ends with \textit{solint}=\textit{int}, specifying a solution for every integration, the shortest possible time interval within each scan. For the first iteration, the pipeline uses one spectral window for self-calibration, and the solution interval is set to the duration of the scan. If the initial \textit{solint}=\textit{inf} fails, then the pipeline goes to fall-back mode and combines all the observational scans and different spectral windows to do the self-calibration. The self-calibration procedure of the fall-back mode follows the same procedure as the self-calibration done in one spectral window.} 

{The minimum useful solution interval is determined by the statistical assessment of solution intervals. We compared the signal-to-noise ratio before and after each self-calibration to determine whether or not to accept the solution. We accept the self-calibration solution if} the signal-to-noise ratio of the image increases and the area of the beam size does not change more than 5\% compared to the original image. If the solution is accepted then for the following iterations {the values of the parameter} \textit{solint} are reduced iteratively by a factor of 2 until the solution no longer satisfies the above criteria. 


The default self-calibration pipeline uses the \verb+auto-masking+ option in \verb+tclean+ to automatically draw cleaning masks during the cleaning process. Upon inspection of the masks, we found that the auto-masking algorithm would add masks in features that are artifacts or noise. This resulted in incorrect models for the self-calibration process which resulted in fake sources {with peak intensities} a few times above the noise level after self-calibration.

To address this issue, we turned off the \verb+auto-masking+ algorithm in \verb+tclean+,  and modified the pipeline such that it takes in a self-calibration mask that can be specified by the user. We compared the post-self-calibration images from the default \verb+auto-masking+ pipeline and the modified pipeline with user-specific input masks, and we found that images from the modified pipeline have a better S/N ratio and a cleaner noise pattern. 

To achieve the best signal-to-noise level, we drew 2 versions of self-calibration masks for all the 125 fields observed by ALMA. In the first version, a mask was created only for the brightest source in the field of view. In the second version, the mask includes all the sources above 4 times the rms as well as the extended emission surrounding it. We ran the two versions of the self-calibration masks with the modified pipeline, and we found that the second version, the one that masks sources and their associated extended emission, performed much better and had a higher S/N ratio.   

The default of the self-calibration pipeline uses \textit{briggs} weighting with the robustness parameter set to 0.5. As the goal of our study was to study the disks we wanted to have the best possible resolution without significantly reducing the signal to noise. For both the self-calibration process and the imaging process, we tried three different weighting schemes (i.e., \textit{natural}, \textit{briggs} with robustness of 0.5, and \textit{uniform}).  Each of the self-calibrated visibilities with different weighted models were imaged them with \verb+tclean+ using three different weighting schemes, in total producing 9 sets of images. We compared the S/N ratio of the 9 sets of images and we found in order to have the best S/N ratio the the weighting used in the self-calibration \verb+tclean+ should match the weighting used in the \verb+tclean+ for {the final imaging procedure}. This is true for all the fields without exception. 

\section{Self-calibration tables for all the CAMPOS fields}
\label{Appendix:Self-cal_fields}
In this section, we present the properties of the Briggs 0.5, Uniform, and Natural weighted images before and after self-calibration using the modified self-calibration pipeline (see \autoref{table:1} and \autoref{table:2} and \autoref{table:3}). Note that the weighting for self-calibration \verb+tclean+ matches the weighting used in \verb+tclean+   for imaging. The tables keep track of how self-calibration is applied for each ALMA field and the corresponding improvements.

\begin{center}
\begin{longtable*}{ c c c c c c c c c c c c c c c c}
\caption{Summary of properties of the Briggs 0.5 weighted images before and after  self-calibration using the modified self-calibration pipeline 
}
\\ %
\label{table:1}
&  &    & \\
\hline
\hline
Field\tablenotemark{a}  & Final\tablenotemark{b}  & Stop\tablenotemark{c} & SN$_0$ & SN$_0$ & RMS$_0$	& RMS$_0$ & SN$_f$ & SN$_f$ & RMS$_f$ & RMS$_f$ & Beam$_f$ & PA & $\Delta$SN & $\Delta$SN  \\
& selfcal & reason & & (near) & & (near)&  & (near) & &(near) & & & & (near)\\
\hline
\endfirsthead
\multicolumn{6}{c}%
{\tablename\ \thetable\ -- \textit{Continued from previous page}} \\
\hline

\hline
\endhead
\hline \multicolumn{4}{r}{\textit{Continued on next page}} \\
\endfoot
\hline
\endlastfoot
\hline                        
Aql01 & inf\_ap & None & 231.49 & 252.84 &  0.12 &  0.11 & 290.26 & 290.26 & 0.10 & 0.08 & 0.15x0.11 & -84.45 & 1.25 & 1.31 \\
Aql02 & inf\_EB & SNd & 110.34 & 101.77 & 0.09 & 0.10 & 110.01 & 110.01 & 0.09 & 0.09 & 0.15x0.11 & -87.16 & 1.0 & 1.06 \\
Aql03 & int & B & 106.11 & 98.78 &  0.09 &  0.10 & 111.57 & 111.57 & 0.09 & 0.09 & 0.15x0.11 & -85.45 & 1.05 & 1.1 \\
Aql04 & int & B & 140.57 & 135.29 &  0.10 &  0.10 & 156.41 & 156.41 & 0.09 & 0.09 & 0.15x0.11 & -86.16 & 1.11 & 1.18 \\
Aql05 & inf & B & 36.82 & 34.31 & 0.09 & 0.09 & 37.71 & 37.71 & 0.09 & 0.08 & 0.15x0.11 & -85.54 & 1.02 & 1.14 \\
Aql06 & int & B & 82.17 & 72.2 &  0.09 &  0.10 & 88.6 & 88.6 & 0.09 & 0.09 & 0.15x0.11 & -81.66 & 1.08 & 1.21 \\
Aql07 & inf & B & 164.96 & 143.85 & 0.09 & 0.11 & 175.98 & 175.98 & 0.09 & 0.09 & 0.15x0.11 & -85.17 & 1.07 & 1.17 \\
Aql08 & FAIL & B & 107.29 & 105.02 & 0.09 & 0.09 & 107.29 & 107.29 & 0.09 & 0.09 & 0.15x0.11 & -86.20 & 1.0 & 1.0 \\
Aql09 & inf & B & 43.29 & 43.84 & 0.09 & 0.09 & 45.2 & 45.2 & 0.09 & 0.08 & 0.15x0.11 & -85.06 & 1.04 & 1.08 \\
Aql10 & int & B & 110.22 & 85.1 &  0.13 &  0.17 & 117.86 & 117.86 & 0.13 & 0.16 & 0.15x0.11 & -85.77 & 1.07 & 1.07 \\
Aql11 & inf & B & 50.25 & 46.4 & 0.09 & 0.10 & 52.8 & 52.8 & 0.09 & 0.09 & 0.15x0.10 & -85.02 & 1.05 & 1.05 \\
Aql12 & inf\_ap & None & 173.25 & 161.91 &  0.10 &  0.10 & 184.71 & 184.71 & 0.09 & 0.09 & 0.15x0.11 & -85.28 & 1.07 & 1.13 \\
Aql13 & FAIL & B & 247.14 & 213.83 & 0.12 & 0.14 & 247.14 & 247.14 & 0.12 & 0.14 & 0.15x0.11 & -85.73 & 1.0 & 1.0 \\
Aql14 & int & B & 138.62 & 122.47 &  0.09 &  0.10 & 150.28 & 150.28 & 0.09 & 0.09 & 0.15x0.11 & -84.86 & 1.08 & 1.22 \\
Aql15 & inf & B & 55.95 & 54.48 & 0.09 & 0.09 & 56.98 & 56.98 & 0.09 & 0.09 & 0.15x0.10 & -84.88 & 1.02 & 1.03 \\
Aql16 & 4.03 & B & 73.46 & 73.27 & 0.09 & 0.09 & 77.96 & 77.96 & 0.09 & 0.09 & 0.15x0.11 & -84.81 & 1.06 & 1.04 \\
Aql17 & 4.03 & B & 75.08 & 69.15 & 0.09 & 0.10 & 80.62 & 80.62 & 0.09 & 0.09 & 0.15x0.11 & -80.65 & 1.07 & 1.1 \\
Aql18 & inf\_ap & None & 116.89 & 112.74 & 0.10 & 0.10 & 124.87 & 124.87 & 0.09 & 0.09 & 0.15x0.11 & -81.92 & 1.07 & 1.15 \\
ChamII01 & inf\_ap & None & 103.81 & 70.12 & 0.67 & 0.99 & 1201.11 & 1099.66 & 0.08 & 0.09 & 0.17x0.10 & 9.44 & 11.57 & 15.68 \\
ChamII02 & FAIL & B & 73.48 & 49.83 & 0.12 & 0.18 & 73.48 & 73.48 & 0.12 & 0.18 & 0.16x0.10 & 8.74 & 1.0 & 1.0 \\
ChamII03 & FAIL & B & 37.6 & 25.28 & 0.10 & 0.15 & 37.6 & 37.6 & 0.10 & 0.15 & 0.16x0.10 & 7.75 & 1.0 & 1.0 \\
ChamI01 & FAIL & B & 35.57 & 25.0 & 0.07 & 0.10 & 35.57 & 35.57 & 0.07 & 0.10 & 0.10x0.07 & 3.20 & 1.0 & 1.0 \\
ChamI02 & FAIL & B & 75.33 & 46.25 & 0.11 & 0.17 & 75.33 & 75.33 & 0.11 & 0.17 & 0.10x0.07 & -1.41 & 1.0 & 1.0 \\
ChamI03 & FAIL & NA & NA & NA & NA & NA & NA & NA & NA & NA & NA & NA & NA & NA \\
ChamI04 & FAIL & NA & NA & NA & NA & NA & NA & NA & NA & NA & NA & NA & NA & NA \\
ChamI05 & FAIL & NA & NA & NA & NA & NA & NA & NA & NA & NA & NA & NA & NA & NA \\
ChamI06 & FAIL & B & 45.33 & 39.18 & 0.08 & 0.09 & 45.33 & 45.33 & 0.08 & 0.09 & 0.10x0.07 & 0.74 & 1.0 & 1.0 \\
ChamI07 & FAIL & B & 62.72 & 46.95 & 0.10 & 0.13 & 62.72 & 62.72 & 0.10 & 0.13 & 0.10x0.07 & 1.03 & 1.0 & 1.0 \\
ChamI08 & FAIL & B & 45.14 & 43.38 & 0.08 & 0.08 & 45.14 & 45.14 & 0.08 & 0.08 & 0.13x0.11 & 37.48 & 1.0 & 1.0 \\
ChamI09 & FAIL & NA & NA & NA & NA & NA & NA & NA & NA & NA & NA & NA & NA & NA \\
ChamI10 & FAIL & NA & NA & NA & NA & NA & NA & NA & NA & NA & NA & NA & NA & NA \\
ChamI11 & FAIL & B & 78.2 & 67.51 & 0.08 & 0.09 & 78.2 & 78.2 & 0.08 & 0.09 & 0.13x0.12 & 42.67 & 1.0 & 1.0 \\
ChamI12 & inf & B & 119.57 & 62.95 & 0.09 & 0.18 & 204.24 & 204.24 & 0.07 & 0.07 & 0.13x0.12 & 34.04 & 1.71 & 3.5 \\
ChamI13 & FAIL & B & 22.07 & 22.04 & 0.11 & 0.11 & 22.07 & 22.07 & 0.11 & 0.11 & 0.17x0.10 & 40.09 & 1.0 & 1.0 \\
CrAus01 & inf\_ap & None & 117.68 & 65.77 &  0.24 &  0.44 & 288.81 & 288.81 & 0.10 & 0.08 & 0.13x0.10 & -81.94 & 2.45 & 5.52 \\
CrAus02 & inf\_ap & None & 77.6 & 36.83 &  0.37 &  0.79 & 203.39 & 203.39 & 0.14 & 0.12 & 0.13x0.10 & -83.18 & 2.62 & 6.53 \\
CrAus03 & inf\_ap & None & 119.77 & 86.78 &  0.11 &  0.15 & 164.75 & 164.75 & 0.09 & 0.09 & 0.13x0.10 & -82.96 & 1.38 & 1.84 \\
CrAus04 & inf\_ap & None & 124.08 & 83.96 &  0.15 &  0.22 & 223.06 & 223.06 & 0.09 & 0.08 & 0.13x0.10 & -84.91 & 1.8 & 2.99 \\
CrAus05 & inf\_ap & None & 199.3 & 105.2 & 0.35 & 0.66 & 881.61 & 881.61 & 0.08 & 0.09 & 0.13x0.10 & -83.07 & 4.42 & 8.11 \\
CrAus06 & FAIL & NA & NA & NA & NA & NA & NA & NA & NA & NA & NA & NA & NA & NA \\
CrAus07 & inf\_ap & None & 98.67 & 73.22 &  0.20 &  0.27 & 235.13 & 235.13 & 0.10 & 0.10 & 0.13x0.10 & -81.89 & 2.38 & 3.24 \\
CrAus08 & inf\_ap & None & 133.52 & 105.32 &  0.34 &  0.43 & 411.84 & 411.84 & 0.12 & 0.12 & 0.13x0.10 & -83.74 & 3.08 & 4.05 \\
CrAus09 & inf\_ap & None & 60.88 & 45.57 & 0.17 & 0.23 & 108.07 & 108.07 & 0.10 & 0.16 & 0.13x0.10 & -85.96 & 1.78 & 1.45 \\
CrAus10 & inf\_ap & None & 149.06 & 85.67 &  0.19 &  0.34 & 336.11 & 336.11 & 0.09 & 0.09 & 0.13x0.10 & -81.86 & 2.25 & 3.91 \\
CrAus11 & inf\_ap & None & 160.19 & 109.82 &  0.39 &  0.56 & 811.4 & 811.4 & 0.09 & 0.09 & 0.13x0.10 & -82.14 & 5.07 & 6.74 \\
OphN01 & inf & SNd & 330.52 & 237.17 & 0.08 & 0.11 & 882.21 & 882.21 & 0.03 & 0.03 & 0.14x0.09 & -86.90 & 2.67 & 3.71 \\
OphN02 & inf\_ap & None & 221.36 & 254.26 & 0.08  & 0.07  & 478.62 & 523.08 & 0.04 & 0.03 & 0.13x0.10 & -87.78 & 2.16 & 2.06 \\
OphN03 & 12.10 & B & 471.62 & 249.37 & 0.04  & 0.08  & 701.06 & 465.4 & 0.03 & 0.05 & 0.14x0.10 & -85.35 & 1.49 & 1.87 \\
Oph01 & inf\_ap & None & 41.54 & 47.23 & 0.09 & 0.08 & 46.76 & 46.76 & 0.09 & 0.08 & 0.16x0.12 & -76.05 & 1.13 & 1.07 \\
Oph02 & inf\_ap & None & 105.71 & 92.12 &  0.10 &  0.11 & 119.24 & 119.24 & 0.09 & 0.09 & 0.15x0.12 & -77.95 & 1.13 & 1.3 \\
Oph03 & inf\_ap & None & 189.13 & 187.38 &  0.11 &  0.11 & 250.96 & 250.96 & 0.09 & 0.08 & 0.15x0.12 & -75.90 & 1.33 & 1.43 \\
Oph04 & FAIL & NA & NA & NA & NA & NA & NA & NA & NA & NA & NA & NA & NA & NA \\
Oph05 & FAIL & NA & NA & NA & NA & NA & NA & NA & NA & NA & NA & NA & NA & NA \\
Oph06 & inf\_ap & None & 112.25 & 91.36 &  0.10 &  0.12 & 134.66 & 134.66 & 0.08 & 0.08 & 0.16x0.12 & -77.43 & 1.2 & 1.48 \\
Oph07 & inf\_ap & None & 110.43 & 86.97 & 0.11 & 0.14 & 139.46 & 139.46 & 0.09 & 0.08 & 0.15x0.12 & -75.53 & 1.26 & 1.79 \\
Oph08 & int & B & 109.6 & 113.65 &  0.19 &  0.18 & 225.58 & 225.58 & 0.09 & 0.07 & 0.15x0.12 & -78.61 & 2.06 & 2.52 \\
Oph09 & inf\_ap & None & 170.3 & 156.31 & 0.09 & 0.10 & 190.11 & 190.11 & 0.08 & 0.08 & 0.15x0.12 & -75.96 & 1.12 & 1.29 \\
Oph10 & inf\_ap & None & 135.15 & 148.96 &  0.18 &  0.16 & 255.49 & 255.49 & 0.10 & 0.12 & 0.15x0.12 & -75.97 & 1.89 & 1.39 \\
Oph11 & 8.06 & SNd & 58.36 & 64.45 & 0.09 & 0.08 & 64.71 & 64.71 & 0.08 & 0.07 & 0.16x0.11 & -76.28 & 1.11 & 1.15 \\
Oph12 & inf\_ap & None & 349.81 & 250.45 & 0.12 & 0.16 & 464.6 & 464.6 & 0.09 & 0.09 & 0.15x0.11 & -75.87 & 1.33 & 1.93 \\
Oph13 & inf\_ap & None & 92.25 & 85.19 &  0.15 &  0.16 & 159.17 & 159.17 & 0.09 & 0.09 & 0.15x0.12 & -78.27 & 1.73 & 1.8 \\
Oph14 & inf\_ap & None & 114.51 & 113.4 & 0.09 & 0.09 & 122.66 & 122.66 & 0.08 & 0.08 & 0.15x0.12 & -79.25 & 1.07 & 1.17 \\
Oph15 & 8.06 & B & 70.71 & 83.38 & 0.10 & 0.09 & 81.51 & 81.51 & 0.09 & 0.08 & 0.15x0.12 & -76.95 & 1.15 & 1.06 \\
Oph16 & inf\_ap & None & 42.94 & 37.23 & 0.09 & 0.10 & 46.08 & 46.08 & 0.09 & 0.10 & 0.15x0.12 & -75.95 & 1.07 & 1.11 \\
Oph17 & FAIL & B & 62.09 & 68.44 & 0.08 & 0.08 & 62.09 & 62.09 & 0.08 & 0.08 & 0.15x0.11 & -76.26 & 1.0 & 1.0 \\
Oph18 & inf\_ap & None & 138.62 & 133.51 & 0.09 & 0.09 & 150.34 & 150.34 & 0.08 & 0.07 & 0.16x0.11 & -75.56 & 1.08 & 1.35 \\
Oph19 & inf\_ap & None & 117.53 & 98.69 & 0.09 & 0.11 & 128.46 & 128.46 & 0.08 & 0.07 & 0.15x0.12 & -75.09 & 1.09 & 1.52 \\
Oph20 & inf\_ap & None & 152.34 & 113.36 &  0.09 &  0.12 & 164.34 & 164.34 & 0.08 & 0.09 & 0.15x0.12 & -75.62 & 1.08 & 1.42 \\
Oph21 & inf\_ap & None & 116.48 & 110.07 & 0.09 & 0.10 & 132.21 & 132.21 & 0.08 & 0.08 & 0.15x0.12 & -75.02 & 1.14 & 1.21 \\
Oph22 & 8.06 & B & 78.38 & 77.34 & 0.09 & 0.09 & 82.55 & 82.55 & 0.08 & 0.08 & 0.15x0.12 & -75.60 & 1.05 & 1.13 \\
Oph23 & inf\_ap & None & 98.79 & 92.48 &  0.09 &  0.10 & 107.82 & 107.82 & 0.09 & 0.09 & 0.15x0.11 & -76.63 & 1.09 & 1.17 \\
Oph24 & FAIL & NA & NA & NA & NA & NA & NA & NA & NA & NA & NA & NA & NA & NA \\
Oph25 & inf & B & 42.18 & 44.14 & 0.09 & 0.08 & 44.97 & 44.97 & 0.09 & 0.09 & 0.15x0.11 & -74.77 & 1.07 & 1.03 \\
Oph26 & inf\_ap & None & 120.85 & 97.24 &  0.09 &  0.11 & 135.09 & 135.09 & 0.09 & 0.10 & 0.15x0.12 & -73.93 & 1.12 & 1.18 \\
Oph27 & inf & B & 107.48 & 101.62 & 0.09 & 0.09 & 113.09 & 113.09 & 0.09 & 0.08 & 0.15x0.11 & -74.57 & 1.05 & 1.13 \\
Oph28 & FAIL & B & 23.42 & 22.71 & 0.08 & 0.09 & 23.42 & 23.42 & 0.08 & 0.09 & 0.15x0.11 & -75.72 & 1.0 & 1.0 \\
Oph29 & 4.03 & B & 60.19 & 54.64 & 0.09 & 0.09 & 65.72 & 65.72 & 0.08 & 0.09 & 0.15x0.12 & -72.58 & 1.09 & 1.18 \\
Oph30 & inf & B & 51.33 & 59.32 & 0.08 & 0.07 & 54.38 & 54.38 & 0.08 & 0.08 & 0.15x0.11 & -74.37 & 1.06 & 0.93 \\
Oph31 & 18.14 & B & 221.08 & 127.0 & 0.12 & 0.21 & 255.71 & 255.71 & 0.10 & 0.10 & 0.15x0.11 & -75.50 & 1.16 & 1.99 \\
Oph32 & inf\_EB & SNd & 89.52 & 77.6 & 0.09 & 0.10 & 91.99 & 91.99 & 0.09 & 0.09 & 0.15x0.11 & -74.79 & 1.03 & 1.17 \\
Oph33 & inf\_EB & lowSN & 24.64 & 27.58 & 0.08 & 0.08 & 26.1 & 26.1 & 0.09 & 0.09 & 0.15x0.11 & -74.88 & 1.06 & 0.95 \\
Oph34 & inf\_ap & None & 199.63 & 211.39 &  0.19 &  0.18 & 464.36 & 464.36 & 0.09 & 0.08 & 0.15x0.10 & -82.66 & 2.33 & 2.5 \\
Oph35 & None & B & 60.33 & 49.99 & 0.07 & 0.09 & 60.33 & 60.33 & 0.07 & 0.09 & 0.15x0.10 & -83.95 & 1.0 & 1.0 \\
Oph36 & 12.10 & B & 92.36 & 66.14 & 0.07 & 0.10 & 105.42 & 105.42 & 0.07 & 0.07 & 0.15x0.10 & -83.18 & 1.14 & 1.64 \\
Oph37 & FAIL & NA & NA & NA & NA & NA & NA & NA & NA & NA & NA & NA & NA & NA \\
Oph38 & inf\_EB & SNd & 43.29 & 39.19 & 0.07 & 0.08 & 49.71 & 49.71 & 0.07 & 0.06 & 0.15x0.10 & -83.78 & 1.15 & 1.62 \\
Oph39 & inf\_ap & None & 256.65 & 127.49 &  0.13 &  0.26 & 440.57 & 440.57 & 0.08 & 0.11 & 0.15x0.10 & -83.48 & 1.72 & 2.49 \\
Oph40 & inf\_ap & None & 135.29 & 94.57 & 1.12 & 1.61 & 282.9 & 282.9 & 0.54 & 1.06 & 0.15x0.10 & -84.14 & 2.09 & 1.52 \\
Oph41 & FAIL & B & 62.35 & 65.07 & 0.08 & 0.08 & 62.35 & 62.35 & 0.08 & 0.08 & 0.15x0.09 & -85.04 & 1.0 & 1.0 \\
Serp01 & FAIL & NA & NA & NA & NA & NA & NA & NA & NA & NA & NA & NA & NA & NA \\
Serp02 & FAIL & NA & NA & NA & NA & NA & NA & NA & NA & NA & NA & NA & NA & NA \\
Serp03 & int & B & 103.11 & 99.31 &  0.09 &  0.09 & 117.71 & 117.71 & 0.09 & 0.09 & 0.15x0.11 & 87.39 & 1.14 & 1.16 \\
Serp04 & FAIL & NA & NA & NA & NA & NA & NA & NA & NA & NA & NA & NA & NA & NA \\
Serp05 & int & B & 137.58 & 117.41 &  0.10 &  0.11 & 157.99 & 157.99 & 0.09 & 0.10 & 0.15x0.11 & 87.98 & 1.15 & 1.26 \\
Serp06 & inf\_EB & B & 63.7 & 59.09 & 0.09 & 0.10 & 66.31 & 66.31 & 0.09 & 0.09 & 0.15x0.11 & 83.65 & 1.04 & 1.12 \\
Serp07 & inf\_ap & None & 55.26 & 48.93 & 0.09 & 0.10 & 57.39 & 57.39 & 0.09 & 0.08 & 0.16x0.10 & 84.89 & 1.04 & 1.22 \\
Serp08 & 18.14 & B & 30.25 & 29.2 & 0.09 & 0.09 & 32.64 & 32.64 & 0.09 & 0.09 & 0.15x0.11 & 84.32 & 1.08 & 1.11 \\
Serp09 & FAIL & NA & NA & NA & NA & NA & NA & NA & NA & NA & NA & NA & NA & NA \\
Serp10 & 18.14 & B & 36.55 & 35.13 & 0.09 & 0.09 & 38.83 & 38.83 & 0.09 & 0.09 & 0.15x0.11 & 84.52 & 1.06 & 1.08 \\
Serp11 & int & B & 102.36 & 87.3 &  0.11 &  0.13 & 124.98 & 124.98 & 0.09 & 0.10 & 0.16x0.11 & -89.49 & 1.22 & 1.39 \\
Serp12 & inf\_ap & None & 100.35 & 98.53 &  0.12 &  0.12 & 122.21 & 122.21 & 0.10 & 0.10 & 0.15x0.11 & 85.18 & 1.22 & 1.2 \\
Serp13 & inf\_ap & None & 309.28 & 266.91 & 0.15 & 0.17 & 446.93 & 446.93 & 0.10 & 0.10 & 0.15x0.11 & 88.63 & 1.45 & 1.66 \\
Serp14 & inf\_EB & SNd & 25.74 & 28.53 & 0.09 & 0.08 & 27.83 & 27.83 & 0.09 & 0.08 & 0.15x0.11 & 85.31 & 1.08 & 1.11 \\
Serp15 & inf\_ap & None & 227.34 & 172.03 &  0.12 &  0.16 & 299.76 & 299.76 & 0.09 & 0.10 & 0.15x0.11 & -73.58 & 1.32 & 1.71 \\
Serp16 & int & B & 147.14 & 126.65 &  0.14 &  0.17 & 204.32 & 204.32 & 0.11 & 0.15 & 0.15x0.11 & -73.59 & 1.39 & 1.15 \\
Serp17 & 4.03 & B & 76.68 & 76.41 &  0.09 &  0.09 & 88.58 & 88.58 & 0.09 & 0.09 & 0.16x0.11 & -76.82 & 1.16 & 1.19 \\
Serp18 & inf\_EB & SNd & 99.99 & 81.13 & 0.76 & 0.94 & 244.08 & 244.08 & 0.30 & 0.36 & 0.15x0.11 & -73.17 & 2.44 & 2.46 \\
Serp19 & inf\_ap & None & 284.57 & 262.53 &  0.13 &  0.14 & 407.79 & 407.79 & 0.09 & 0.10 & 0.15x0.11 & -74.69 & 1.43 & 1.5 \\
Serp20 & inf\_ap & None & 100.26 & 91.9 &  0.11 &  0.12 & 129.45 & 129.45 & 0.09 & 0.09 & 0.16x0.11 & -73.22 & 1.29 & 1.4 \\
Serp21 & FAIL & B & 34.53 & 34.15 & 0.09 & 0.09 & 34.53 & 34.53 & 0.09 & 0.09 & 0.15x0.11 & 85.82 & 1.0 & 1.0 \\
Serp22 & int & B & 159.23 & 157.85 &  0.10 &  0.10 & 186.85 & 186.85 & 0.09 & 0.09 & 0.15x0.11 & -75.08 & 1.17 & 1.21 \\
Serp23 & 8.06 & B & 158.17 & 163.42 & 0.10 & 0.10 & 188.78 & 188.78 & 0.09 & 0.08 & 0.15x0.11 & 88.40 & 1.19 & 1.26 \\
Serp24 & FAIL & NA & NA & NA & NA & NA & NA & NA & NA & NA & NA & NA & NA & NA \\
Serp25 & inf & SNd,B & 50.26 & 53.9 & 0.17 & 0.16 & 52.43 & 52.43 & 0.16 & 0.15 & 0.15x0.11 & -75.20 & 1.04 & 1.06 \\
Serp26 & inf\_ap & None & 232.56 & 225.08 &  0.15 &  0.16 & 312.08 & 312.08 & 0.11 & 0.10 & 0.15x0.11 & -71.76 & 1.34 & 1.58 \\
Serp27 & int & B & 126.09 & 120.84 &  0.10 &  0.10 & 144.2 & 144.2 & 0.09 & 0.09 & 0.15x0.11 & -75.68 & 1.14 & 1.27 \\
Serp28 & FAIL & NA & NA & NA & NA & NA & NA & NA & NA & NA & NA & NA & NA & NA \\
Serp29 & inf\_ap & None & 121.6 & 96.54 &  0.21 &  0.26 & 252.32 & 252.32 & 0.11 & 0.11 & 0.16x0.11 & -76.10 & 2.08 & 2.64 \\
Serp30 & int & B & 111.75 & 104.92 &  0.14 &  0.15 & 137.76 & 137.76 & 0.12 & 0.13 & 0.15x0.11 & -71.69 & 1.23 & 1.25 \\
Serp31 & 18.14 & B & 27.18 & 25.92 & 0.09 & 0.10 & 31.1 & 31.1 & 0.09 & 0.09 & 0.15x0.11 & -71.71 & 1.14 & 1.15 \\
Serp32 & FAIL & NA & NA & NA & NA & NA & NA & NA & NA & NA & NA & NA & NA & NA \\
Serp33 & 18.14 & B & 25.93 & 25.79 & 0.09 & 0.09 & 30.26 & 30.26 & 0.09 & 0.09 & 0.15x0.11 & -71.13 & 1.17 & 1.14 \\
Serp34 & inf\_ap & None & 165.29 & 134.97 &  0.16 &  0.19 & 299.36 & 299.36 & 0.10 & 0.09 & 0.16x0.11 & -69.21 & 1.81 & 2.33 \\
Serp35 & None & B & 29.23 & 28.24 & 0.09 & 0.09 & 29.23 & 29.23 & 0.09 & 0.09 & 0.15x0.11 & 86.30 & 1.0 & 1.0 \\
Serp36 & None & B & 61.84 & 59.65 & 0.09 & 0.09 & 61.84 & 61.84 & 0.09 & 0.09 & 0.15x0.10 & 89.39 & 1.0 & 1.0 \\
\hline
\hline
\end{longtable*}
\tablenotetext{a}{Field name in the CAMPOS ALMA observation. The characters represent the name of the cloud (eg. Aql: Aquila, ChamI: Chamaeloeon I, ChamII: Chamaeloeon II, Oph: Ophiuchus, OphN: Ophiuchus North, Serp: Serpens)}
\tablenotetext{b}{Final integration time adopted for self-calibration solution (solint). The characters ``ap" represent amplitude calibration applied, and ``inf\_EB" represents the fall-back mode where both the spectral window and observational scans are combined. A number in this column represents the solution interval (in seconds). ``FAIL" denotes the self-calibration process failed. ``None" indicates that self-calibration solutions are found, but are not adopted because the beam size changed by more than 5\%.}
\tablenotetext{c}{Reason for the modified self-calibration pipeline to stop. B represents the area of the beam size changed by more than 5\%, and SNd represents the S/N ratio decreased after the next round of self-calibration.}
\tablenotetext{}{ \textbf{Note}: Column title SN represents the signal-to-noise ratio, and RMS represents the noise level in mJy. The subscript 0 and f represent before and after self-calibration, respectively.  ``Near" represents the near field, the central region within the map with the outer extent around 5 times the major axis of the beam. $\Delta$SN represents the change in signal-to-noise ratio after self-calibration. It is calculated by dividing the final S/N ratio by the initial S/N ratio before self-calibration. The final beam size is given in arcseconds, and the position angles (PA) are in degrees.}
\end{center} 

\begin{center}
\begin{longtable*}{ c c c c c c c c c c c c c c c c}
\caption{Summary of properties of the Uniform weighted images before and after  self-calibration using the modified self-calibration pipeline 
}\\ %
\label{table:2}
&  &    & \\
\hline
\hline
Field\tablenotemark{a}  & Final\tablenotemark{b}  & Stop\tablenotemark{c} & SN$_0$ & SN$_0$ & RMS$_0$	& RMS$_0$ & SN$_f$ & SN$_f$ & RMS$_f$ & RMS$_f$ & Beam$_f$ & PA & $\Delta$SN & $\Delta$SN  \\
& selfcal & reason & & (near) & & (near)&  & (near) & &(near) & & & & (near)\\
\hline
\endfirsthead
\multicolumn{6}{c}%
{\tablename\ \thetable\ -- \textit{Continued from previous page}} \\
\hline

\hline
\endhead
\hline \multicolumn{4}{r}{\textit{Continued on next page}} \\
\endfoot
\hline
\endlastfoot
\hline                        
Aql01 & FAIL & NA & NA & NA & NA & NA & NA & NA & NA & NA & NA & NA & NA & NA \\
Aql02 & inf\_EB & SNd & 32.78 & 34.51 & 0.26 & 0.25 & 32.01 & 32.01 & 0.26 & 0.25 & 0.13x0.09 & 89.93 & 0.98 & 0.96 \\
Aql03 & inf\_EB & SNd & 39.71 & 41.66 & 0.22 & 0.21 & 40.34 & 40.34 & 0.22 & 0.21 & 0.13x0.09 & -88.22 & 1.02 & 1.0 \\
Aql04 & 18.14 & SNd & 41.13 & 42.61 & 0.27 & 0.26 & 42.15 & 42.15 & 0.27 & 0.26 & 0.13x0.09 & -89.88 & 1.02 & 1.03 \\
Aql05 & FAIL & NA & NA & NA & NA & NA & NA & NA & NA & NA & NA & NA & NA & NA \\
Aql06 & inf\_EB & lowSN & 26.02 & 25.19 & 0.26 & 0.26 & 27.09 & 27.09 & 0.26 & 0.26 & 0.13x0.09 & -89.39 & 1.04 & 1.06 \\
Aql07 & 4.03 & SNd & 57.01 & 56.63 & 0.22 & 0.22 & 59.4 & 59.4 & 0.22 & 0.21 & 0.13x0.09 & -87.86 & 1.04 & 1.1 \\
Aql08 & inf\_EB & SNd & 33.41 & 34.64 & 0.26 & 0.25 & 33.57 & 33.57 & 0.26 & 0.24 & 0.13x0.09 & -86.45 & 1.0 & 1.05 \\
Aql09 & FAIL & NA & NA & NA & NA & NA & NA & NA & NA & NA & NA & NA & NA & NA \\
Aql10 & 18.14 & SNd & 48.95 & 46.58 & 0.27 & 0.28 & 50.18 & 50.18 & 0.26 & 0.28 & 0.13x0.09 & -89.49 & 1.03 & 1.01 \\
Aql11 & FAIL & NA & NA & NA & NA & NA & NA & NA & NA & NA & NA & NA & NA & NA \\
Aql12 & inf\_EB & SNd & 49.82 & 50.54 & 0.29 & 0.29 & 49.34 & 49.34 & 0.28 & 0.29 & 0.13x0.09 & -88.49 & 0.99 & 0.97 \\
Aql13 & FAIL & B & 86.98 & 79.28 & 0.27 & 0.30 & 86.98 & 86.98 & 0.27 & 0.30 & 0.13x0.09 & -88.70 & 1.0 & 1.0 \\
Aql14 & 18.14 & SNd & 41.85 & 43.33 & 0.28 & 0.27 & 43.16 & 43.16 & 0.28 & 0.26 & 0.13x0.09 & -88.12 & 1.03 & 1.07 \\
Aql15 & FAIL & NA & NA & NA & NA & NA & NA & NA & NA & NA & NA & NA & NA & NA \\
Aql16 & inf\_EB & lowSN & 22.36 & 24.36 & 0.29 & 0.27 & 23.35 & 23.35 & 0.29 & 0.26 & 0.13x0.09 & -87.81 & 1.04 & 1.08 \\
Aql17 & inf\_EB & lowSN & 22.6 & 20.04 & 0.28 & 0.31 & 24.1 & 24.1 & 0.27 & 0.30 & 0.13x0.09 & -87.42 & 1.07 & 1.11 \\
Aql18 & inf & SNd,B & 35.04 & 35.01 & 0.24 & 0.24 & 34.94 & 34.94 & 0.24 & 0.24 & 0.13x0.08 & -88.22 & 1.0 & 1.0 \\
ChamI01 & FAIL & NA & NA & NA & NA & NA & NA & NA & NA & NA & NA & NA & NA & NA \\
ChamI02 & FAIL & NA & NA & NA & NA & NA & NA & NA & NA & NA & NA & NA & NA & NA \\
ChamI03 & FAIL & NA & NA & NA & NA & NA & NA & NA & NA & NA & NA & NA & NA & NA \\
ChamI04 & FAIL & NA & NA & NA & NA & NA & NA & NA & NA & NA & NA & NA & NA & NA \\
ChamI05 & FAIL & NA & NA & NA & NA & NA & NA & NA & NA & NA & NA & NA & NA & NA \\
ChamI06 & FAIL & NA & NA & NA & NA & NA & NA & NA & NA & NA & NA & NA & NA & NA \\
ChamI07 & FAIL & NA & NA & NA & NA & NA & NA & NA & NA & NA & NA & NA & NA & NA \\
ChamI08 & FAIL & NA & NA & NA & NA & NA & NA & NA & NA & NA & NA & NA & NA & NA \\
ChamI09 & FAIL & NA & NA & NA & NA & NA & NA & NA & NA & NA & NA & NA & NA & NA \\
ChamI10 & FAIL & NA & NA & NA & NA & NA & NA & NA & NA & NA & NA & NA & NA & NA \\
ChamI11 & FAIL & NA & NA & NA & NA & NA & NA & NA & NA & NA & NA & NA & NA & NA \\
ChamI12 & inf & SNd & 29.56 & 25.84 & 0.30 & 0.34 & 39.59 & 39.59 & 0.30 & 0.31 & 0.12x0.09 & 36.56 & 1.34 & 1.46 \\
ChamI13 & FAIL & NA & NA & NA & NA & NA & NA & NA & NA & NA & NA & NA & NA & NA \\
ChamII01 & int & SNd & 99.66 & 68.56 & 0.52 & 0.75 & 256.25 & 287.3 & 0.31 & 0.28 & 0.15x0.08 & 9.24 & 2.57 & 4.19 \\
ChamII02 & FAIL & NA & NA & NA & NA & NA & NA & NA & NA & NA & NA & NA & NA & NA \\
ChamII03 & FAIL & NA & NA & NA & NA & NA & NA & NA & NA & NA & NA & NA & NA & NA \\
CrAus01 & 4.03 & SNd & 67.0 & 60.11 & 0.34 & 0.38 & 83.52 & 83.52 & 0.30 & 0.27 & 0.11x0.08 & -83.37 & 1.25 & 1.55 \\
CrAus02 & inf\_EB & SNd & 60.83 & 41.61 & 0.37 & 0.54 & 78.5 & 78.5 & 0.29 & 0.24 & 0.11x0.08 & -83.88 & 1.29 & 2.28 \\
CrAus03 & inf & SNd & 47.73 & 46.53 & 0.23 & 0.24 & 52.96 & 52.96 & 0.22 & 0.20 & 0.11x0.08 & -83.05 & 1.11 & 1.25 \\
CrAus04 & 8.06 & SNd & 50.62 & 50.51 & 0.31 & 0.31 & 54.95 & 54.95 & 0.29 & 0.25 & 0.11x0.09 & -83.87 & 1.09 & 1.27 \\
CrAus05 & int & SNd & 162.74 & 104.15 &  0.34 &  0.53 & 258.65 & 258.65 & 0.24 & 0.23 & 0.12x0.08 & -82.61 & 1.59 & 2.51 \\
CrAus06 & FAIL & NA & NA & NA & NA & NA & NA & NA & NA & NA & NA & NA & NA & NA \\
CrAus07 & 4.03 & SNd & 57.37 & 52.63 & 0.28 & 0.30 & 80.61 & 80.61 & 0.24 & 0.24 & 0.11x0.08 & -82.34 & 1.41 & 1.51 \\
CrAus08 & int & SNd & 97.85 & 80.14 &  0.38 &  0.46 & 142.25 & 142.25 & 0.30 & 0.31 & 0.11x0.08 & -84.19 & 1.45 & 1.72 \\
CrAus09 & FAIL & B & 31.82 & 30.27 & 0.26 & 0.27 & 31.82 & 31.82 & 0.26 & 0.27 & 0.11x0.08 & -82.64 & 1.0 & 1.0 \\
CrAus10 & 4.03 & SNd & 73.89 & 54.69 &  0.33 &  0.45 & 91.05 & 91.05 & 0.29 & 0.28 & 0.11x0.08 & -83.60 & 1.23 & 1.72 \\
CrAus11 & 4.03 & SNd & 134.86 & 99.19 &  0.37 &  0.51 & 238.22 & 238.22 & 0.24 & 0.22 & 0.11x0.08 & -82.89 & 1.77 & 2.6 \\
OphN01 & 6.05 & SNd & 147.77 & 156.83 & 0.14 & 0.13 & 165.34 & 192.42 & 0.13 & 0.11 & 0.13x0.07 & 87.96 & 1.12 & 1.23 \\
OphN02 & 26.21 & SNd & 92.76 & 92.06 & 0.14 & 0.14 & 99.66 & 99.66 & 0.14 & 0.12 & 0.12x0.07 & 86.83 & 1.07 & 1.22 \\
OphN03 & inf & SNd & 121.58 & 124.22 & 0.13 & 0.13 & 125.24 & 125.24 & 0.13 & 0.12 & 0.12x0.07 & -89.65 & 1.03 & 1.05 \\
Oph01 & FAIL & NA & NA & NA & NA & NA & NA & NA & NA & NA & NA & NA & NA & NA \\
Oph02 & inf\_EB & SNd & 33.58 & 34.6 & 0.28 & 0.27 & 32.38 & 32.38 & 0.27 & 0.27 & 0.13x0.10 & -72.31 & 0.96 & 0.96 \\
Oph03 & 4.03 & SNd & 74.64 & 69.85 & 0.24 & 0.25 & 79.99 & 79.99 & 0.23 & 0.24 & 0.13x0.10 & -72.04 & 1.07 & 1.09 \\
Oph04 & FAIL & NA & NA & NA & NA & NA & NA & NA & NA & NA & NA & NA & NA & NA \\
Oph05 & inf & SNd & 58.7 & 54.85 & 0.24 & 0.26 & 59.07 & 59.07 & 0.23 & 0.25 & 0.13x0.10 & -72.53 & 1.01 & 1.01 \\
Oph06 & 16.13 & SNd & 35.94 & 35.45 & 0.29 & 0.30 & 37.54 & 37.54 & 0.29 & 0.28 & 0.13x0.10 & -72.55 & 1.04 & 1.09 \\
Oph07 & FAIL & B & 41.78 & 44.18 & 0.24 & 0.23 & 41.78 & 41.78 & 0.24 & 0.23 & 0.13x0.10 & -72.96 & 1.0 & 1.0 \\
Oph08 & 8.06 & SNd & 53.87 & 62.28 & 0.31 & 0.27 & 59.76 & 59.76 & 0.28 & 0.25 & 0.13x0.10 & -72.94 & 1.11 & 1.07 \\
Oph09 & 8.06 & SNd & 53.66 & 50.22 & 0.24 & 0.25 & 56.4 & 56.4 & 0.24 & 0.25 & 0.13x0.10 & -73.68 & 1.05 & 1.11 \\
Oph10 & 8.06 & SNd & 68.33 & 72.46 & 0.30 & 0.28 & 77.75 & 77.75 & 0.27 & 0.25 & 0.13x0.10 & -73.42 & 1.14 & 1.15 \\
Oph11 & FAIL & NA & NA & NA & NA & NA & NA & NA & NA & NA & NA & NA & NA & NA \\
Oph12 & 4.03 & SNd & 148.85 & 124.5 &  0.24 &  0.28 & 159.47 & 159.47 & 0.23 & 0.23 & 0.13x0.10 & -72.38 & 1.07 & 1.24 \\
Oph13 & FAIL & B & 37.48 & 36.31 & 0.28 & 0.29 & 37.48 & 37.48 & 0.28 & 0.29 & 0.13x0.10 & -72.18 & 1.0 & 1.0 \\
Oph14 & inf & SNd & 37.07 & 42.01 & 0.23 & 0.21 & 37.32 & 37.32 & 0.23 & 0.21 & 0.13x0.10 & -71.51 & 1.01 & 0.99 \\
Oph15 & inf\_EB & lowSN & 21.87 & 22.37 & 0.30 & 0.30 & 22.24 & 22.24 & 0.29 & 0.28 & 0.13x0.10 & -72.66 & 1.02 & 1.04 \\
Oph16 & FAIL & NA & NA & NA & NA & NA & NA & NA & NA & NA & NA & NA & NA & NA \\
Oph17 & FAIL & NA & NA & NA & NA & NA & NA & NA & NA & NA & NA & NA & NA & NA \\
Oph18 & inf\_EB & SNd & 38.21 & 39.67 & 0.29 & 0.28 & 38.13 & 38.13 & 0.29 & 0.28 & 0.13x0.10 & -72.13 & 1.0 & 0.97 \\
Oph19 & inf & SNd & 37.23 & 38.81 & 0.24 & 0.23 & 37.58 & 37.58 & 0.23 & 0.22 & 0.13x0.10 & -71.97 & 1.01 & 1.03 \\
Oph20 & 8.06 & SNd & 47.56 & 42.15 & 0.28 & 0.31 & 49.98 & 49.98 & 0.27 & 0.30 & 0.13x0.10 & -72.87 & 1.05 & 1.1 \\
Oph21 & inf & SNd & 39.42 & 37.23 & 0.23 & 0.25 & 41.17 & 41.17 & 0.23 & 0.22 & 0.13x0.10 & -71.34 & 1.04 & 1.15 \\
Oph22 & inf\_EB & lowSN & 20.56 & 21.7 & 0.29 & 0.27 & 20.87 & 20.87 & 0.28 & 0.27 & 0.13x0.10 & -72.36 & 1.01 & 1.01 \\
Oph23 & inf & SNd & 38.26 & 35.96 & 0.23 & 0.24 & 40.17 & 40.17 & 0.23 & 0.23 & 0.13x0.10 & -72.05 & 1.05 & 1.09 \\
Oph24 & FAIL & NA & NA & NA & NA & NA & NA & NA & NA & NA & NA & NA & NA & NA \\
Oph25 & FAIL & NA & NA & NA & NA & NA & NA & NA & NA & NA & NA & NA & NA & NA \\
Oph26 & inf & SNd & 42.77 & 38.83 & 0.24 & 0.27 & 44.91 & 44.91 & 0.24 & 0.26 & 0.13x0.10 & -70.57 & 1.05 & 1.06 \\
Oph27 & inf & SNd & 40.5 & 38.3 & 0.24 & 0.25 & 40.76 & 40.76 & 0.24 & 0.26 & 0.13x0.10 & -71.17 & 1.01 & 0.96 \\
Oph28 & FAIL & NA & NA & NA & NA & NA & NA & NA & NA & NA & NA & NA & NA & NA \\
Oph29 & FAIL & NA & NA & NA & NA & NA & NA & NA & NA & NA & NA & NA & NA & NA \\
Oph30 & FAIL & NA & NA & NA & NA & NA & NA & NA & NA & NA & NA & NA & NA & NA \\
Oph31 & inf\_EB & SNd & 69.34 & 63.43 & 0.30 & 0.33 & 70.6 & 70.6 & 0.29 & 0.28 & 0.13x0.10 & -71.21 & 1.02 & 1.17 \\
Oph32 & inf & SNd & 30.86 & 27.08 & 0.24 & 0.28 & 31.7 & 31.7 & 0.24 & 0.26 & 0.13x0.10 & -70.53 & 1.03 & 1.08 \\
Oph33 & FAIL & NA & NA & NA & NA & NA & NA & NA & NA & NA & NA & NA & NA & NA \\
Oph34 & int & SNd & 123.13 & 132.27 &  0.26 &  0.24 & 155.65 & 155.65 & 0.22 & 0.20 & 0.13x0.08 & -84.51 & 1.26 & 1.31 \\
Oph35 & FAIL & NA & NA & NA & NA & NA & NA & NA & NA & NA & NA & NA & NA & NA \\
Oph36 & inf\_EB & lowSN & 22.61 & 22.55 & 0.28 & 0.29 & 24.21 & 24.21 & 0.29 & 0.29 & 0.13x0.08 & -85.98 & 1.07 & 1.08 \\
Oph37 & FAIL & NA & NA & NA & NA & NA & NA & NA & NA & NA & NA & NA & NA & NA \\
Oph38 & FAIL & NA & NA & NA & NA & NA & NA & NA & NA & NA & NA & NA & NA & NA \\
Oph39 & 6.05 & SNd & 101.17 & 88.1 &  0.30 &  0.34 & 117.2 & 117.2 & 0.28 & 0.25 & 0.13x0.08 & -86.64 & 1.16 & 1.46 \\
Oph40 & FAIL & B & 180.15 & 140.39 & 0.62 & 0.80 & 180.15 & 180.15 & 0.62 & 0.80 & 0.13x0.08 & -87.60 & 1.0 & 1.0 \\
Oph41 & FAIL & NA & NA & NA & NA & NA & NA & NA & NA & NA & NA & NA & NA & NA \\
Serp01 & FAIL & NA & NA & NA & NA & NA & NA & NA & NA & NA & NA & NA & NA & NA \\
Serp02 & FAIL & NA & NA & NA & NA & NA & NA & NA & NA & NA & NA & NA & NA & NA \\
Serp03 & 8.06 & SNd & 39.94 & 39.56 & 0.23 & 0.23 & 43.96 & 43.96 & 0.22 & 0.23 & 0.13x0.09 & 85.80 & 1.1 & 1.07 \\
Serp04 & FAIL & NA & NA & NA & NA & NA & NA & NA & NA & NA & NA & NA & NA & NA \\
Serp05 & inf & SNd & 54.22 & 52.51 & 0.23 & 0.24 & 57.99 & 57.99 & 0.22 & 0.22 & 0.13x0.09 & 86.22 & 1.07 & 1.14 \\
Serp06 & FAIL & NA & NA & NA & NA & NA & NA & NA & NA & NA & NA & NA & NA & NA \\
Serp07 & FAIL & NA & NA & NA & NA & NA & NA & NA & NA & NA & NA & NA & NA & NA \\
Serp08 & FAIL & NA & NA & NA & NA & NA & NA & NA & NA & NA & NA & NA & NA & NA \\
Serp09 & FAIL & NA & NA & NA & NA & NA & NA & NA & NA & NA & NA & NA & NA & NA \\
Serp10 & FAIL & NA & NA & NA & NA & NA & NA & NA & NA & NA & NA & NA & NA & NA \\
Serp11 & inf & SNd & 43.94 & 42.95 & 0.23 & 0.24 & 45.85 & 45.85 & 0.22 & 0.22 & 0.13x0.09 & 87.49 & 1.04 & 1.07 \\
Serp12 & 18.14 & SNd & 36.97 & 37.91 & 0.29 & 0.28 & 37.86 & 37.86 & 0.28 & 0.27 & 0.13x0.09 & 83.66 & 1.02 & 1.04 \\
Serp13 & 8.06 & SNd & 152.03 & 140.6 & 0.24 & 0.26 & 165.67 & 165.67 & 0.23 & 0.23 & 0.13x0.09 & 87.26 & 1.09 & 1.18 \\
Serp14 & FAIL & NA & NA & NA & NA & NA & NA & NA & NA & NA & NA & NA & NA & NA \\
Serp15 & 18.14 & SNd & 87.68 & 80.5 & 0.27 & 0.30 & 93.48 & 93.48 & 0.26 & 0.25 & 0.13x0.09 & -77.95 & 1.07 & 1.18 \\
Serp16 & 4.03 & SNd & 65.56 & 63.57 & 0.29 & 0.30 & 71.81 & 71.81 & 0.27 & 0.27 & 0.13x0.09 & -78.26 & 1.1 & 1.13 \\
Serp17 & inf & SNd & 28.51 & 29.34 & 0.22 & 0.22 & 32.31 & 32.31 & 0.22 & 0.21 & 0.13x0.09 & -79.01 & 1.13 & 1.15 \\
Serp18 & int & SNd & 100.73 & 81.56 &  0.56 &  0.69 & 182.23 & 182.23 & 0.31 & 0.31 & 0.13x0.09 & -77.81 & 1.81 & 2.21 \\
Serp19 & 8.06 & SNd & 137.59 & 138.15 & 0.23 & 0.23 & 153.46 & 153.46 & 0.21 & 0.20 & 0.13x0.09 & -78.89 & 1.12 & 1.16 \\
Serp20 & 18.14 & SNd & 37.44 & 36.36 & 0.28 & 0.29 & 40.38 & 40.38 & 0.27 & 0.26 & 0.13x0.09 & -77.22 & 1.08 & 1.13 \\
Serp21 & FAIL & NA & NA & NA & NA & NA & NA & NA & NA & NA & NA & NA & NA & NA \\
Serp22 & inf & SNd & 64.6 & 65.72 & 0.23 & 0.23 & 68.67 & 68.67 & 0.23 & 0.22 & 0.13x0.09 & -79.29 & 1.06 & 1.08 \\
Serp23 & inf\_EB & SNd & 52.24 & 52.05 & 0.27 & 0.27 & 53.64 & 53.64 & 0.27 & 0.27 & 0.13x0.09 & 86.72 & 1.03 & 1.01 \\
Serp24 & FAIL & NA & NA & NA & NA & NA & NA & NA & NA & NA & NA & NA & NA & NA \\
Serp25 & inf & SNd & 28.98 & 29.52 & 0.24 & 0.24 & 29.22 & 29.22 & 0.24 & 0.23 & 0.13x0.09 & -79.68 & 1.01 & 1.03 \\
Serp26 & inf\_EB & SNd & 100.7 & 98.28 & 0.30 & 0.30 & 103.75 & 103.75 & 0.28 & 0.28 & 0.13x0.09 & -77.02 & 1.03 & 1.08 \\
Serp27 & 8.06 & SNd & 43.71 & 42.28 & 0.23 & 0.23 & 47.55 & 47.55 & 0.22 & 0.22 & 0.13x0.09 & -79.75 & 1.09 & 1.12 \\
Serp28 & FAIL & NA & NA & NA & NA & NA & NA & NA & NA & NA & NA & NA & NA & NA \\
Serp29 & 8.06 & SNd & 76.13 & 73.67 & 0.27 & 0.27 & 93.67 & 93.67 & 0.23 & 0.22 & 0.13x0.09 & -79.49 & 1.23 & 1.36 \\
Serp30 & 8.06 & SNd & 46.36 & 44.17 & 0.29 & 0.31 & 53.01 & 53.01 & 0.29 & 0.29 & 0.13x0.09 & -76.67 & 1.14 & 1.17 \\
Serp31 & FAIL & NA & NA & NA & NA & NA & NA & NA & NA & NA & NA & NA & NA & NA \\
Serp32 & FAIL & NA & NA & NA & NA & NA & NA & NA & NA & NA & NA & NA & NA & NA \\
Serp33 & FAIL & NA & NA & NA & NA & NA & NA & NA & NA & NA & NA & NA & NA & NA \\
Serp34 & 4.03 & SNd & 75.87 & 72.22 &  0.28 &  0.30 & 93.75 & 93.75 & 0.26 & 0.26 & 0.14x0.09 & -75.88 & 1.24 & 1.3 \\
Serp35 & FAIL & NA & NA & NA & NA & NA & NA & NA & NA & NA & NA & NA & NA & NA \\
Serp36 & FAIL & B & 21.78 & 23.45 & 0.23 & 0.21 & 21.78 & 21.78 & 0.23 & 0.21 & 0.13x0.08 & 87.77 & 1.0 & 1.0 \\
\hline
\hline
\end{longtable*}
\tablenotetext{a}{Field name in the CAMPOS ALMA observation. The characters represent the name of the cloud (eg. Aql: Aquila, ChamI: Chamaeloeon I, ChamII: Chamaeloeon II, Oph: Ophiuchus, OphN: Ophiuchus North, Serp: Serpens)}
\tablenotetext{b}{Final integration time adopted for self-calibration solution (solint). The characters ``ap" represent amplitude calibration applied, and ``inf\_EB" represents the fall-back mode where both the spectral window and observational scans are combined. A number in this column represents the solution interval (in seconds). ``FAIL" denotes the self-calibration process failed. ``None" indicates that self-calibration solutions are found, but are not adopted because the beam size changed by more than 5\%.}
\tablenotetext{c}{Reason for the modified self-calibration pipeline to stop. B represents the area of the beam size changed by more than 5\%, and SNd represents the S/N ratio decreased after the next round of self-calibration.}
\tablenotetext{}{ \textbf{Note}: Column title SN represents the signal-to-noise ratio, and RMS represents the noise level in mJy. The subscript 0 and f represent before and after self-calibration, respectively.  ``Near" represents the near field, the central region within the map with the outer extent around 5 times the major axis of the beam. $\Delta$SN represents the change in signal-to-noise ratio after self-calibration. It is calculated by dividing the final S/N ratio by the initial S/N ratio before self-calibration. The final beam size is given in arcseconds, and the position angles (PA) are in degrees.}
\end{center}

\begin{center}
\begin{longtable*}{ c c c c c c c c c c c c c c c c}
\caption{Summary of properties of the Natural weighted images before and after  self-calibration using the modified self-calibration pipeline 
}\\ %
\label{table:3}
&  &    & \\
\hline
\hline
Field\tablenotemark{a}  & Final\tablenotemark{b}  & Stop\tablenotemark{c} & SN$_0$ & SN$_0$ & RMS$_0$	& RMS$_0$ & SN$_f$ & SN$_f$ & RMS$_f$ & RMS$_f$ & Beam$_f$ & PA & $\Delta$SN & $\Delta$SN  \\
& selfcal & reason & & (near) & & (near)&  & (near) & &(near) & & & & (near)\\
\hline
\endfirsthead
\multicolumn{6}{c}%
{\tablename\ \thetable\ -- \textit{Continued from previous page}} \\
\hline

\hline
\endhead
\hline \multicolumn{4}{r}{\textit{Continued on next page}} \\
\endfoot
\hline
\endlastfoot
\hline                        
Aql01 & 8.06 & B & 268.77 & 289.44 & 0.14 & 0.13 & 365.25 & 365.25 & 0.10 & 0.10 & 0.20x0.15 & -81.30 & 1.36 & 1.28 \\
Aql02 & inf\_EB & SNd & 168.08 & 140.28 & 0.10 & 0.11 & 178.22 & 178.22 & 0.09 & 0.09 & 0.20x0.14 & -82.12 & 1.06 & 1.16 \\
Aql03 & int & B & 122.77 & 94.31 &  0.10 &  0.13 & 134.89 & 134.89 & 0.09 & 0.11 & 0.19x0.14 & -80.08 & 1.1 & 1.11 \\
Aql04 & 18.14 & B & 187.89 & 166.46 & 0.10 & 0.12 & 222.98 & 222.98 & 0.09 & 0.09 & 0.19x0.14 & -82.01 & 1.19 & 1.29 \\
Aql05 & inf & B & 42.5 & 43.52 & 0.08 & 0.08 & 44.06 & 44.06 & 0.08 & 0.07 & 0.19x0.14 & -80.22 & 1.04 & 1.15 \\
Aql06 & 4.03 & B & 95.22 & 89.1 &  0.08 &  0.09 & 101.96 & 101.96 & 0.08 & 0.08 & 0.19x0.15 & -79.28 & 1.07 & 1.2 \\
Aql07 & inf & B & 218.99 & 177.69 & 0.09 & 0.11 & 245.91 & 245.91 & 0.08 & 0.09 & 0.19x0.14 & -79.72 & 1.12 & 1.21 \\
Aql08 & 18.14 & B & 142.24 & 136.11 & 0.08 & 0.09 & 146.31 & 146.31 & 0.08 & 0.08 & 0.19x0.15 & -79.41 & 1.03 & 1.07 \\
Aql09 & inf & B & 48.54 & 42.04 & 0.08 & 0.09 & 50.42 & 50.42 & 0.08 & 0.09 & 0.19x0.14 & -79.91 & 1.04 & 1.12 \\
Aql10 & int & SNd,B & 87.7 & 55.34 &  0.20 &  0.32 & 94.09 & 94.09 & 0.19 & 0.30 & 0.19x0.14 & -80.49 & 1.07 & 1.07 \\
Aql11 & inf & B & 57.4 & 48.49 & 0.08 & 0.09 & 60.79 & 60.79 & 0.08 & 0.10 & 0.19x0.14 & -79.68 & 1.06 & 1.0 \\
Aql12 & 4.03 & B & 224.31 & 185.44 &  0.10 &  0.12 & 248.31 & 248.31 & 0.08 & 0.08 & 0.19x0.14 & -79.89 & 1.11 & 1.36 \\
Aql13 & FAIL & B & 305.8 & 300.48 & 0.14 & 0.14 & 305.8 & 305.8 & 0.14 & 0.14 & 0.19x0.14 & -80.42 & 1.0 & 1.0 \\
Aql14 & int & B & 162.15 & 157.64 &  0.08 &  0.09 & 179.61 & 179.61 & 0.08 & 0.08 & 0.19x0.14 & -80.39 & 1.11 & 1.15 \\
Aql15 & inf & B & 70.83 & 74.2 & 0.08 & 0.07 & 72.36 & 72.36 & 0.08 & 0.08 & 0.19x0.14 & -79.58 & 1.02 & 0.98 \\
Aql16 & int & B & 85.29 & 87.23 &  0.08 &  0.08 & 92.45 & 92.45 & 0.08 & 0.09 & 0.19x0.14 & -79.53 & 1.08 & 0.97 \\
Aql17 & 4.03 & B & 90.56 & 77.43 &  0.08 &  0.09 & 96.64 & 96.64 & 0.08 & 0.09 & 0.20x0.15 & -76.23 & 1.07 & 1.11 \\
Aql18 & inf & B & 177.72 & 134.82 & 0.10 & 0.13 & 195.64 & 195.64 & 0.09 & 0.09 & 0.19x0.14 & -79.05 & 1.1 & 1.41 \\
ChamII01 & 4.03 & SNd & 83.93 & 66.12 & 1.22  & 1.55  & 1569.74 & 1323.17 & 0.09 & 0.10 & 0.19x0.13 & 10.55 & 18.7 & 20.01 \\
ChamII02 & FAIL & B & 87.33 & 64.38 & 0.15 & 0.20 & 87.33 & 87.33 & 0.15 & 0.20 & 0.19x0.13 & 10.11 & 1.0 & 1.0 \\
ChamII03 & FAIL & B & 51.93 & 35.64 & 0.09 & 0.13 & 51.93 & 51.93 & 0.09 & 0.13 & 0.19x0.13 & 9.17 & 1.0 & 1.0 \\
ChamI01 & FAIL & B & 53.51 & 28.35 & 0.07 & 0.14 & 53.51 & 53.51 & 0.07 & 0.14 & 0.13x0.10 & 4.21 & 1.0 & 1.0 \\
ChamI02 & FAIL & B & 82.23 & 46.07 & 0.16 & 0.28 & 82.23 & 82.23 & 0.16 & 0.28 & 0.13x0.10 & 3.09 & 1.0 & 1.0 \\
ChamI03 & FAIL & NA & NA & NA & NA & NA & NA & NA & NA & NA & NA & NA & NA & NA \\
ChamI04 & FAIL & NA & NA & NA & NA & NA & NA & NA & NA & NA & NA & NA & NA & NA \\
ChamI05 & FAIL & B & 33.24 & 23.43 & 0.08 & 0.11 & 33.24 & 33.24 & 0.08 & 0.11 & 0.13x0.10 & 3.87 & 1.0 & 1.0 \\
ChamI06 & FAIL & B & 56.0 & 48.72 & 0.10 & 0.12 & 56.0 & 56.0 & 0.10 & 0.12 & 0.13x0.10 & 4.06 & 1.0 & 1.0 \\
ChamI07 & FAIL & B & 65.67 & 56.37 & 0.13 & 0.15 & 65.67 & 65.67 & 0.13 & 0.15 & 0.13x0.10 & 5.03 & 1.0 & 1.0 \\
ChamI08 & FAIL & B & 58.32 & 53.29 & 0.07 & 0.08 & 58.32 & 58.32 & 0.07 & 0.08 & 0.16x0.14 & 37.53 & 1.0 & 1.0 \\
ChamI09 & FAIL & NA & NA & NA & NA & NA & NA & NA & NA & NA & NA & NA & NA & NA \\
ChamI10 & FAIL & NA & NA & NA & NA & NA & NA & NA & NA & NA & NA & NA & NA & NA \\
ChamI11 & FAIL & B & 97.54 & 83.35 & 0.08 & 0.09 & 97.54 & 97.54 & 0.08 & 0.09 & 0.15x0.15 & -78.58 & 1.0 & 1.0 \\
ChamI12 & inf & B & 126.53 & 81.66 & 0.11 & 0.17 & 271.35 & 271.35 & 0.06 & 0.06 & 0.15x0.15 & -28.88 & 2.14 & 3.68 \\
ChamI13 & FAIL & B & 31.25 & 31.64 & 0.10 & 0.10 & 31.25 & 31.25 & 0.10 & 0.10 & 0.20x0.12 & 41.28 & 1.0 & 1.0 \\
CrAus01 & int & B & 109.2 & 62.97 &  0.36 &  0.62 & 333.42 & 333.42 & 0.12 & 0.08 & 0.16x0.12 & -84.33 & 3.05 & 7.56 \\
CrAus02 & 8.06 & B & 64.01 & 28.3 & 0.67 & 1.52 & 181.92 & 181.92 & 0.23 & 0.18 & 0.16x0.12 & -84.61 & 2.84 & 7.89 \\
CrAus03 & 4.03 & B & 137.76 & 85.14 &  0.13 &  0.21 & 229.05 & 229.05 & 0.08 & 0.08 & 0.16x0.12 & -86.92 & 1.66 & 2.75 \\
CrAus04 & int & B & 124.3 & 87.68 &  0.20 &  0.28 & 290.8 & 290.8 & 0.09 & 0.08 & 0.17x0.12 & -89.34 & 2.34 & 3.79 \\
CrAus05 & inf & SNd,B & 181.39 & 109.31 & 0.53 & 0.89 & 1259.63 & 1259.63 & 0.08 & 0.10 & 0.16x0.12 & -84.67 & 6.94 & 9.27 \\
CrAus06 & FAIL & NA & NA & NA & NA & NA & NA & NA & NA & NA & NA & NA & NA & NA \\
CrAus07 & inf\_ap & None & 85.19 & 63.98 & 0.33 & 0.44 & 234.95 & 234.95 & 0.13 & 0.14 & 0.16x0.13 & -83.13 & 2.76 & 3.55 \\
CrAus08 & int & B & 119.4 & 126.21 &  0.52 &  0.49 & 411.36 & 411.36 & 0.16 & 0.13 & 0.16x0.12 & -84.72 & 3.45 & 3.97 \\
CrAus09 & inf & B & 43.46 & 45.15 & 0.37 & 0.36 & 110.07 & 110.07 & 0.14 & 0.31 & 0.16x0.12 & -84.60 & 2.53 & 1.15 \\
CrAus10 & 4.03 & B & 151.46 & 89.67 &  0.24 &  0.40 & 382.83 & 382.83 & 0.10 & 0.14 & 0.16x0.13 & -83.23 & 2.53 & 3.07 \\
CrAus11 & inf & B & 152.16 & 85.18 & 0.56 & 1.00 & 1106.74 & 1106.74 & 0.08 & 0.09 & 0.16x0.12 & -85.10 & 7.27 & 11.81 \\
OphN01 & inf & SNd,B & 303.14 & 258.51 & 0.12 & 0.14 & 1355.97 & 1355.97 & 0.03 & 0.03 & 0.17x0.13 & -79.32 & 4.47 & 4.4 \\
OphN02 & 300-ap & B & 170.5 & 200.68 & 0.14 & 0.12 & 467.38 & 819.0 & 0.05 & 0.03 & 0.17x0.13 & -82.43 & 2.74 & 4.08 \\
OphN03 & 26.21 & B & 501.62 & 209.09 & 0.06 & 0.13 & 932.39 & 348.57 & 0.03 & 0.08 & 0.20x0.14 & -78.85 & 1.86 & 1.67 \\
Oph01 & 18.14 & B & 62.26 & 73.38 & 0.09 & 0.08 & 73.21 & 73.21 & 0.08 & 0.07 & 0.21x0.14 & -81.41 & 1.18 & 1.12 \\
Oph02 & 4.03 & SNd & 149.75 & 138.42 &  0.11 &  0.11 & 188.45 & 188.45 & 0.08 & 0.08 & 0.21x0.14 & -81.41 & 1.26 & 1.43 \\
Oph03 & inf & B & 206.89 & 204.22 & 0.14 & 0.14 & 330.08 & 330.08 & 0.09 & 0.07 & 0.21x0.14 & -79.96 & 1.6 & 1.97 \\
Oph04 & FAIL & NA & NA & NA & NA & NA & NA & NA & NA & NA & NA & NA & NA & NA \\
Oph05 & 8.06 & B & 203.65 & 181.91 & 0.11 & 0.13 & 280.73 & 280.73 & 0.08 & 0.07 & 0.21x0.14 & -80.54 & 1.38 & 1.71 \\
Oph06 & int & B & 143.49 & 118.09 &  0.10 &  0.13 & 186.34 & 186.34 & 0.08 & 0.07 & 0.21x0.14 & -82.11 & 1.3 & 1.76 \\
Oph07 & inf & B & 126.52 & 105.03 & 0.15 & 0.18 & 212.67 & 212.67 & 0.09 & 0.07 & 0.21x0.14 & -80.20 & 1.68 & 2.38 \\
Oph08 & 4.03 & B & 107.77 & 95.77 &  0.29 &  0.33 & 304.01 & 304.01 & 0.10 & 0.08 & 0.22x0.14 & -81.47 & 2.82 & 3.93 \\
Oph09 & inf & B & 225.79 & 189.94 & 0.09 & 0.11 & 260.22 & 260.22 & 0.08 & 0.07 & 0.21x0.14 & -80.32 & 1.15 & 1.56 \\
Oph10 & 16.13 & B & 118.97 & 149.62 & 0.26 & 0.21 & 255.18 & 255.18 & 0.12 & 0.19 & 0.21x0.14 & -79.96 & 2.14 & 1.12 \\
Oph11 & 8.06 & SNd & 77.05 & 77.7 & 0.10 & 0.10 & 101.62 & 101.62 & 0.08 & 0.07 & 0.22x0.14 & -81.41 & 1.32 & 1.46 \\
Oph12 & int & B & 399.52 & 288.44 &  0.12 &  0.17 & 633.79 & 633.79 & 0.08 & 0.08 & 0.21x0.14 & -80.05 & 1.59 & 2.02 \\
Oph13 & inf\_ap & None & 86.36 & 106.58 & 0.23 & 0.18 & 209.38 & 209.38 & 0.09 & 0.10 & 0.21x0.14 & -82.28 & 2.42 & 1.89 \\
Oph14 & inf & B & 163.65 & 128.74 & 0.09 & 0.11 & 184.52 & 184.52 & 0.08 & 0.07 & 0.21x0.14 & -79.24 & 1.13 & 1.46 \\
Oph15 & 18.14 & B & 96.53 & 131.13 & 0.13 & 0.09 & 128.79 & 128.79 & 0.09 & 0.08 & 0.21x0.14 & -79.92 & 1.33 & 1.15 \\
Oph16 & inf\_ap & None & 47.46 & 41.61 & 0.08 & 0.09 & 51.19 & 51.19 & 0.08 & 0.08 & 0.20x0.15 & -80.08 & 1.08 & 1.15 \\
Oph17 & FAIL & B & 90.54 & 94.26 & 0.08 & 0.07 & 90.54 & 90.54 & 0.08 & 0.07 & 0.21x0.14 & -81.21 & 1.0 & 1.0 \\
Oph18 & 8.06 & B & 176.69 & 145.6 & 0.09 & 0.11 & 202.17 & 202.17 & 0.08 & 0.07 & 0.21x0.15 & -79.23 & 1.14 & 1.57 \\
Oph19 & inf & B & 180.47 & 141.79 & 0.09 & 0.12 & 204.21 & 204.21 & 0.08 & 0.07 & 0.21x0.14 & -79.52 & 1.13 & 1.62 \\
Oph20 & int & B & 169.42 & 106.37 &  0.08 &  0.13 & 181.62 & 181.62 & 0.08 & 0.09 & 0.21x0.14 & -80.01 & 1.07 & 1.45 \\
Oph21 & inf & B & 163.66 & 142.92 & 0.09 & 0.10 & 192.92 & 192.92 & 0.08 & 0.08 & 0.21x0.14 & -79.34 & 1.18 & 1.27 \\
Oph22 & 8.06 & B & 98.27 & 95.51 & 0.08 & 0.08 & 103.81 & 103.81 & 0.08 & 0.07 & 0.21x0.15 & -82.07 & 1.06 & 1.24 \\
Oph23 & 4.03 & B & 117.81 & 130.83 &  0.08 &  0.07 & 126.57 & 126.57 & 0.08 & 0.07 & 0.21x0.14 & -79.05 & 1.07 & 1.05 \\
Oph24 & FAIL & NA & NA & NA & NA & NA & NA & NA & NA & NA & NA & NA & NA & NA \\
Oph25 & inf & B & 48.1 & 52.51 & 0.08 & 0.07 & 51.25 & 51.25 & 0.08 & 0.07 & 0.20x0.14 & -78.92 & 1.07 & 1.03 \\
Oph26 & inf & B & 134.06 & 82.42 & 0.09 & 0.15 & 147.66 & 147.66 & 0.08 & 0.11 & 0.20x0.14 & -79.01 & 1.1 & 1.3 \\
Oph27 & inf & B & 115.48 & 101.05 & 0.08 & 0.10 & 121.9 & 121.9 & 0.08 & 0.09 & 0.20x0.14 & -78.98 & 1.06 & 1.1 \\
Oph28 & FAIL & NA & NA & NA & NA & NA & NA & NA & NA & NA & NA & NA & NA & NA \\
Oph29 & 4.03 & B & 70.85 & 77.06 & 0.08 & 0.07 & 76.6 & 76.6 & 0.08 & 0.07 & 0.21x0.15 & -77.97 & 1.08 & 1.1 \\
Oph30 & inf & B & 58.85 & 77.5 & 0.08 & 0.06 & 62.46 & 62.46 & 0.08 & 0.06 & 0.20x0.14 & -78.74 & 1.06 & 1.0 \\
Oph31 & 18.14 & B & 275.5 & 168.54 & 0.13 & 0.22 & 343.8 & 371.72 & 0.10 & 0.10 & 0.22x0.14 & -79.67 & 1.25 & 2.21 \\
Oph32 & inf\_EB & SNd & 112.55 & 93.78 & 0.08 & 0.10 & 116.19 & 116.19 & 0.08 & 0.08 & 0.20x0.14 & -78.86 & 1.03 & 1.19 \\
Oph33 & FAIL & NA & NA & NA & NA & NA & NA & NA & NA & NA & NA & NA & NA & NA \\
Oph34 & inf\_ap & None & 166.77 & 197.8 &  0.31 &  0.27 & 477.02 & 477.02 & 0.12 & 0.08 & 0.19x0.13 & -80.40 & 2.86 & 3.56 \\
Oph35 & FAIL & B & 81.69 & 75.04 & 0.07 & 0.08 & 81.69 & 81.69 & 0.07 & 0.08 & 0.18x0.13 & -81.62 & 1.0 & 1.0 \\
Oph36 & 6.05 & B & 111.81 & 80.59 &  0.07 &  0.09 & 126.66 & 126.66 & 0.06 & 0.06 & 0.18x0.13 & -80.52 & 1.13 & 1.66 \\
Oph37 & FAIL & NA & NA & NA & NA & NA & NA & NA & NA & NA & NA & NA & NA & NA \\
Oph38 & inf\_EB & SNd & 51.04 & 46.24 & 0.06 & 0.07 & 57.33 & 57.33 & 0.06 & 0.05 & 0.19x0.13 & -81.30 & 1.12 & 1.46 \\
Oph39 & int & B & 252.17 & 102.99 &  0.15 &  0.36 & 409.13 & 409.13 & 0.10 & 0.20 & 0.18x0.13 & -81.70 & 1.62 & 1.85 \\
Oph40 & 4.03 & B & 91.29 & 55.64 &  2.58 &  4.23 & 173.36 & 173.36 & 1.31 & 2.44 & 0.19x0.13 & -81.36 & 1.9 & 1.67 \\
Oph41 & FAIL & B & 86.66 & 75.51 & 0.09 & 0.10 & 86.66 & 86.66 & 0.09 & 0.10 & 0.19x0.13 & -81.65 & 1.0 & 1.0 \\
Serp01 & FAIL & NA & NA & NA & NA & NA & NA & NA & NA & NA & NA & NA & NA & NA \\
Serp02 & FAIL & NA & NA & NA & NA & NA & NA & NA & NA & NA & NA & NA & NA & NA \\
Serp03 & int & B & 121.67 & 101.69 &  0.08 &  0.10 & 138.08 & 138.08 & 0.08 & 0.09 & 0.19x0.15 & 89.47 & 1.13 & 1.2 \\
Serp04 & FAIL & NA & NA & NA & NA & NA & NA & NA & NA & NA & NA & NA & NA & NA \\
Serp05 & inf & B & 136.31 & 106.05 & 0.11 & 0.14 & 163.57 & 163.57 & 0.09 & 0.14 & 0.19x0.15 & 89.70 & 1.2 & 1.03 \\
Serp06 & inf\_EB & B & 80.08 & 73.58 & 0.08 & 0.09 & 85.06 & 85.06 & 0.08 & 0.08 & 0.19x0.15 & 87.83 & 1.06 & 1.11 \\
Serp07 & inf & SNd & 75.21 & 62.07 & 0.08 & 0.10 & 79.18 & 79.18 & 0.08 & 0.08 & 0.19x0.15 & -89.92 & 1.05 & 1.28 \\
Serp08 & 18.14 & B & 35.93 & 32.02 & 0.08 & 0.09 & 38.53 & 38.53 & 0.08 & 0.09 & 0.20x0.15 & 88.57 & 1.07 & 1.11 \\
Serp09 & FAIL & B & 27.56 & 28.35 & 0.08 & 0.08 & 27.56 & 27.56 & 0.08 & 0.08 & 0.19x0.15 & -89.90 & 1.0 & 1.0 \\
Serp10 & 18.14 & B & 44.61 & 40.61 & 0.08 & 0.09 & 47.41 & 47.41 & 0.08 & 0.08 & 0.19x0.15 & 88.78 & 1.06 & 1.09 \\
Serp11 & 8.06 & B & 107.48 & 98.54 & 0.15 & 0.17 & 157.98 & 157.98 & 0.11 & 0.12 & 0.20x0.15 & -87.32 & 1.47 & 1.42 \\
Serp12 & int & B & 108.06 & 106.66 &  0.15 &  0.15 & 147.8 & 147.8 & 0.11 & 0.13 & 0.19x0.15 & 89.51 & 1.37 & 1.18 \\
Serp13 & inf & B & 315.25 & 289.41 & 0.19 & 0.21 & 605.78 & 605.78 & 0.10 & 0.15 & 0.19x0.15 & -89.24 & 1.92 & 1.34 \\
Serp14 & inf\_ap & None & 28.68 & 35.74 & 0.08 & 0.06 & 31.75 & 31.75 & 0.08 & 0.06 & 0.18x0.14 & 85.63 & 1.11 & 1.22 \\
Serp15 & int & B & 256.96 & 205.8 &  0.13 &  0.16 & 368.65 & 368.65 & 0.09 & 0.08 & 0.20x0.14 & -69.92 & 1.43 & 1.99 \\
Serp16 & int & B & 108.91 & 80.81 &  0.23 &  0.30 & 170.01 & 170.01 & 0.15 & 0.26 & 0.20x0.14 & -69.88 & 1.56 & 1.17 \\
Serp17 & 8.06 & B & 91.27 & 87.57 & 0.09 & 0.10 & 105.52 & 105.52 & 0.09 & 0.10 & 0.21x0.14 & -72.03 & 1.16 & 1.11 \\
Serp18 & 18.14 & B & 94.27 & 90.4 & 1.33 & 1.39 & 236.64 & 236.64 & 0.50 & 0.62 & 0.20x0.14 & -69.69 & 2.51 & 2.11 \\
Serp19 & 4.03 & B & 257.43 & 260.29 &  0.19 &  0.18 & 483.23 & 483.23 & 0.10 & 0.10 & 0.20x0.14 & -68.91 & 1.88 & 1.81 \\
Serp20 & 4.03 & B & 94.58 & 87.05 &  0.14 &  0.16 & 144.08 & 144.08 & 0.09 & 0.10 & 0.20x0.14 & -69.59 & 1.52 & 1.54 \\
Serp21 & FAIL & B & 51.19 & 48.61 & 0.08 & 0.09 & 51.19 & 51.19 & 0.08 & 0.09 & 0.19x0.15 & -89.94 & 1.0 & 1.0 \\
Serp22 & inf & B & 186.31 & 183.39 & 0.09 & 0.10 & 222.27 & 222.27 & 0.08 & 0.08 & 0.20x0.14 & -68.79 & 1.19 & 1.24 \\
Serp23 & 18.14 & B & 195.98 & 195.01 & 0.11 & 0.11 & 267.63 & 267.63 & 0.08 & 0.08 & 0.20x0.15 & -88.39 & 1.37 & 1.49 \\
Serp25 & inf\_EB & SNd & 34.23 & 33.75 & 0.31 & 0.31 & 35.86 & 35.86 & 0.30 & 0.28 & 0.20x0.14 & -68.77 & 1.05 & 1.12 \\
Serp26 & int & B & 251.14 & 237.43 &  0.19 &  0.20 & 352.14 & 352.14 & 0.13 & 0.10 & 0.20x0.14 & -69.39 & 1.4 & 1.89 \\
Serp27 & int & B & 162.66 & 151.86 &  0.10 &  0.11 & 199.31 & 199.31 & 0.08 & 0.08 & 0.21x0.14 & -70.93 & 1.23 & 1.35 \\
Serp29 & int & B & 102.77 & 111.3 &  0.35 &  0.32 & 278.43 & 278.43 & 0.14 & 0.13 & 0.21x0.14 & -70.82 & 2.71 & 2.54 \\
Serp30 & int & B & 105.03 & 102.37 &  0.17 &  0.17 & 128.45 & 128.45 & 0.15 & 0.15 & 0.20x0.14 & -68.79 & 1.22 & 1.3 \\
Serp31 & 18.14 & B & 31.07 & 29.78 & 0.08 & 0.09 & 35.62 & 35.62 & 0.08 & 0.09 & 0.20x0.14 & -68.66 & 1.15 & 1.15 \\
Serp33 & 18.14 & B & 28.5 & 28.62 & 0.09 & 0.09 & 33.11 & 33.11 & 0.09 & 0.09 & 0.20x0.14 & -68.49 & 1.16 & 1.14 \\
Serp34 & int & B & 194.35 & 180.89 &  0.18 &  0.19 & 398.97 & 398.97 & 0.09 & 0.08 & 0.21x0.14 & -67.03 & 2.05 & 2.74 \\
Serp35 & FAIL & B & 40.93 & 36.2 & 0.08 & 0.09 & 40.93 & 40.93 & 0.08 & 0.09 & 0.19x0.15 & -89.54 & 1.0 & 1.0 \\
Serp36 & FAIL & B & 91.92 & 90.29 & 0.08 & 0.09 & 91.92 & 91.92 & 0.08 & 0.09 & 0.19x0.15 & -88.56 & 1.0 & 1.0 \\
\hline
\hline
\end{longtable*}
\tablenotetext{a}{Field name in the CAMPOS ALMA observation. The characters represent the name of the cloud (eg. Aql: Aquila, ChamI: Chamaeloeon I, ChamII: Chamaeloeon II, Oph: Ophiuchus, OphN: Ophiuchus North, Serp: Serpens)}
\tablenotetext{b}{Final integration time adopted for self-calibration solution (solint). The characters ``ap" represent amplitude calibration applied, and ``inf\_EB" represents the fall-back mode where both the spectral window and observational scans are combined. A number in this column represents the solution interval (in seconds). ``FAIL" denotes the self-calibration process failed. ``None" indicates that self-calibration solutions are found, but are not adopted because the beam size changed by more than 5\%.}
\tablenotetext{c}{Reason for the modified self-calibration pipeline to stop. B represents the area of the beam size changed by more than 5\%, and SNd represents the S/N ratio decreased after the next round of self-calibration.}
\tablenotetext{}{ \textbf{Note}: Column title SN represents the signal-to-noise ratio, and RMS represents the noise level in mJy. The subscript 0 and f represent before and after self-calibration, respectively. ``Near" represents the near field, the central region within the map with the outer extent around 5 times the major axis of the beam. $\Delta$SN represents the change in signal-to-noise ratio after self-calibration. It is calculated by dividing the final S/N ratio by the initial S/N ratio before self-calibration. The final beam size is given in arcseconds, and the position angles (PA) are in degrees.}

\end{center}

\clearpage



\section{Summary of sources detected in the CAMPOS survey}
\label{Appendix:Source_summary}
In this section, we present all of the sources detected in this survey in \autoref{table:source_cross_match} and \autoref{table:source}. We cross-matched the CAMPOS sources with the literature and the names for each source are organized in \autoref{table:source_cross_match}. In \autoref{table:source} we  report the following properties: source position,  disk size  (major and minor axes), and position angle. Note that for multiple systems that are unresolved in \citet{2015ApJS..220...11D}, we assume the same bolometric temperature and class for all sources in the multiple system. The images of the sources identified in \autoref{table:source} are shown in \autoref{Appendix:Image_gallery}.


\begin{center}
\begin{longtable*}{ c c c c c c}
\caption{Names of all the sources detected in this survey}\\ %
\label{table:source_cross_match}
&  &    & \\
\hline
\hline

CAMPOS ID & Source Name & Alternative Source Names  \\ 
\hline
\endfirsthead
\multicolumn{6}{c}%
{\tablename\ \thetable\ -- \textit{Continued from previous page}} \\
\hline

\hline
\endhead
\hline \multicolumn{4}{r}{\textit{Continued on next page}} \\
\endfoot
\hline
\endlastfoot
\hline                        
Aql-01-0 & Aqu-MM3 & IRAS 18264-0143, SSTgbs J1829053-014156, eHOPS-aql-31 \\
Aql-02-0 & Aqu-MM5 & IRAS 18268-0140, SSTgbs J1829234-013855, eHOPS-aql-48 \\
Aql-03-0 & SSTgbs J1829381-015100 & NA \\
Aql-03-1 & SerpS-MM1 & NA \\
Aql-04-0 & SerpS-MM2 a & eHOPS-aql-54, SerpS-MM2 \\
Aql-04-1 & SerpS-MM2 b & NA \\
Aql-04-2 & SSTgbs J1829386-015100 & NA \\
Aql-05-0 & SSTgbs J1829419-015011 & NA \\
Aql-06-0 & SerpS-MM4 & SSTgbs J1829433-015651, eHOPS-aql-57 \\
Aql-07-0 & SerpS-MM6b & NA \\
Aql-07-1 & SerpS-MM6a & IRAS 18271-0157, SSTgbs J1829470-015548, eHOPS-aql-60, SerpS-MM6 \\
Aql-08-0 & eHOPS-aql-87 & SSTgbs J1829594-020106, serps7 \\
Aql-09-0 & eHOPS-aql-96 a & SSTgbs J1830010-020608, eHOPS-aql-96 \\
Aql-09-1 & eHOPS-aql-96 b & NA \\
Aql-10-0 & MIRES G028.6593+03.8185 & NA \\
Aql-10-1 & SSTgbs J1830027-020259 & SSTgbs J1830027-020259, serp 20 \\
Aql-10-2 & SSTgbs J1830024-020257 & SSTgbs J1830024-020257, serp 17 \\
Aql-10-3 & SerpS10-mm & NA \\
Aql-10-4 & SerpS-MM16 & serp16, CARMA-3, eHOPS-aql-102 \\
Aql-10-5 & eHOPS-aql-107 & serp 29, CARMA-5 \\
Aql-11-0 & eHOPS-aql-122 & SSTgbs J1830170-020958, SerpS-MM24 \\
Aql-11-1 & eHOPS-aql-123 & SSTgbs J1830174-020958 \\
Aql-12-0 & IRAS 18278-0156 & SSTgbs J1830245-01541, eHOPS-aql-127 \\
Aql-13-0 & SerpS-MM25 & SSTgbs J1830258-021042, eHOPS-aql-129 \\
Aql-14-0 & IRAS 18278-0158 b & NA \\
Aql-14-1 & IRAS 18278-0158 a & SSTgbs J1830292-015642, eHOPS-aql-135, IRAS 18278-0158 \\
Aql-15-0 & SSTgbs J1830469-015651 & NA \\
Aql-15-1 & eHOPS-aql-138 & SSTgbs J1830469-015645 \\
Aql-16-0 & eHOPS-aql-139 a & SSTgbs J1830487-015601, eHOPS-aql-139 \\
Aql-16-1 & eHOPS-aql-139 b & NA \\
Aql-17-0 & W40-MM30 & SSTgbs J1831521-020126, eHOPS-aql-151 \\
Aql-18-0 & W40-MM36 & SSTgbs J1832131-015730, eHOPS-aql-154 \\
ChamI-01-0 & IRAS 11030-7702 & SSTgbs J1104227-771808 \\
ChamI-02-0 & TIC 454291385 & SSTgbs J1106464-772232 \\
ChamI-05-0 & ISO-ChaI 101 & SSTgbs J1107213-772211 \\
ChamI-06-0 & Ass Cha T 1-15 & SSTgbs J1107435-773941 \\
ChamI-07-0 & V* HO Cha a & NA \\
ChamI-07-1 & V* HO Cha b & SSTgbs J1108029-773842 \\
ChamI-08-0 & V* GM Cha & SSTgbs J1109285-763328 \\
ChamI-09-0 & ChamI-9 mm & NA \\
ChamI-09-1 & ISO-ChaI 204 & SSTgbs J1109461-763446 \\
ChamI-10-0 & ISO-ChaI 207 & SSTgbs J1109472-772629 \\
ChamI-11-0 & IR Cha INa4 & SSTgbs J1110033-763311 \\
ChamI-12-0 & ISO-ChaI 237 & SSTgbs J1110113-763529 \\
ChamI-13-0 & TIC 454329229 & SSTgbs J1111107-764157 \\
ChamII-01-0 & V* DK Cha & SSTc2d J125317.2-770710 \\
ChamII-02-0 & IRAS 12500-7658 & SSTc2d J125342.9-771511 \\
ChamII-03-0 & IRAS 12553-7651 & STc2d J125906.6-770740 \\
CrAus-01-0 & V* S CrA B & V* S CrA \\
CrAus-01-1 & V* S CrA A & SSTgbs J1901086-365720, ISO-CrA 116, IRAS 18577-3701, V* S CrA \\
CrAus-02-0 & IRS 2 & SSTgbs J1901415-365831, R CrA 1, eHOPS-cra-2 \\
CrAus-03-0 & IRS 5a & IRS 5a \\
CrAus-03-1 & IRS 5b & SSTgbs J1901480-365722, eHOPS-cra-3, IRS5A \\
CrAus-04-0 & IRS 5N & SSTgbs J1901484-365714, eHOPS-cra-4 \\
CrAus-05-0 & V* V710 CrA & SSTgbs J1901506-365809, IRS 1, eHOPS-cra-6 \\
CrAus-07-0 & IRS 7A & SSTgbs J1901553-365721, eHOPS-cra-7, RCrA IRS 7A \\
CrAus-07-1 & SMM1C & NA \\
CrAus-07-2 & CrAus7-mm & NA \\
CrAus-08-0 & CrAus8-mm1 & NA \\
CrAus-08-1 & IRS7B-a & SSTgbs J1901564-365728, SMM 1B, RCrA IRS7B, eHOPS-cra-8 \\
CrAus-08-2 & CXO 34 & J190155.76-365727.7 \\
CrAus-08-3 & IRS7B-b & SSTgbs J1901564-365728 \\
CrAus-09-0 & SMM 2 & SSTgbs J1901585-365708, eHOPS-cra-9 \\
CrAus-10-0 & IRAS 32 A & SSTgbs J1902586-370735, eHOPS-cra-10, IRAS 18595-3712, R CrA IRAS 32 A \\
CrAus-10-1 & IRAS 32 B & R CrA IRAS 32 B \\
CrAus-11-0 & V* VV CrA A & SSTgbs J1903068-371249, eHOPS-cra-11, IRAS 33 \\
CrAus-11-1 & V* VV CrA B & NA \\
OphN-01-0 & IRAS 16442-0930 & SSTgbs J1646582-093519 \\
OphN-02-0 & IRAS 16459-1411 & SSTgbs J1648456-141636 \\
OphN-03-0 & CB 68 SMM 1 & IRAS 16544-1604 \\
Oph-01-0 & ISO-Oph 2a & NA \\
Oph-01-1 & ISO-Oph 2b & NA \\
Oph-02-0 & DoAr 20 & IRAS 16229-2413, HH 312, ISO-Oph 6 \\
Oph-03-0 & ISO-Oph 17 & NA \\
Oph-05-0 & Elias 2-20 & IRAS 16233-2421, ISO-Oph 24 \\
Oph-06-0 & Oph-emb 8 & eHOPS-oph-2, ISO-Oph 29 \\
Oph-06-1 & ISO-Oph 31 & NA \\
Oph-07-0 & DoAr 25 & IRAS 16234-2436, ISO-Oph 38 \\
Oph-08-0 & Elias 2-24 & IRAS 16233-2409, ISO-Oph 40 \\
Oph-09-0 & Oph-emb 9 & eHOPS-oph-4 \\
Oph-10-0 & VLA 1623B & NA \\
Oph-10-1 & VLA 1623Ab & NA \\
Oph-10-2 & VLA 1623Aa & HH 313, eHOPS-oph-5 \\
Oph-10-3 & VLA 1623W & Oph-emb 3 \\
Oph-11-0 & Oph-emb 22 & eHOPS-oph-7, eHOPS-oph-8, ISO-Oph 54 \\
Oph-12-0 & IRAS 16237-2428 & eHOPS-oph-9, ISO-Oph 65 \\
Oph-13-0 & Elia 2-27 & eHOPS-oph-10, ISO-Oph 67 \\
Oph-14-0 & Oph-emb 23 & ISO-Oph 70 \\
Oph-14-1 & CFHTWIR-Oph 43 & NA \\
Oph-15-0 & DoAr 29 & IRAS 16239-2438, ISO-Oph 88 \\
Oph-16-0 & Oph-emb 21 & IRAS 16240-2430(W), eHOPS-oph-12, ISO-Oph 92 \\
Oph-17-0 & ISO-Oph 93 & NA \\
Oph-18-0 & Oph-emb 6 & eHOPS-oph-16, ISO-Oph 99 \\
Oph-19-0 & Oph-emb 20 & eHOPS-oph-17, ISO-Oph 103 \\
Oph-20-0 & Oph-emb 16 & IRAS 16240-2430(E), Elia 2-29, eHOPS-oph-19, ISO-Oph 108 \\
Oph-21-0 & eHOPS-oph-20a & CFHTWIR-Oph 67, ISO-Oph 121, WL 20S \\
Oph-21-1 & eHOPS-oph-20b & WL 20E \\
Oph-21-2 & eHOPS-oph-20c & NA \\
Oph-22-0 & Oph-emb 11 & eHOPS-oph-23, ISO-Oph 124 \\
Oph-23-0 & Oph-emb 28 & ISO-Oph 132 \\
Oph-25-0 & Oph-emb 12 & eHOPS-oph-26, ISO-Oph 132 \\
Oph-26-0 & Oph-emb 14 VLA 1 & IRAS 16244-2434, eHOPS-oph-29, ISO-Oph 141 \\
Oph-26-1 & Oph-emb 14 VLA 2 & NA \\
Oph-26-2 & CFHTWIR-Oph 79 & NA \\
Oph-27-0 & Oph-emb 13 & IRAS 16244-2432, eHOPS-oph-30, ISO-Oph 143 \\
Oph-28-0 & Oph-emb 19 & Elia 2-32, ISO-Oph 144 \\
Oph-29-0 & Oph-emb 26a & Elia 2-33, ISO-Oph 147 \\
Oph-29-1 & Oph-emb 26b & NA \\
Oph-30-0 & Oph-emb 24 & eHOPS-oph-35, ISO-Oph 161 \\
Oph-31-0 & Oph-emb 27 & IRAS 16246-2436, ISO-Oph 167 \\
Oph-31-1 & F-MM7 & NA \\
Oph-32-0 & Oph-emb 1 & eHOPS-oph-42 \\
Oph-33-0 & Oph-emb 18 & eHOPS-oph-43 \\
Oph-34-0 & Oph-emb 17 & IRAS 16285-2355, IRS 63, eHOPS-oph-44 \\
Oph-35-0 & Oph-emb 4 & eHOPS-oph-46 \\
Oph-36-0 & Oph-emb 25 & eHOPS-oph-47, ISO-Oph 200 \\
Oph-38-0 & Oph-emb 15 & eHOPS-oph-48, ISO-Oph 203 \\
Oph-39-0 & Oph-emb 10a & IRAS 16288-2450(E), eHOPS-oph-49, ISO-Oph 209 \\
Oph-39-1 & Oph-emb 10b & NA \\
Oph-40-0 & IRAS 16293-2422A & Oph-emb 2, eHOPS-oph-51 \\
Oph-40-1 & IRAS 16293-2422B & NA \\
Oph-41-0 & EDJ 1013 & NA \\
Serp-01-0 & Ser-emb 28 & eHOPS-aql-13 \\
Serp-02-0 & Ser-emb 16 & eHOPS-aql-14 \\
Serp-03-0 & Ser-emb 10A & eHOPS-aql-15 \\
Serp-03-1 & Ser-emb 10B & NA \\
Serp-04-0 & Ser-emb 25 & eHOPS-aql-17 \\
Serp-05-0 & Ser-emb 7 & eHOPS-aql-18 \\
Serp-06-0 & Ser-emb 3 & eHOPS-aql-19 \\
Serp-07-0 & Ser-emb 5 & eHOPS-aql-21 \\
Serp-08-0 & Ser-emb 9A & eHOPS-aql-22 \\
Serp-08-1 & Ser-emb 9B & NA \\
Serp-09-0 & Ser-emb 13 & eHOPS-aql-26 \\
Serp-10-0 & Ser-emb 34A & eHOPS-aql-28 \\
Serp-10-1 & Ser-emb 34B & SSTc2d J182902.8+003009 \\
Serp-11-0 & Ser-emb 17B & Ser-emb ALMA 1 \\
Serp-11-1 & Ser-emb 17A & eHOPS-aql-34 \\
Serp-12-0 & Ser-emb 11(W) & eHOPS-aql-35 \\
Serp-12-1 & Ser-emb 11(E) & SSTc2d J182906.7+003034 \\
Serp-12-2 & SSTc2d J182907.1+003043 & NA \\
Serp-13-0 & Ser-emb 1B & SSTc2d J182908.6+003130 \\
Serp-13-1 & Ser-emb 1A & eHOPS-aql-40 \\
Serp-14-0 & Ser-emb 33 & IRAS 18267+0016 \\
Serp-15-0 & Ser-emb 31 & eHOPS-aql-50 \\
Serp-16-0 & Ser-emb 8A & eHOPS-aql-62 \\
Serp-16-1 & Ser-emb 8B & NA \\
Serp-16-2 & Ser-emb 8C & NA \\
Serp-16-3 & Ser-emb 8(N) & NA \\
Serp-17-0 & Ser-emb 20 & eHOPS-aql-64 \\
Serp-17-1 & V* V370 Ser & eHOPS-aql-65 \\
Serp-18-0 & Ser-emb 6A & eHOPS-aql-67, SMM1a \\
Serp-18-1 & Ser-emb 6B & SMM1b(E) \\
Serp-18-2 & Ser-emb 6C & SMM1b(W) \\
Serp-18-3 & Ser-emb 6D & SMM1c \\
Serp-18-4 & Ser-emb 6E & SMM1d \\
Serp-19-0 & Ser-emb 21 & eHOPS-aql-68 \\
Serp-20-0 & Ser-emb 12B & eHOPS-aql-71 \\
Serp-20-1 & Ser-emb 12A & NA \\
Serp-20-2 & Ser-emb 12C & NA \\
Serp-21-0 & eHOPS-aql-72 & SSTc2d J182952.3+003553 \\
Serp-21-1 & Ser-emb 2 & eHOPS-aql-73 \\
Serp-22-0 & Ser-emb 18 & eHOPS-aql-74 \\
Serp-23-0 & Ser-emb 15A & eHOPS-aql-77 \\
Serp-23-1 & Ser-emb 15B & NA \\
Serp-24-0 & Ser-emb 27 & HBC 672 \\
Serp-25-0 & Serpens SMM4B & eHOPS-aql-78 \\
Serp-26-0 & Ser-emb 30C & NA \\
Serp-26-1 & Ser-emb 30B & eHOPS-aql-81 \\
Serp-26-2 & Ser-emb 30A & NA \\
Serp-27-0 & eHOPS-aql-82 & SSTc2d J182957.8+011237 \\
Serp-27-1 & SSTc2d J182957.8+011246 & NA \\
Serp-27-2 & Ser-emb 23 & SSTc2d J182957.8+011251 \\
Serp-27-3 & Ser-emb 22 & eHOPS-aql-80 \\
Serp-28-0 & Ser-emb 26 & eHOPS-aql-84 \\
Serp-29-0 & eHOPS-aql-86A & SSTc2d J182959.2+011401 \\
Serp-29-1 & eHOPS-aql-86B & NA \\
Serp-30-0 & Ser-emb 24B & eHOPS-aql-88 \\
Serp-30-1 & Ser-emb 24A & NA \\
Serp-31-0 & Ser-emb 19 & eHOPS-aql-89 \\
Serp-33-0 & Ser-emb 4(E) & NA \\
Serp-33-1 & Ser-emb 4 & eHOPS-aql-92 \\
Serp-34-0 & Ser-emb 29 & eHOPS-aql-103 \\
Serp-35-0 & Ser-emb 14 & eHOPS-aql-112 \\
Serp-36-0 & Ser-emb 32 & SSTc2d J183005.7+003931 \\
\hline
\hline
\end{longtable*}
\end{center}

\begin{center}
\begin{longtable*}{ c c c c c c c c c c c c c c c c c}
\caption{Properties of all the sources detected in this survey}\\ %
\label{table:source}
&  &    & \\
\hline
\hline
Name\tablenotemark{a} & Cloud & RA  & DEC & CL & ${\rm T}_{\rm bol}$ & ${\rm R}_{\rm maj}$ & E${\rm R}_{\rm maj}$ & ${\rm R}_{\rm min}$ & E${\rm R}_{\rm min}$ & PA & EPA & map\tablenotemark{b}\\
& & (J2000)  & (J2000) & & (K) & ($\as$)& ($\as$)& ($\as$)& ($\as$)& ($^\circ$)&($^\circ$) &\\
\hline
\endfirsthead
\multicolumn{6}{c}%
{\tablename\ \thetable\ -- \textit{Continued from previous page}} \\
\hline

\hline
\endhead
\hline \multicolumn{4}{r}{\textit{Continued on next page}} \\
\endfoot
\hline
\endlastfoot
\hline                        
Aqu-MM3 & Aquila & 18:29:5.33 & -1:41:57.01 & 0 & 56 & 0.323 & 0.003 & 0.117 & 0.003 & 132.0 & 0.7 & U \\
Aqu-MM5 & Aquila & 18:29:23.41 & -1:38:55.73 & I & 195 & 0.229 & 0.017 & 0.203 & 0.017 & 174.9 & 40.0 & U \\
SSTgbs J1829381-015100 & Aquila & 18:29:38.12 & -1:51:0.74 & I & 240 & 0.087 & 0.008 & 0.062 & 0.008 & 55.3 & 14.8 & U \\
SerpS-MM1 & Aquila & 18:29:37.62 & -1:50:59.55 & I & 240 & 0.131 & 0.010 & 0.078 & 0.010 & 64.9 & 10.4 & U \\
SerpS-MM2 a & Aquila & 18:29:38.93 & -1:51:6.94 & I & 84 & 0.162 & 0.006 & 0.101 & 0.006 & 125.5 & 5.4 & U \\
SerpS-MM2 b & Aquila & 18:29:38.83 & -1:51:5.58 & I & 84 & 0.200 & 0.043 & 0.034 & 0.043 & 44.3 & 16.6 & U \\
SSTgbs J1829386-015100 & Aquila & 18:29:38.78 & -1:51:0.19 & I & 84 & 0.078 & 0.032 & 0.046 & 0.032 & 158.8 & 25.7 & U \\
SSTgbs J1829419-015011 & Aquila & 18:29:41.92 & -1:50:11.78 & I & 110 & 0.195 & 0.031 & 0.097 & 0.031 & 103.9 & 21.4 & B \\
SerpS-MM4 & Aquila & 18:29:43.34 & -1:56:52.04 & Flat & 223 & 0.044 & 0.014 & 0.024 & 0.014 & 175.0 & 16.2 & B \\
SerpS-MM6b & Aquila & 18:29:46.93 & -1:55:49.53 & I & 165 & 0.061 & 0.025 & 0.047 & 0.025 & 5.8 & 57.4 & U \\
SerpS-MM6a & Aquila & 18:29:47.02 & -1:55:48.3 & I & 165 & 0.132 & 0.004 & 0.083 & 0.004 & 70.9 & 4.7 & U \\
eHOPS-aql-87 & Aquila & 18:29:59.48 & -2:1:6.51 & I & 81 & 0.142 & 0.005 & 0.073 & 0.005 & 108.8 & 3.6 & U \\
eHOPS-aql-96 a & Aquila & 18:30:1.07 & -2:6:9.15 & 0 & 53 & 0.052 & 0.019 & 0.013 & 0.019 & 63.2 & 33.1 & U \\
eHOPS-aql-96 b & Aquila & 18:30:1.09 & -2:6:8.96 & 0 & 53 & NA & NA & NA & NA & NA & NA & N \\
MIRES G028.6593+03.8185 & Aquila & 18:30:2.48 & -2:3:4.29 & I & 390 & 0.312 & 0.036 & 0.062 & 0.036 & 160.7 & 69.1 & B \\
SSTgbs J1830027-020259 & Aquila & 18:30:2.77 & -2:2:59.79 & I & 390 & 0.084 & 0.043 & 0.042 & 0.043 & 148.8 & 50.0 & B \\
SSTgbs J1830024-020257 & Aquila & 18:30:2.45 & -2:2:58.04 & I & 390 & 0.061 & 0.008 & 0.044 & 0.008 & 101.4 & 31.1 & U \\
SerpS10-mm & Aquila & 18:30:2.6 & -2:2:57.23 & I & 390 & 0.181 & 0.054 & 0.134 & 0.054 & 18.6 & 61.7 & B \\
SerpS-MM16 & Aquila & 18:30:2.41 & -2:2:49.31 & I & 390 & 0.107 & 0.016 & 0.093 & 0.016 & 126.8 & 52.7 & U \\
eHOPS-aql-107 & Aquila & 18:30:3.37 & -2:2:45.75 & I & 390 & 0.062 & 0.034 & 0.048 & 0.034 & 59.6 & 74.8 & B \\
eHOPS-aql-122 & Aquila & 18:30:17.01 & -2:9:58.85 & 0 & 63 & 0.057 & 0.014 & 0.019 & 0.014 & 79.0 & 18.8 & N \\
eHOPS-aql-123 & Aquila & 18:30:17.47 & -2:9:58.59 & 0 & 63 & 0.043 & 0.023 & 0.017 & 0.023 & 143.4 & 41.2 & U \\
IRAS 18278-0156 & Aquila & 18:30:24.57 & -1:54:10.78 & Flat & 322 & 0.161 & 0.015 & 0.118 & 0.015 & 36.7 & 16.2 & U \\
SerpS-MM25 & Aquila & 18:30:25.88 & -2:10:43.04 & I & 88 & 0.330 & 0.004 & 0.121 & 0.004 & 158.3 & 0.9 & U \\
IRAS 18278-0158 b & Aquila & 18:30:29.36 & -1:56:43.24 & Flat & 362 & 0.037 & 0.016 & 0.028 & 0.016 & 23.8 & 85.0 & U \\
IRAS 18278-0158 a & Aquila & 18:30:29.3 & -1:56:42.61 & Flat & 362 & 0.056 & 0.013 & 0.044 & 0.013 & 35.0 & 40.5 & U \\
SSTgbs J1830469-015651 & Aquila & 18:30:46.91 & -1:56:51.4 & I & 88 & 0.098 & 0.013 & 0.039 & 0.013 & 108.3 & 10.4 & U \\
eHOPS-aql-138 & Aquila & 18:30:46.94 & -1:56:45.79 & I & 88 & 0.099 & 0.031 & 0.032 & 0.031 & 21.0 & 16.9 & U \\
eHOPS-aql-139 a & Aquila & 18:30:48.72 & -1:56:2.12 & I & 114 & 0.173 & 0.062 & 0.077 & 0.062 & 124.9 & 40.9 & U \\
eHOPS-aql-139 b & Aquila & 18:30:48.74 & -1:56:1.94 & I & 114 & 0.050 & 0.018 & 0.019 & 0.018 & 128.1 & 37.7 & U \\
W40-MM30 & Aquila & 18:31:52.18 & -2:1:26.36 & 0 & 58 & 0.070 & 0.015 & 0.061 & 0.015 & 115.7 & 73.0 & U \\
W40-MM36 & Aquila & 18:32:13.25 & -1:57:30.97 & 0 & 39 & 0.231 & 0.007 & 0.148 & 0.007 & 63.8 & 4.5 & U \\
IRAS 11030-7702 & Cham I & 11:4:22.56 & -77:18:8.31 & I & 308 & 0.457 & 0.017 & 0.170 & 0.017 & 12.2 & 3.5 & U \\
TIC 454291385 & Cham I & 11:6:46.36 & -77:22:32.93 & I & 74 & 0.346 & 0.011 & 0.128 & 0.011 & 102.3 & 2.9 & U \\
ISO-ChaI 101 & Cham I & 11:7:21.24 & -77:22:11.57 & Flat & 650 & 0.412 & 0.036 & 0.256 & 0.036 & 9.2 & 2.0 & B \\
Ass Cha T 1-15 & Cham I & 11:7:43.53 & -77:39:41.08 & II & 1400 & 0.260 & 0.020 & 0.241 & 0.020 & 175.2 & 49.0 & U \\
V* HO Cha a & Cham I & 11:8:2.78 & -77:38:42.72 & Flat & 710 & 0.114 & 0.012 & 0.040 & 0.012 & 63.4 & 6.0 & U \\
V* HO Cha b & Cham I & 11:8:2.84 & -77:38:42.54 & Flat & 710 & 0.123 & 0.018 & 0.048 & 0.018 & 58.6 & 8.5 & U \\
V* GM Cha & Cham I & 11:9:28.38 & -76:33:28.25 & I & 219 & 0.177 & 0.032 & 0.108 & 0.032 & 22.1 & 87.4 & U \\
ChamI-9 mm & Cham I & 11:9:48.04 & -76:34:55.55 & II & 720 & NA & NA & NA & NA & NA & NA & N \\
ISO-ChaI 204 & Cham I & 11:9:46.05 & -76:34:46.53 & II & 720 & 0.067 & 0.039 & 0.042 & 0.039 & 35.2 & 88.2 & B \\
ISO-ChaI 207 & Cham I & 11:9:47.24 & -77:26:29.14 & II & 1100 & 0.506 & 0.036 & 0.284 & 0.036 & 30.9 & 2.0 & B \\
IR Cha INa4 & Cham I & 11:10:3.2 & -76:33:11.16 & I & 186 & 0.175 & 0.012 & 0.083 & 0.012 & 162.6 & 5.9 & U \\
ISO-ChaI 237 & Cham I & 11:10:11.26 & -76:35:29.31 & II & 1100 & 0.144 & 0.004 & 0.074 & 0.004 & 45.9 & 3.0 & U \\
TIC 454329229 & Cham I & 11:11:10.71 & -76:41:57.29 & Flat & 300 & 0.639 & 0.046 & 0.133 & 0.046 & 143.5 & 4.0 & U \\
V* DK Cha & Cham II & 12:53:17.06 & -77:7:10.87 & II & 660 & 0.313 & 0.004 & 0.277 & 0.004 & 120.6 & 5.3 & U \\
IRAS 12500-7658 & Cham II & 12:53:42.69 & -77:15:11.84 & I & 187 & 0.334 & 0.008 & 0.150 & 0.008 & 170.4 & 2.4 & U \\
IRAS 12553-7651 & Cham II & 12:59:6.43 & -77:7:40.33 & I & 260 & 0.229 & 0.025 & 0.089 & 0.025 & 36.3 & 9.7 & U \\
V* S CrA B & CrAus & 19:1:8.65 & -36:57:21.45 & II & 1000 & 0.238 & 0.011 & 0.222 & 0.011 & 164.5 & 32.3 & U \\
V* S CrA A & CrAus & 19:1:8.6 & -36:57:20.28 & II & 1000 & 0.210 & 0.010 & 0.201 & 0.010 & 168.1 & 88.9 & U \\
IRS 2 & CrAus & 19:1:41.58 & -36:58:31.82 & I & 235 & 0.498 & 0.036 & 0.425 & 0.036 & 138.5 & 22.2 & B \\
IRS 5a & CrAus & 19:1:48.03 & -36:57:23.03 & Flat & 208 & 0.030 & 0.013 & 0.013 & 0.013 & 109.7 & 84.0 & B \\
IRS 5b & CrAus & 19:1:48.08 & -36:57:22.43 & Flat & 208 & 0.228 & 0.041 & 0.094 & 0.041 & 60.3 & 13.9 & N \\
IRS 5N & CrAus & 19:1:48.47 & -36:57:15.38 & 0 & 40 & 0.301 & 0.006 & 0.137 & 0.006 & 81.0 & 2.0 & U \\
V* V710 CrA & CrAus & 19:1:50.69 & -36:58:10.21 & II & 120 & 0.205 & 0.005 & 0.187 & 0.005 & 108.7 & 17.4 & U \\
IRS 7A & CrAus & 19:1:55.33 & -36:57:22.69 & 0 & 54 & 0.058 & 0.010 & 0.038 & 0.010 & 70.7 & 21.6 & U \\
SMM1C & CrAus & 19:1:55.3 & -36:57:17.32 & 0 & 54 & 0.750 & 0.003 & 0.163 & 0.003 & 174.2 & 0.3 & U \\
CrAus7-mm & CrAus & 19:1:54.92 & -36:57:12.86 & 0 & 54 & 0.109 & 0.023 & 0.029 & 0.023 & 107.0 & 7.7 & U \\
CrAus8-mm1 & CrAus & 19:1:56.63 & -36:57:40.57 & 0 & 56 & 0.243 & 0.047 & 0.110 & 0.047 & 72.7 & 20.8 & N \\
IRS7B-a & CrAus & 19:1:56.41 & -36:57:28.67 & 0 & 56 & 0.393 & 0.004 & 0.149 & 0.004 & 115.0 & 0.9 & U \\
CXO 34 & CrAus & 19:1:55.78 & -36:57:28.33 & 0 & 56 & 0.105 & 0.012 & 0.047 & 0.012 & 59.8 & 8.5 & U \\
IRS7B-b & CrAus & 19:1:56.38 & -36:57:28.12 & 0 & 56 & 0.148 & 0.005 & 0.062 & 0.005 & 118.7 & 2.7 & U \\
SMM 2 & CrAus & 19:1:58.56 & -36:57:9.13 & I & 72 & 0.846 & 0.036 & 0.554 & 0.036 & 12.1 & 2.0 & B \\
IRAS 32 A & CrAus & 19:2:58.72 & -37:7:37.37 & 0 & 58 & 0.180 & 0.003 & 0.069 & 0.003 & 135.2 & 1.3 & U \\
IRAS 32 B & CrAus & 19:2:58.64 & -37:7:36.38 & 0 & 58 & 0.152 & 0.003 & 0.055 & 0.003 & 131.2 & 1.5 & U \\
V* VV CrA A & CrAus & 19:3:6.74 & -37:12:50.23 & I & 37 & 0.152 & 0.005 & 0.145 & 0.005 & 51.5 & 30.2 & U \\
V* VV CrA B & CrAus & 19:3:6.87 & -37:12:48.74 & I & 37 & 0.136 & 0.004 & 0.134 & 0.004 & 62.5 & 58.6 & U \\
IRAS 16442-0930 & OphN & 16:46:58.26 & -9:35:20.08 & I & 230 & 0.260 & 0.003 & 0.113 & 0.003 & 93.8 & 1.1 & U \\
IRAS 16459-1411 & OphN & 16:48:45.62 & -14:16:36.34 & II & 1400 & 0.932 & 0.036 & 0.582 & 0.036 & 170.5 & 2.0 & B \\
CB 68 SMM 1 & OphN & 16:57:19.64 & -16:9:24.01 & 0 & 38 & 0.174 & 0.003 & 0.055 & 0.003 & 45.4 & 1.2 & U \\
ISO-Oph 2a & Oph & 16:25:38.12 & -24:22:36.82 & II & 1100 & 0.592 & 0.036 & 0.497 & 0.036 & 6.5 & 2.0 & B \\
ISO-Oph 2b & Oph & 16:25:38.15 & -24:22:38.49 & II & 1100 & 0.082 & 0.057 & 0.070 & 0.057 & 58.8 & 75.9 & B \\
DoAr 20 & Oph & 16:25:56.15 & -24:20:48.8 & II & 1100 & 0.288 & 0.023 & 0.253 & 0.023 & 4.5 & 30.1 & U \\
ISO-Oph 17 & Oph & 16:26:10.33 & -24:20:55.38 & II & 290 & 0.566 & 0.036 & 0.434 & 0.036 & 133.8 & 2.0 & B \\
Elias 2-20 & Oph & 16:26:18.86 & -24:28:20.28 & II & 990 & 0.279 & 0.015 & 0.182 & 0.015 & 154.3 & 7.5 & U \\
Oph-emb 8 & Oph & 16:26:21.35 & -24:23:5.01 & I & 97 & 0.048 & 0.011 & 0.020 & 0.011 & 120.4 & 28.1 & U \\
ISO-Oph 31 & Oph & 16:26:21.72 & -24:22:51.08 & I & 97 & 0.375 & 0.009 & 0.143 & 0.009 & 111.4 & 2.1 & U \\
DoAr 25 & Oph & 16:26:23.68 & -24:43:14.45 & II & 1500 & 1.224 & 0.036 & 0.474 & 0.036 & 114.6 & 2.0 & B \\
Elias 2-24 & Oph & 16:26:24.08 & -24:16:13.98 & II & 980 & 1.033 & 0.036 & 0.968 & 0.036 & 39.8 & 2.0 & B \\
Oph-emb 9 & Oph & 16:26:25.47 & -24:23:1.96 & I & 76 & 0.210 & 0.004 & 0.086 & 0.004 & 28.8 & 1.4 & U \\
VLA 1623B & Oph & 16:26:26.3 & -24:24:30.82 & 0 & 30 & 0.259 & 0.004 & 0.078 & 0.004 & 41.5 & 0.9 & U \\
VLA 1623Ab & Oph & 16:26:26.38 & -24:24:30.98 & 0 & 30 & 0.111 & 0.012 & 0.061 & 0.012 & 40.1 & 8.8 & U \\
VLA 1623Aa & Oph & 16:26:26.39 & -24:24:30.92 & 0 & 30 & 0.129 & 0.016 & 0.066 & 0.016 & 48.0 & 9.7 & U \\
VLA 1623W & Oph & 16:26:25.63 & -24:24:29.7 & I & 84 & 0.608 & 0.007 & 0.101 & 0.007 & 11.3 & 0.6 & U \\
Oph-emb 22 & Oph & 16:26:40.47 & -24:27:15.09 & I & 224 & 0.794 & 0.036 & 0.626 & 0.036 & 158.7 & 2.0 & B \\
IRAS 16237-2428 & Oph & 16:26:44.2 & -24:34:49.03 & Flat & 221 & 0.105 & 0.002 & 0.074 & 0.002 & 58.7 & 3.3 & U \\
Elia 2-27 & Oph & 16:26:45.02 & -24:23:8.38 & Flat & 410 & 1.595 & 0.036 & 0.804 & 0.036 & 123.1 & 2.0 & B \\
Oph-emb 23 & Oph & 16:26:48.48 & -24:28:39.44 & Flat & 440 & 0.353 & 0.007 & 0.130 & 0.007 & 133.0 & 1.8 & U \\
CFHTWIR-Oph 43 & Oph & 16:26:48.38 & -24:28:35.41 & Flat & 440 & 0.031 & 0.011 & 0.029 & 0.011 & 99.9 & 47.6 & U \\
DoAr 29 & Oph & 16:26:58.52 & -24:45:37.34 & II & 840 & 0.710 & 0.036 & 0.456 & 0.036 & 28.4 & 2.0 & B \\
Oph-emb 21 & Oph & 16:27:2.32 & -24:37:27.81 & I & 420 & 0.051 & 0.014 & 0.022 & 0.014 & 110.9 & 18.4 & U \\
ISO-Oph 93 & Oph & 16:27:2.99 & -24:26:15.31 & Flat & 380 & 0.326 & 0.011 & 0.104 & 0.011 & 140.8 & 2.5 & U \\
Oph-emb 6 & Oph & 16:27:5.25 & -24:36:30.29 & 0 & 67 & 0.306 & 0.010 & 0.102 & 0.010 & 168.5 & 2.2 & U \\
Oph-emb 20 & Oph & 16:27:6.76 & -24:38:15.61 & Flat & 310 & 0.270 & 0.036 & 0.227 & 0.036 & 50.2 & 2.0 & B \\
Oph-emb 16 & Oph & 16:27:9.41 & -24:37:19.36 & I & 74 & 0.040 & 0.009 & 0.024 & 0.009 & 119.9 & 23.7 & B \\
eHOPS-oph-20a & Oph & 16:27:15.68 & -24:38:46.17 & I & 57 & 0.584 & 0.036 & 0.113 & 0.036 & 75.3 & 2.0 & B \\
eHOPS-oph-20b & Oph & 16:27:15.65 & -24:38:43.94 & I & 57 & 0.084 & 0.038 & 0.025 & 0.038 & 148.3 & 46.2 & U \\
eHOPS-oph-20c & Oph & 16:27:15.89 & -24:38:43.92 & I & 57 & 0.075 & 0.024 & 0.014 & 0.024 & 96.5 & 35.8 & U \\
Oph-emb 11 & Oph & 16:27:17.58 & -24:28:56.96 & Flat & 163 & 0.112 & 0.025 & 0.037 & 0.025 & 12.2 & 9.0 & U \\
Oph-emb 28 & Oph & 16:27:21.45 & -24:41:43.66 & Flat & 610 & 0.072 & 0.011 & 0.015 & 0.011 & 65.3 & 4.3 & B \\
Oph-emb 12 & Oph & 16:27:24.59 & -24:41:3.83 & I & 115 & 0.044 & 0.015 & 0.030 & 0.015 & 80.5 & 45.1 & U \\
Oph-emb 14 VLA 1 & Oph & 16:27:26.91 & -24:40:51.4 & I & 112 & 0.136 & 0.046 & 0.099 & 0.046 & 85.4 & 72.7 & N \\
Oph-emb 14 VLA 2 & Oph & 16:27:26.91 & -24:40:50.82 & I & 112 & 0.084 & 0.005 & 0.038 & 0.005 & 123.8 & 5.0 & U \\
CFHTWIR-Oph 79 & Oph & 16:27:26.6 & -24:40:45.7 & I & 112 & 0.326 & 0.036 & 0.155 & 0.036 & 127.9 & 2.0 & B \\
Oph-emb-13 & Oph & 16:27:27.99 & -24:39:34.07 & I & 129 & 0.031 & 0.012 & 0.017 & 0.012 & 11.1 & 66.8 & U \\
Oph-emb 19 & Oph & 16:27:28.44 & -24:27:21.78 & Flat & 310 & 0.305 & 0.058 & 0.094 & 0.058 & 39.4 & 17.1 & U \\
Oph-emb 26a & Oph & 16:27:30.17 & -24:27:43.97 & Flat & 500 & 0.193 & 0.035 & 0.067 & 0.035 & 54.3 & 9.6 & U \\
Oph-emb 26b & Oph & 16:27:30.17 & -24:27:43.97 & Flat & 500 & 0.193 & 0.035 & 0.067 & 0.035 & 53.3 & 9.5 & U \\
Oph-emb 24 & Oph & 16:27:37.24 & -24:42:38.55 & Flat & 251 & 0.044 & 0.019 & 0.008 & 0.019 & 114.9 & 17.5 & U \\
Oph-emb 27 & Oph & 16:27:39.82 & -24:43:15.63 & Flat & 570 & 0.173 & 0.003 & 0.121 & 0.003 & 113.0 & 2.7 & U \\
F-MM7 & Oph & 16:27:39.84 & -24:43:14.01 & Flat & 570 & 0.187 & 0.023 & 0.042 & 0.023 & 61.4 & 6.6 & U \\
Oph-emb 1 & Oph & 16:28:21.62 & -24:36:24.33 & 0 & 42 & 0.091 & 0.007 & 0.045 & 0.007 & 119.1 & 6.8 & U \\
Oph-emb 18 & Oph & 16:28:57.87 & -24:40:55.5 & I & 263 & 0.126 & 0.060 & 0.007 & 0.060 & 21.8 & 18.6 & U \\
Oph-emb 17 & Oph & 16:31:35.66 & -24:1:30.03 & Flat & 328 & 0.261 & 0.023 & 0.208 & 0.023 & 145.0 & 25.9 & U \\
Oph-emb 4 & Oph & 16:31:36.78 & -24:4:20.5 & I & 135 & 0.222 & 0.009 & 0.067 & 0.009 & 82.4 & 3.0 & U \\
Oph-emb 25 & Oph & 16:31:43.75 & -24:55:25.04 & Flat & 429 & 0.052 & 0.027 & 0.039 & 0.027 & 145.8 & 48.2 & U \\
Oph-emb 15 & Oph & 16:31:52.44 & -24:55:36.61 & I & 231 & 0.056 & 0.008 & 0.032 & 0.008 & 110.6 & 18.1 & B \\
Oph-emb 10a & Oph & 16:32:0.98 & -24:56:43.59 & I & 78 & 0.076 & 0.002 & 0.051 & 0.002 & 92.4 & 4.1 & U \\
Oph-emb 10b & Oph & 16:32:0.99 & -24:56:42.86 & I & 78 & 0.046 & 0.016 & 0.029 & 0.016 & 134.1 & 38.4 & B \\
IRAS 16293-2422A & Oph & 16:32:22.88 & -24:28:36.76 & 0 & 31 & 0.465 & 0.022 & 0.271 & 0.022 & 33.5 & 5.7 & U \\
IRAS 16293-2422B & Oph & 16:32:22.61 & -24:28:32.68 & 0 & 31 & 0.287 & 0.019 & 0.265 & 0.019 & 170.3 & 47.9 & U \\
EDJ 1013 & Oph & 16:33:55.61 & -24:42:5.49 & II & 1500 & 0.518 & 0.036 & 0.330 & 0.036 & 71.1 & 2.0 & B \\
Ser-emb 28 & Serpens & 18:28:44.03 & 0:53:37.6 & I & 319 & 0.058 & 0.056 & 0.011 & 0.056 & 138.4 & 80.8 & U \\
Ser-emb 16 & Serpens & 18:28:44.8 & 0:51:25.36 & 0 & 65 & 0.115 & 0.032 & 0.060 & 0.032 & 109.2 & 64.1 & B \\
Ser-emb 10A & Serpens & 18:28:45.08 & 0:52:1.86 & I & 82 & 0.054 & 0.011 & 0.026 & 0.011 & 59.2 & 14.2 & U \\
Ser-emb 10B & Serpens & 18:28:44.97 & 0:52:3.12 & I & 82 & 0.064 & 0.018 & 0.032 & 0.018 & 71.0 & 28.7 & B \\
Ser-emb 25 & Serpens & 18:28:51.24 & 0:19:27.05 & I & 253 & NA & NA & NA & NA & NA & NA & N \\
Ser-emb 7 & Serpens & 18:28:54.06 & 0:29:29.29 & 0 & 57 & 0.081 & 0.002 & 0.038 & 0.002 & 84.6 & 2.6 & U \\
Ser-emb 3 & Serpens & 18:28:54.87 & 0:29:51.99 & I & 81 & 0.117 & 0.024 & 0.023 & 0.024 & 155.1 & 7.4 & U \\
Ser-emb 5 & Serpens & 18:28:54.92 & 0:18:32.31 & 0 & 66 & 0.129 & 0.020 & 0.055 & 0.020 & 157.7 & 8.5 & U \\
Ser-emb 9A & Serpens & 18:28:55.77 & 0:29:44.1 & 0 & 55 & 0.092 & 0.009 & 0.058 & 0.009 & 78.5 & 16.4 & B \\
Ser-emb 9B & Serpens & 18:28:55.82 & 0:29:44.32 & 0 & 55 & 0.077 & 0.022 & 0.024 & 0.022 & 103.6 & 27.7 & U \\
Ser-emb 13 & Serpens & 18:29:2.12 & 0:31:20.22 & I & 77 & 0.394 & 0.036 & 0.118 & 0.036 & 3.2 & 2.0 & B \\
Ser-emb 34A & Serpens & 18:29:2.96 & 0:30:7.84 & Flat & 204 & 0.142 & 0.023 & 0.085 & 0.023 & 93.1 & 32.9 & U \\
Ser-emb 34B & Serpens & 18:29:2.83 & 0:30:9.26 & Flat & 204 & 0.057 & 0.022 & 0.029 & 0.022 & 95.2 & 80.0 & B \\
Ser-emb 17B & Serpens & 18:29:5.61 & 0:30:34.85 & 0 & 67 & 0.069 & 0.040 & 0.026 & 0.040 & 153.5 & 49.8 & U \\
Ser-emb 17A & Serpens & 18:29:6.21 & 0:30:43.02 & 0 & 67 & 0.263 & 0.023 & 0.242 & 0.023 & 172.1 & 89.4 & U \\
Ser-emb 11(W) & Serpens & 18:29:6.62 & 0:30:33.9 & 0 & 48 & 0.199 & 0.017 & 0.185 & 0.017 & 61.3 & 70.1 & U \\
Ser-emb 11(E) & Serpens & 18:29:6.78 & 0:30:34.15 & 0 & 48 & 0.088 & 0.024 & 0.057 & 0.024 & 157.3 & 21.0 & U \\
SSTc2d J182907.1+003043 & Serpens & 18:29:7.09 & 0:30:43.02 & 0 & 48 & 0.126 & 0.029 & 0.036 & 0.029 & 63.4 & 17.4 & B \\
Ser-emb 1B & Serpens & 18:29:8.62 & 0:31:30.04 & 0 & 35 & 0.219 & 0.026 & 0.025 & 0.026 & 106.0 & 7.3 & U \\
Ser-emb 1A & Serpens & 18:29:9.09 & 0:31:30.86 & 0 & 35 & 0.175 & 0.003 & 0.096 & 0.003 & 98.5 & 1.9 & U \\
Ser-emb 33 & Serpens & 18:29:16.19 & 0:18:22.23 & II & 455 & 0.058 & 0.016 & 0.021 & 0.016 & 87.8 & 52.3 & N \\
Ser-emb 31 & Serpens & 18:29:31.98 & 1:18:42.42 & Flat & 504 & 0.125 & 0.017 & 0.119 & 0.017 & 134.3 & 83.1 & U \\
Ser-emb 8A & Serpens & 18:29:48.09 & 1:16:43.25 & 0 & 33 & 0.097 & 0.010 & 0.085 & 0.010 & 87.9 & 87.8 & U \\
Ser-emb 8B & Serpens & 18:29:48.13 & 1:16:44.5 & 0 & 33 & 0.091 & 0.027 & 0.047 & 0.027 & 77.6 & 46.8 & U \\
Ser-emb 8C & Serpens & 18:29:48.03 & 1:16:42.62 & 0 & 33 & 0.149 & 0.039 & 0.100 & 0.039 & 121.3 & 71.4 & U \\
Ser-emb 8(N) & Serpens & 18:29:48.73 & 1:16:55.54 & 0 & 33 & 0.206 & 0.033 & 0.111 & 0.033 & 47.1 & 17.1 & U \\
Ser-emb 20 & Serpens & 18:29:49.15 & 1:16:19.64 & I & 109 & 0.107 & 0.008 & 0.074 & 0.008 & 73.6 & 13.0 & U \\
V* V370 Ser & Serpens & 18:29:49.26 & 1:16:31.32 & I & 114 & 0.095 & 0.045 & 0.084 & 0.045 & 81.0 & 82.2 & B \\
Ser-emb 6A & Serpens & 18:29:49.8 & 1:15:20.34 & 0 & 42 & 0.408 & 0.008 & 0.356 & 0.008 & 166.5 & 6.9 & U \\
Ser-emb 6B & Serpens & 18:29:49.68 & 1:15:21.04 & 0 & 42 & 0.121 & 0.011 & 0.078 & 0.011 & 102.6 & 18.7 & U \\
Ser-emb 6C & Serpens & 18:29:49.66 & 1:15:21.11 & 0 & 42 & 0.138 & 0.031 & 0.128 & 0.031 & 17.9 & 43.6 & U \\
Ser-emb 6D & Serpens & 18:29:49.93 & 1:15:21.91 & 0 & 42 & 0.082 & 0.036 & 0.059 & 0.036 & 122.4 & 18.2 & B \\
Ser-emb 6E & Serpens & 18:29:49.99 & 1:15:22.92 & 0 & 42 & 0.134 & 0.074 & 0.102 & 0.074 & 151.3 & 85.3 & U \\
Ser-emb 21 & Serpens & 18:29:51.17 & 1:16:40.34 & I & 90 & 0.125 & 0.004 & 0.107 & 0.004 & 58.3 & 10.1 & U \\
Ser-emb 12B & Serpens & 18:29:52.22 & 1:15:47.24 & I & 74 & 0.096 & 0.024 & 0.073 & 0.024 & 81.5 & 72.1 & U \\
Ser-emb 12A & Serpens & 18:29:52.02 & 1:15:50.14 & I & 74 & 0.126 & 0.015 & 0.103 & 0.015 & 91.2 & 80.4 & U \\
Ser-emb 12C & Serpens & 18:29:52.22 & 1:15:47.34 & I & 74 & 0.111 & 0.036 & 0.081 & 0.036 & 52.5 & 4.0 & B \\
eHOPS-aql-72 & Serpens & 18:29:52.4 & 0:35:52.66 & II & 462 & 0.248 & 0.066 & 0.159 & 0.066 & 101.6 & 35.7 & U \\
Ser-emb 2 & Serpens & 18:29:52.53 & 0:36:11.5 & 0 & 55 & 0.447 & 0.036 & 0.146 & 0.036 & 164.6 & 2.0 & B \\
Ser-emb 18 & Serpens & 18:29:52.86 & 1:14:55.77 & I & 101 & 0.059 & 0.015 & 0.046 & 0.015 & 136.5 & 45.7 & U \\
Ser-emb 15A & Serpens & 18:29:54.31 & 0:36:0.69 & II & 51 & 0.249 & 0.004 & 0.108 & 0.004 & 115.8 & 1.5 & U \\
Ser-emb 15B & Serpens & 18:29:54.31 & 0:36:0.71 & II & 51 & 0.101 & 0.036 & 0.083 & 0.036 & 108.8 & 1.0 & B \\
Ser-emb 27 & Serpens & 18:29:56.89 & 1:14:46.11 & I & 330 & 0.150 & 0.058 & 0.062 & 0.058 & 44.3 & 30.9 & U \\
Serpens SMM4B & Serpens & 18:29:56.53 & 1:13:11.48 & 0 & 16 & 0.095 & 0.030 & 0.080 & 0.030 & 81.9 & 76.7 & U \\
Ser-emb 30C & Serpens & 18:29:58.02 & 1:14:2.82 & II & 390 & 0.204 & 0.059 & 0.152 & 0.059 & 95.1 & 55.4 & B \\
Ser-emb 30B & Serpens & 18:29:57.74 & 1:14:5.15 & II & 390 & 0.063 & 0.018 & 0.024 & 0.018 & 107.5 & 17.5 & B \\
Ser-emb 30A & Serpens & 18:29:57.76 & 1:14:6.69 & II & 390 & 0.173 & 0.012 & 0.166 & 0.012 & 121.2 & 79.0 & U \\
eHOPS-aql-82 & Serpens & 18:29:57.85 & 1:12:37.65 & I & 147 & 0.167 & 0.025 & 0.077 & 0.025 & 91.6 & 19.7 & U \\
SSTc2d J182957.8+011246 & Serpens & 18:29:57.9 & 1:12:46.06 & I & 147 & 0.044 & 0.026 & 0.033 & 0.026 & 112.1 & 89.7 & U \\
Ser-emb 23 & Serpens & 18:29:57.85 & 1:12:51.13 & I & 147 & 0.128 & 0.007 & 0.081 & 0.007 & 58.7 & 7.0 & U \\
Ser-emb 22 & Serpens & 18:29:57.61 & 1:13:0.16 & I & 147 & 0.088 & 0.026 & 0.044 & 0.026 & 74.3 & 22.2 & U \\
Ser-emb 26 & Serpens & 18:29:58.79 & 1:14:25.93 & Flat & 247 & 0.147 & 0.055 & 0.091 & 0.055 & 122.5 & 70.7 & U \\
eHOPS-aql-86A & Serpens & 18:29:59.31 & 1:14:0.25 & 0 & 46 & 0.704 & 0.005 & 0.275 & 0.005 & 61.7 & 0.7 & U \\
eHOPS-aql-86B & Serpens & 18:29:59.26 & 1:13:59.92 & 0 & 46 & 0.107 & 0.036 & 0.077 & 0.036 & 64.4 & 11.7 & B \\
Ser-emb 24B & Serpens & 18:29:59.59 & 1:11:58.16 & Flat & 202 & 0.087 & 0.018 & 0.040 & 0.018 & 108.8 & 22.5 & U \\
Ser-emb 24A & Serpens & 18:29:59.62 & 1:11:59.44 & Flat & 202 & 0.097 & 0.004 & 0.064 & 0.004 & 85.8 & 6.9 & U \\
Ser-emb 19 & Serpens & 18:29:59.95 & 1:13:11.26 & I & 143 & 0.057 & 0.018 & 0.047 & 0.018 & 26.7 & 69.8 & B \\
Ser-emb 4(E) & Serpens & 18:30:0.73 & 1:12:56.2 & 0 & 28 & 0.280 & 0.029 & 0.014 & 0.029 & 114.2 & 3.9 & U \\
Ser-emb 4 & Serpens & 18:30:0.67 & 1:13:0.04 & 0 & 28 & 0.094 & 0.018 & 0.039 & 0.018 & 103.0 & 75.6 & U \\
Ser-emb 29 & Serpens & 18:30:2.75 & 1:12:27.76 & Flat & 314 & 0.163 & 0.006 & 0.127 & 0.006 & 125.0 & 9.4 & U \\
Ser-emb 14 & Serpens & 18:30:5.29 & 0:41:3.96 & I & 71 & 0.162 & 0.023 & 0.048 & 0.023 & 135.4 & 9.6 & U \\
Ser-emb 32 & Serpens & 18:30:5.72 & 0:39:31.41 & II & 540 & 0.224 & 0.020 & 0.152 & 0.020 & 172.1 & 14.3 & U \\
\hline
\hline
\end{longtable*}

\tablenotetext{a}{Field name of the CAMPOS ALMA observation. The characters represent the name of the cloud (eg. Aql: Aquila, ChamI: Chamaeloeon I, ChamII: Chamaeloeon II, Oph: Ophiuchus, OphN: Ophiuchus North, Serp: Serpens)}
\tablenotetext{b}{Map used for the measurement (eg. U: Uniformly weighted maps, B: Briggs 0.5 weighted maps, N: Naturally weighted maps)}
\tablenotetext{}{\textbf{Note}: The characters in  the table header represent the following. CL: evolutionary class; ${\rm L}_{\rm bol}$: bolometric luminosity; ${\rm T}_{\rm bol}$: bolometric temperature; ${\rm R}_{\rm maj}$: deconvolved major axis of the disk in arcseconds; E${\rm R}_{\rm maj}$: uncertainty of the deconvolved major axis in arcseconds; ${\rm R}_{\rm min}$: deconvolved minor axis of the disk in arcseconds; E${\rm R}_{\rm min}$: uncertainty in the deconvolved minor axis of the disk in arcseconds; PA: Position angles in degrees; EPA: uncertainty in the position angle in degrees.}
\end{center} 


\clearpage

\section{Image Gallery of all the sources detected in the CAMPOS survey}
\label{Appendix:Image_gallery}
\begin{figure*}[tbh!]
    \includegraphics[width=.99\textwidth]{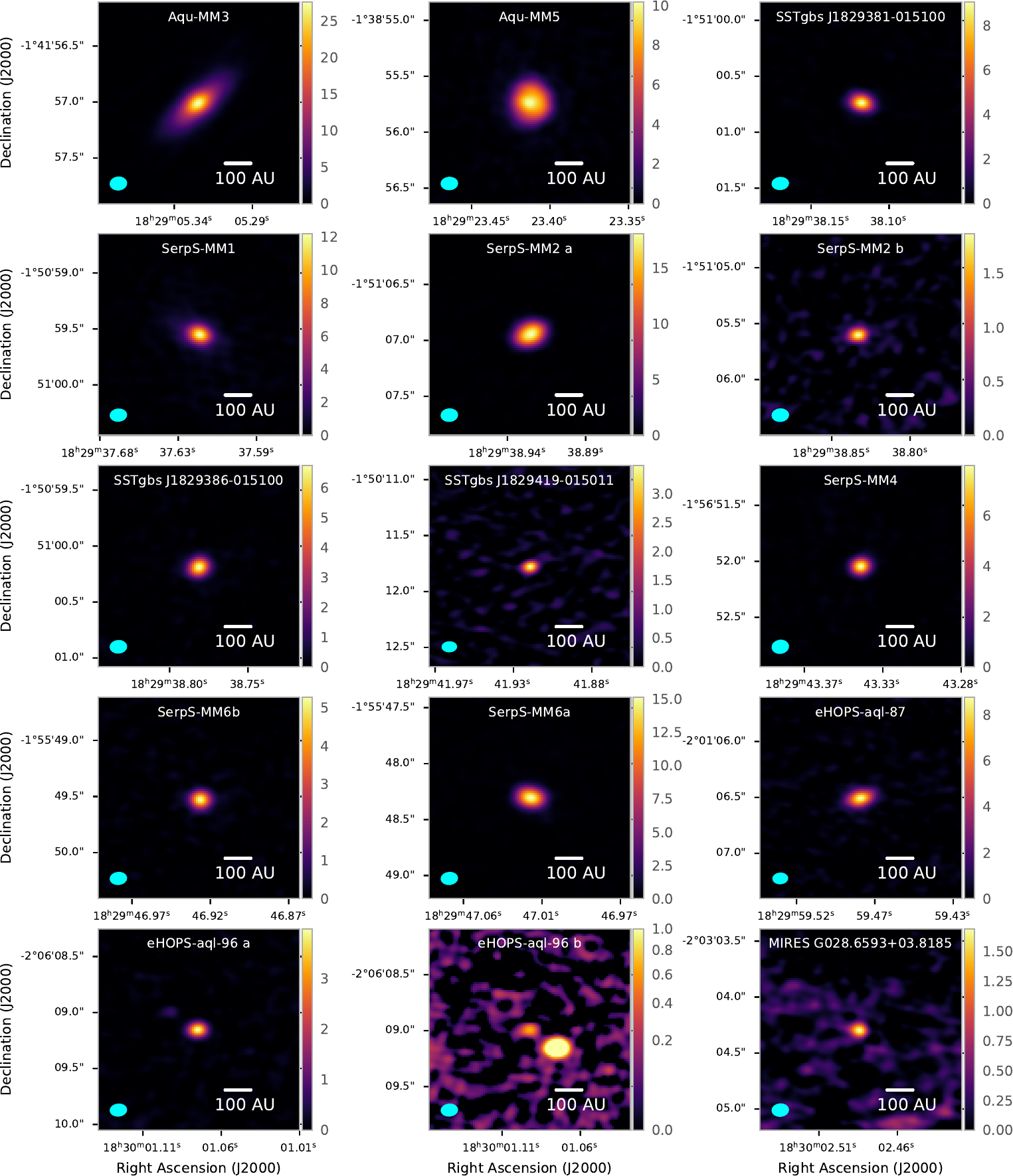}
    \caption{Part 1 of the Aquila disks detected in our ALMA Band 6 CAMPOS dust continuum survey. The cyan-filled ellipse represents the synthesized beam size. All the color scales are in units of mJy/beam. The white line marks a scale of 100\,au.} 
\label{fig:Aql1}
\end{figure*}

\begin{figure*}[tbh!]
    \includegraphics[width=.99\textwidth]{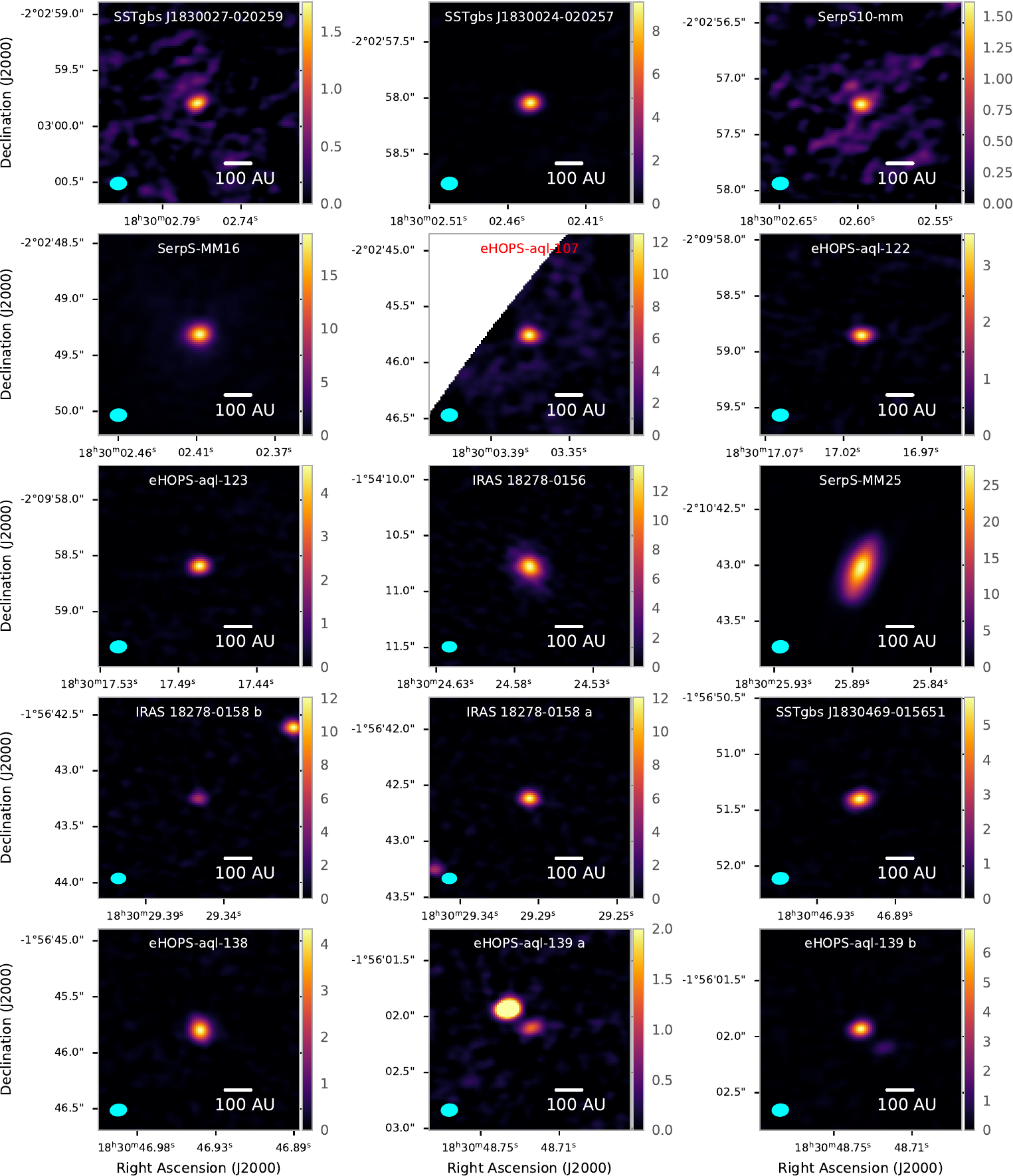}
    \caption{Part 2 of the Aquila disks detected in our ALMA Band 6 CAMPOS dust continuum survey. The cyan-filled circle represents the synthesized beam size. All the color scales are in units of mJy/beam. The white line marks a scale of 100\,au.} 
\label{fig:Aql2}
\end{figure*}

\begin{figure*}[tbh!]
    \includegraphics[width=.66\textwidth]{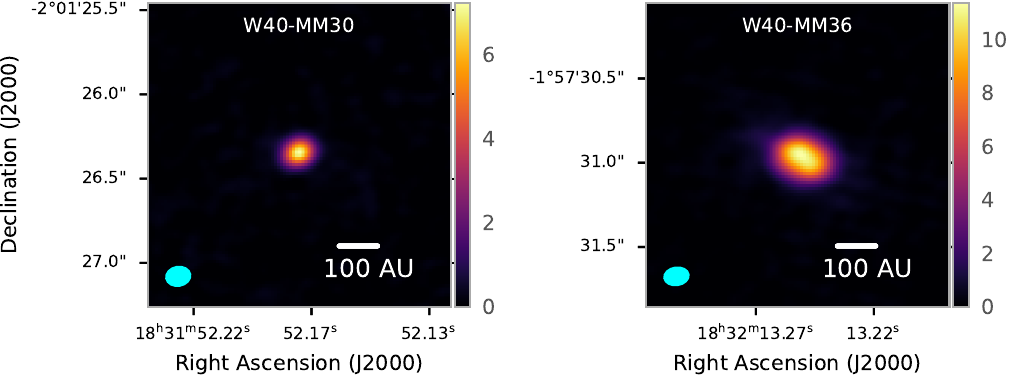}
    \caption{Part 3 of the Aquila disks detected in our ALMA Band 6 CAMPOS dust continuum survey. The cyan-filled ellipse represents the synthesized beam size. All the color scales are in units of mJy/beam. The white line marks a scale of 100\,au.} 
\label{fig:Aql3}
\end{figure*}

\begin{figure*}[tbh!]
    \includegraphics[width=.99\textwidth]{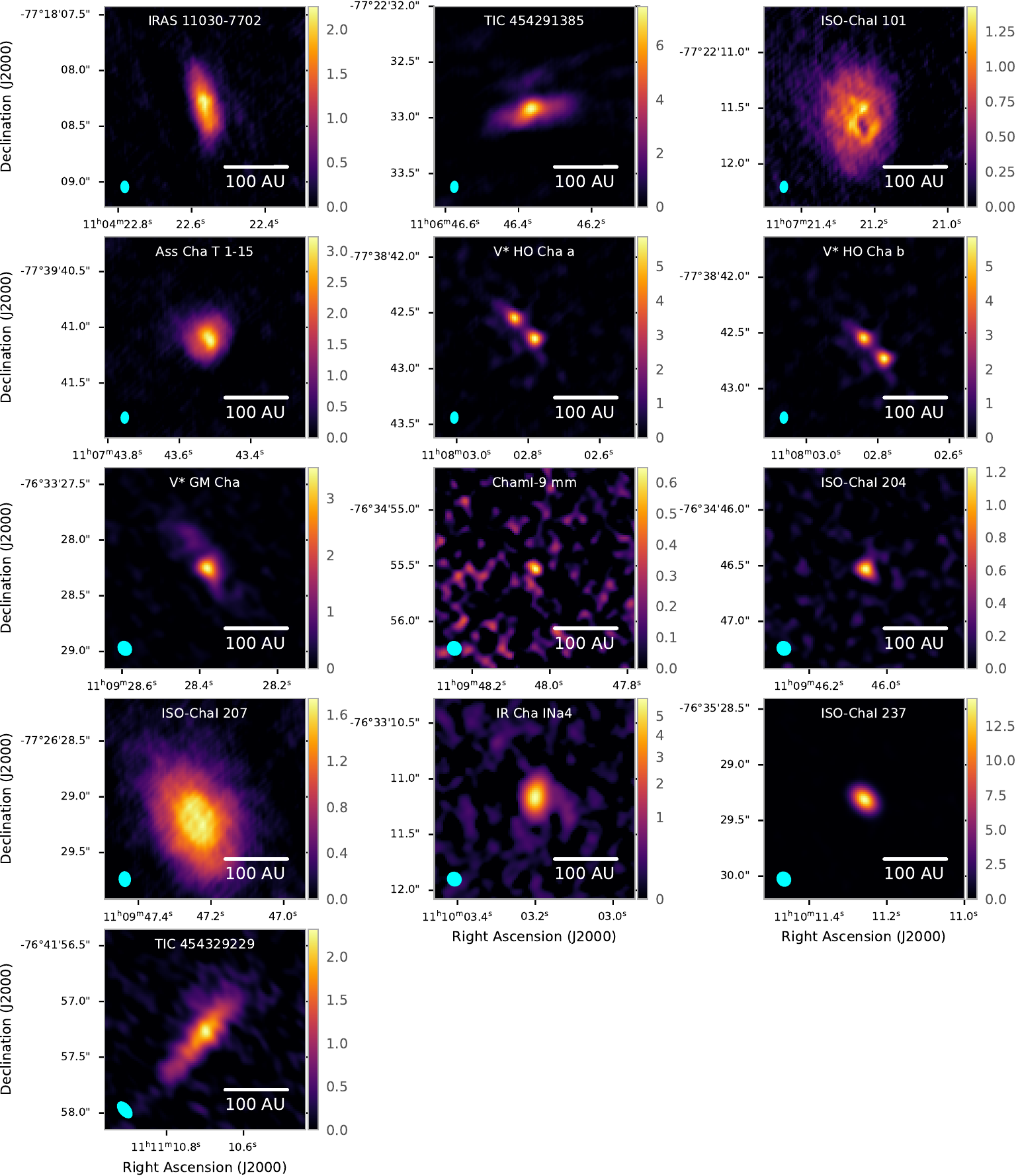}
    \caption{Chamaeleon I disks detected in our ALMA Band 6 CAMPOS dust continuum survey. The cyan-filled ellipse represents the synthesized beam size. All the color scales are in units of mJy/beam. The white line marks a scale of 100\,au.} 
\label{fig:ChamI1}
\end{figure*}

\begin{figure*}[tbh!]
    \includegraphics[width=.99\textwidth]{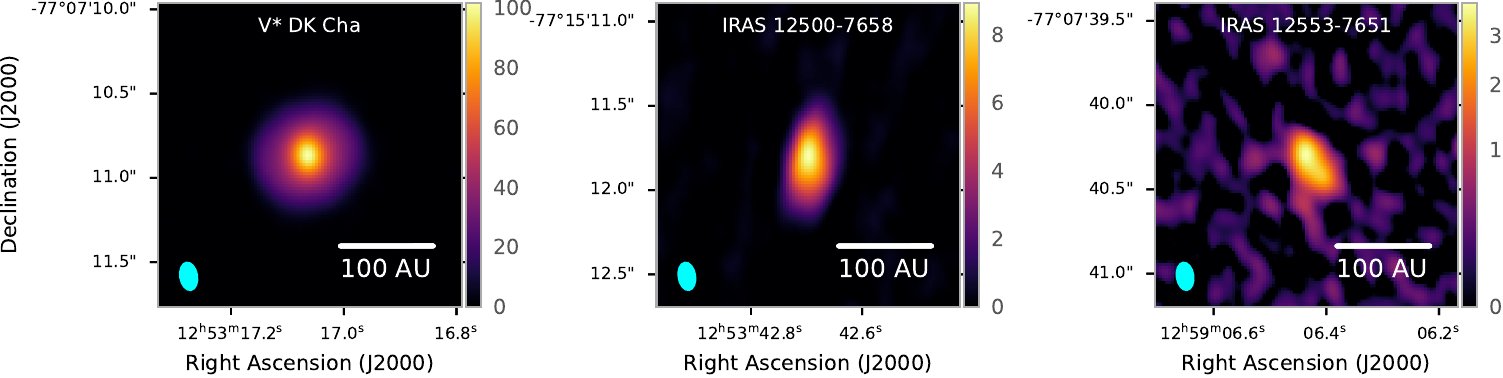}
    \caption{Chamaeleon II disks detected in our ALMA Band 6 CAMPOS dust continuum survey. The cyan-filled ellipse represents the synthesized beam size. All the color scales are in units of mJy/beam. The white line marks a scale of 100\,au.} 
\label{fig:ChamII1}
\end{figure*}

\begin{figure*}[tbh!]
    \includegraphics[width=.99\textwidth]{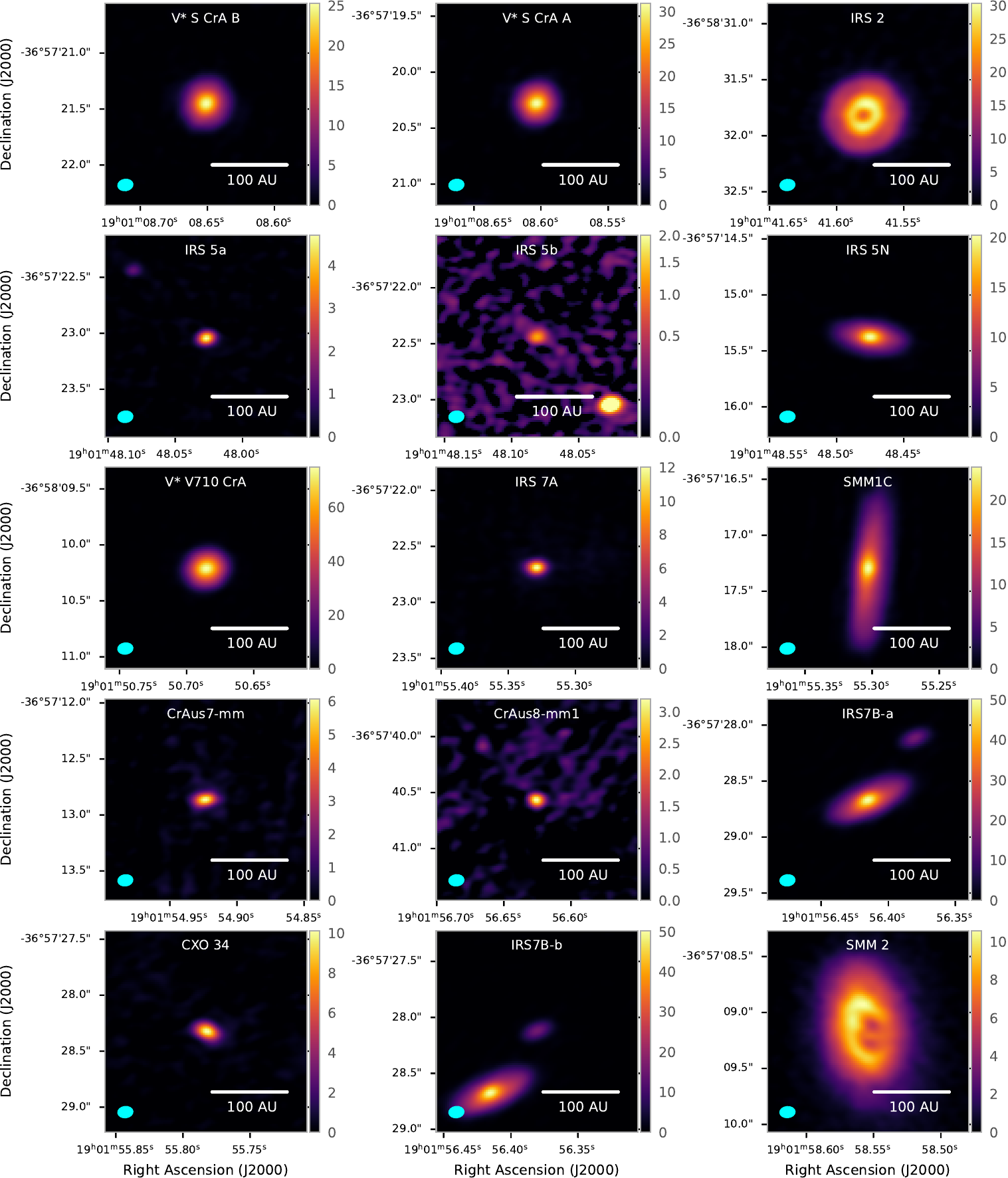}
    \caption{Part 1 of the Corona Australis disks detected in our ALMA Band 6 CAMPOS dust continuum survey. The cyan-filled ellipse represents the synthesized beam size. All the color scales are in units of mJy/beam. The white line marks a scale of 100\,au.}
\label{fig:CrAus1}
\end{figure*}

\begin{figure*}[tbh!]
    \includegraphics[width=.99\textwidth]{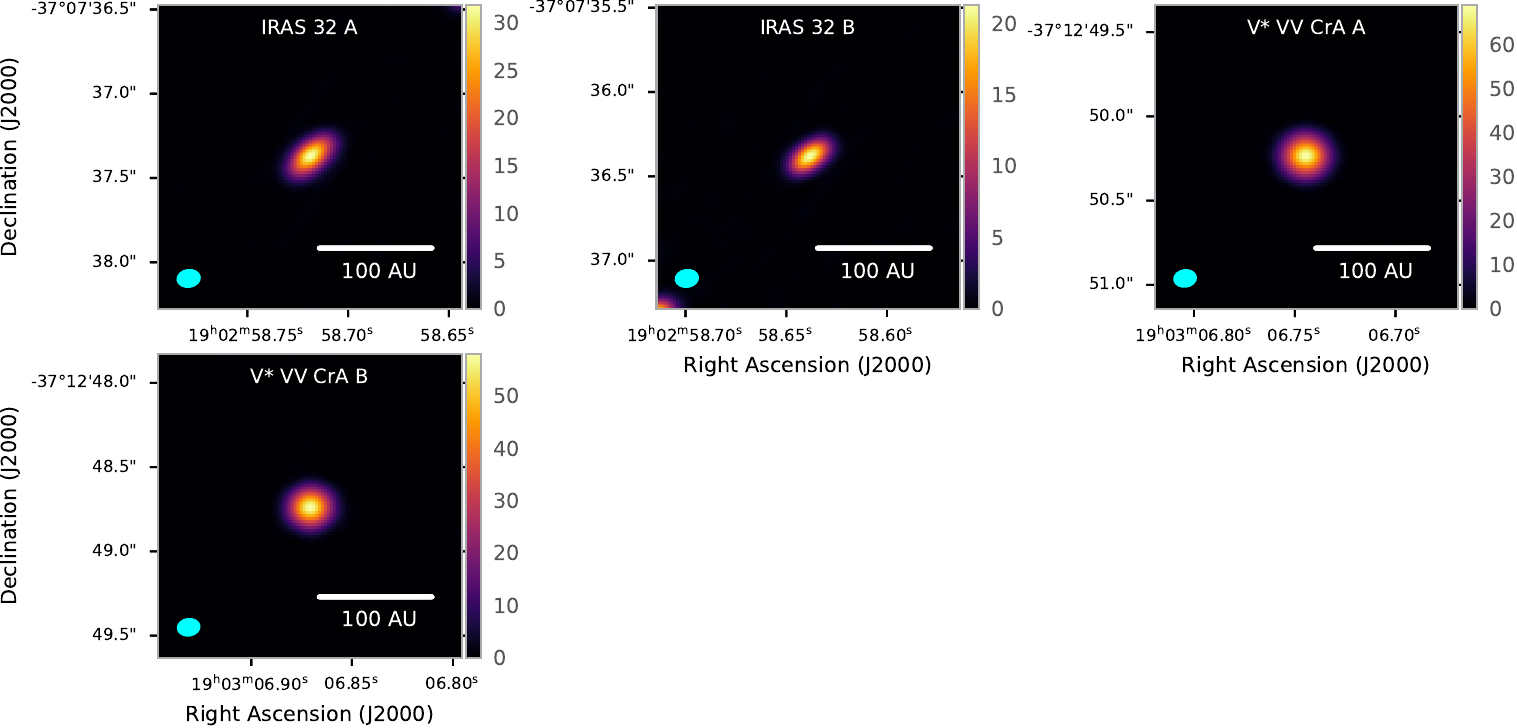}
    \caption{Part 2 of the Corona Australis disks detected in our ALMA Band 6 CAMPOS dust continuum survey. The cyan-filled ellipse represents the synthesized beam size. All the color scales are in units of mJy/beam. The white line marks a scale of 100\,au.}
\label{fig:CrAus2}
\end{figure*}

\begin{figure*}[tbh!]
    \includegraphics[width=.99\textwidth]{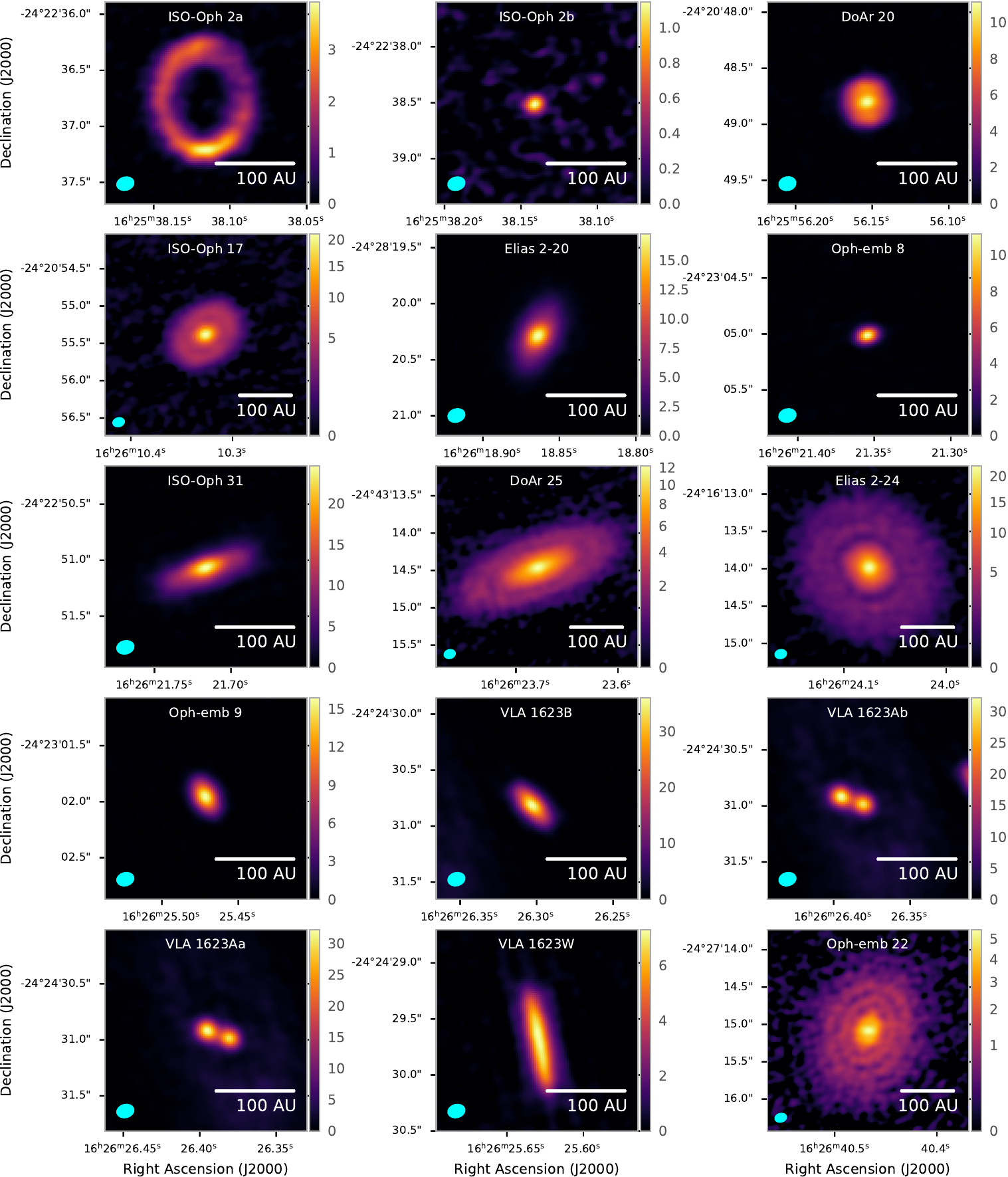}
    \caption{Part 1 of the Ophiuchus disks detected in our ALMA Band 6 CAMPOS dust continuum survey. The cyan-filled ellipse represents the synthesized beam size. All the color scales are in units of mJy/beam. The white line marks a scale of 100\,au.}
\label{fig:Oph1}
\end{figure*}

\begin{figure*}[tbh!]
    \includegraphics[width=.99\textwidth]{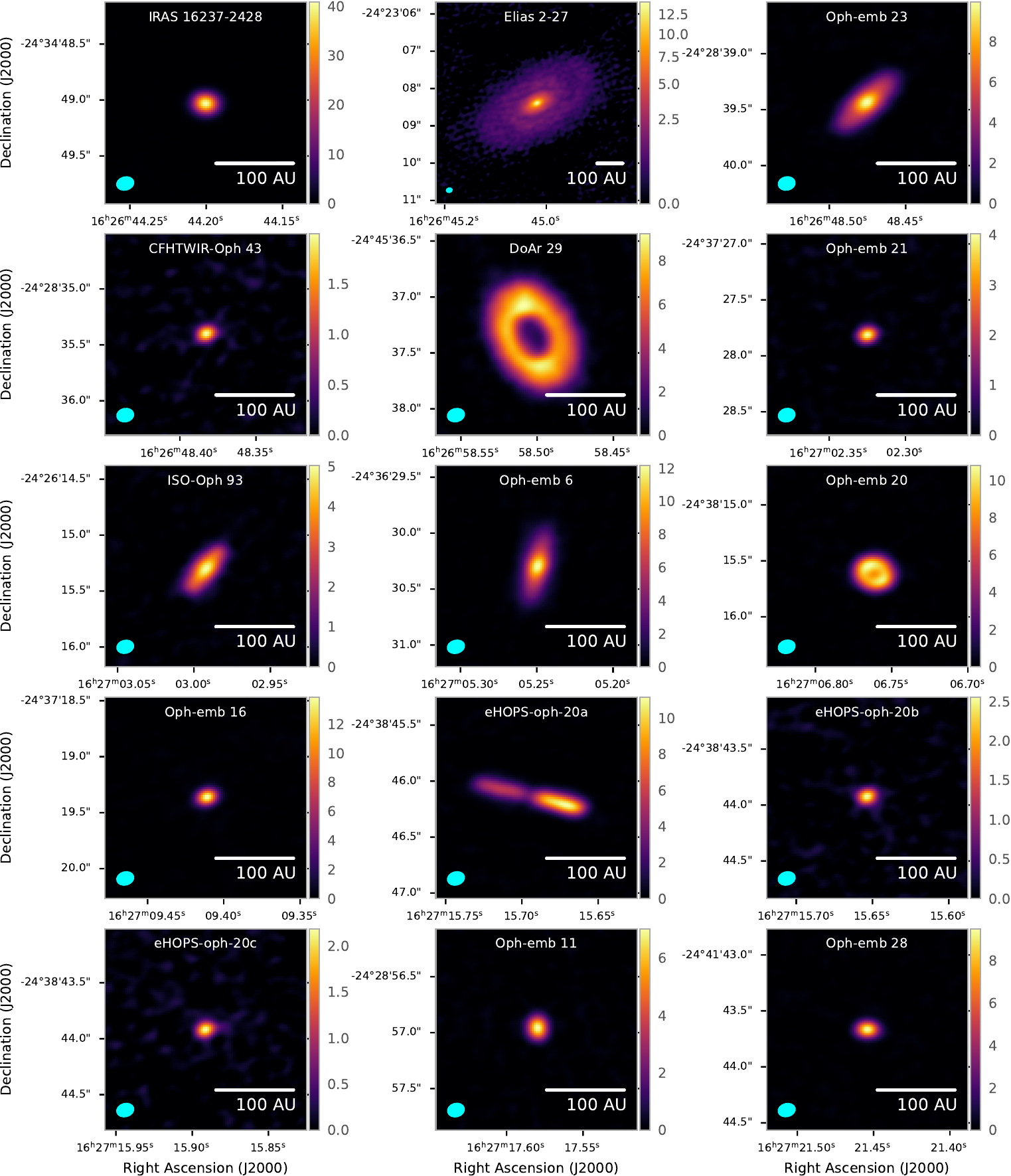}
    \caption{Part 2 of the Ophiuchus disks detected in our ALMA Band 6 CAMPOS dust continuum survey. The cyan-filled ellipse represents the synthesized beam size. All the color scales are in units of mJy/beam. The white line marks a scale of 100\,au.}
\label{fig:Oph2}
\end{figure*}

\begin{figure*}[tbh!]
    \includegraphics[width=.99\textwidth]{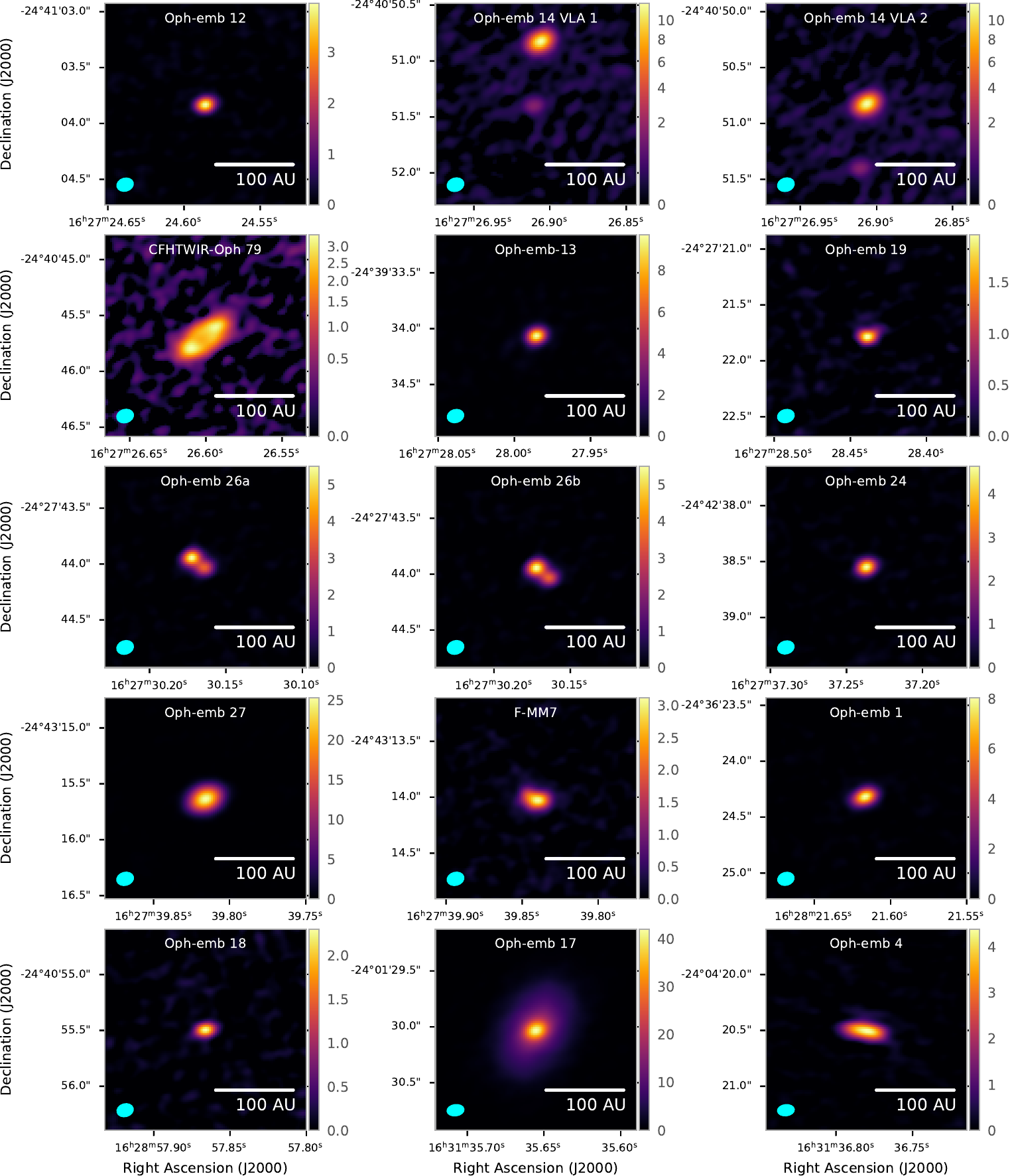}
    \caption{Part 3 of the Ophiuchus disks detected in our ALMA Band 6 CAMPOS dust continuum survey. The cyan-filled ellipse represents the synthesized beam size. All the color scales are in units of mJy/beam. The white line marks a scale of 100\,au.}
\label{fig:Oph3}
\end{figure*}

\begin{figure*}[tbh!]
    \includegraphics[width=.99\textwidth]{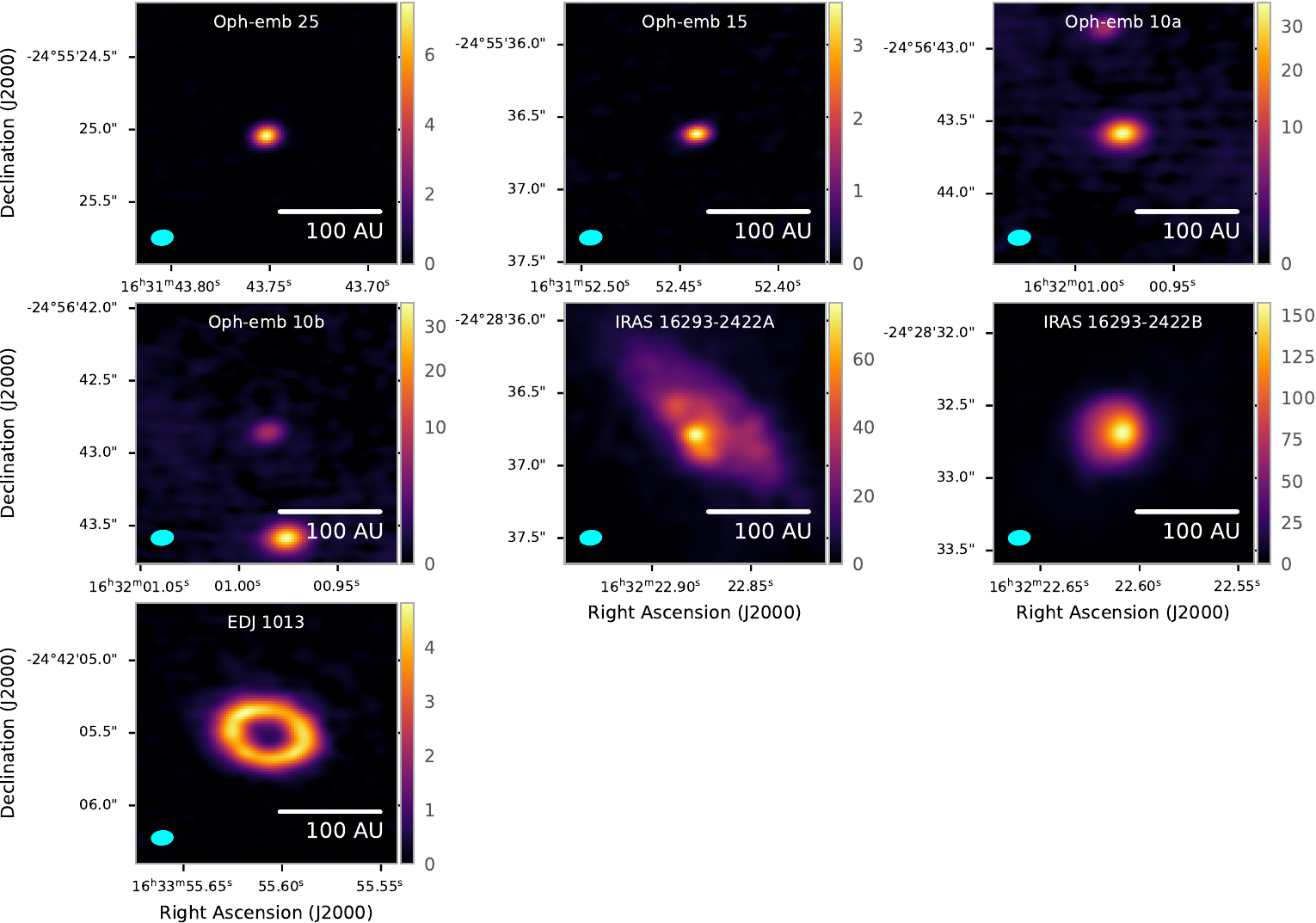}
    \caption{Part 4 of the Ophiuchus disks detected in our ALMA Band 6 CAMPOS dust continuum survey. The cyan-filled ellipse represents the synthesized beam size. All the color scales are in units of mJy/beam. The white line marks a scale of 100\,au.}
\label{fig:Oph4}
\end{figure*}

In this section, we present the image gallery of all the continuum sources detected in our CAMPOS survey. We present a zoomed-in image cutout around each protostar with the source centered in the panel. \autoref{fig:Aql1}, \autoref{fig:Aql2}, and \autoref{fig:Aql3} show the sources detected in the Aquila molecular cloud. \autoref{fig:ChamI1} shows the disks in the Chamaeleon I molecular cloud. \autoref{fig:ChamII1} shows the disks in the Chamaeleon II molecular cloud. The dust disks in the Corona Australis molecular cloud are shown in \autoref{fig:CrAus1} and \autoref{fig:CrAus2}. \autoref{fig:Oph1} to \autoref{fig:Oph4} show all the detected sources in the Ophiuchus molecular cloud. From \autoref{fig:Serp1} to \autoref{fig:Serp5} we present all the sources detected in the Serpens molecular cloud. All the sources shown in the image gallery can be found in the source Table shown in \autoref{Appendix:Source_summary}. 

\begin{figure*}[tbh!]
    \includegraphics[width=.99\textwidth]{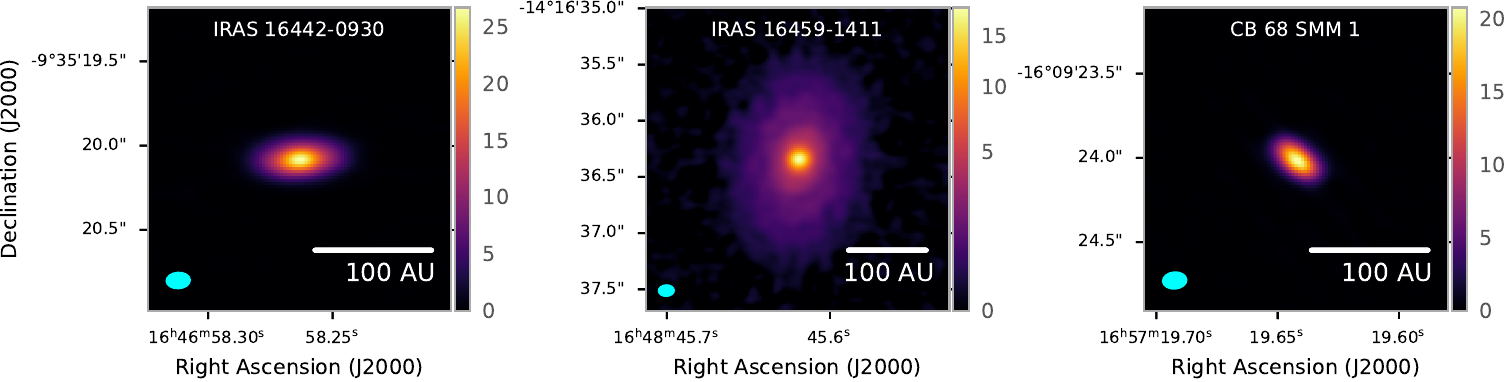}
    \caption{Ophiuchus North disks detected in our ALMA Band 6 CAMPOS dust continuum survey. The cyan-filled ellipse represents the synthesized beam size. All the color scales are in units of mJy/beam. The white line marks a scale of 100\,au.} 
\label{fig:OphN}
\end{figure*}

\begin{figure*}[tbh!]
    \includegraphics[width=.99\textwidth]{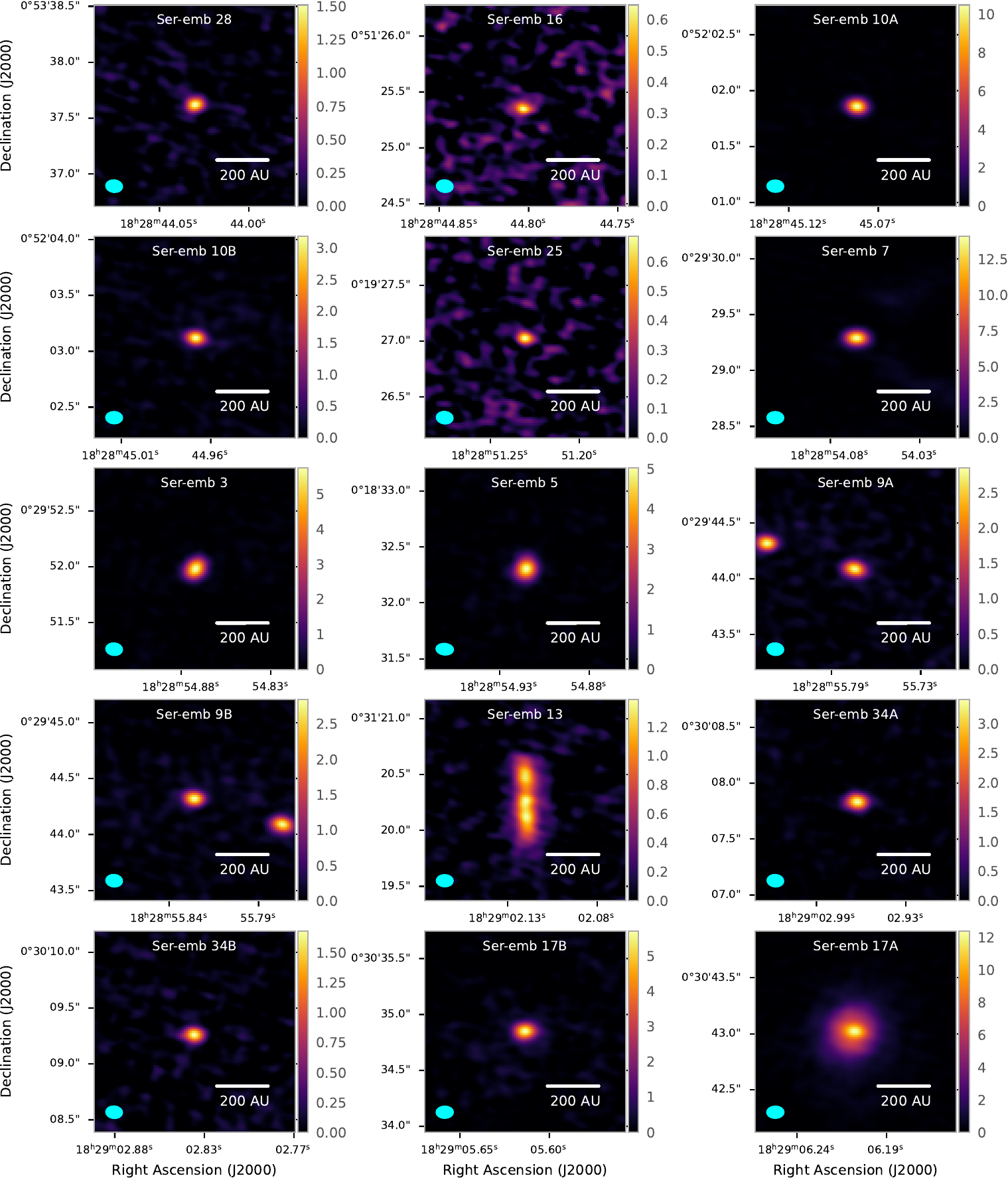}
    \caption{Part 1 of the Serpens disks detected in our ALMA Band 6 CAMPOS dust continuum survey. The cyan-filled ellipse represents the synthesized beam size. All the color scales are in units of mJy/beam. The white line marks a scale of 200\,au.}
\label{fig:Serp1}
\end{figure*}

\begin{figure*}[tbh!]
    \includegraphics[width=.99\textwidth]{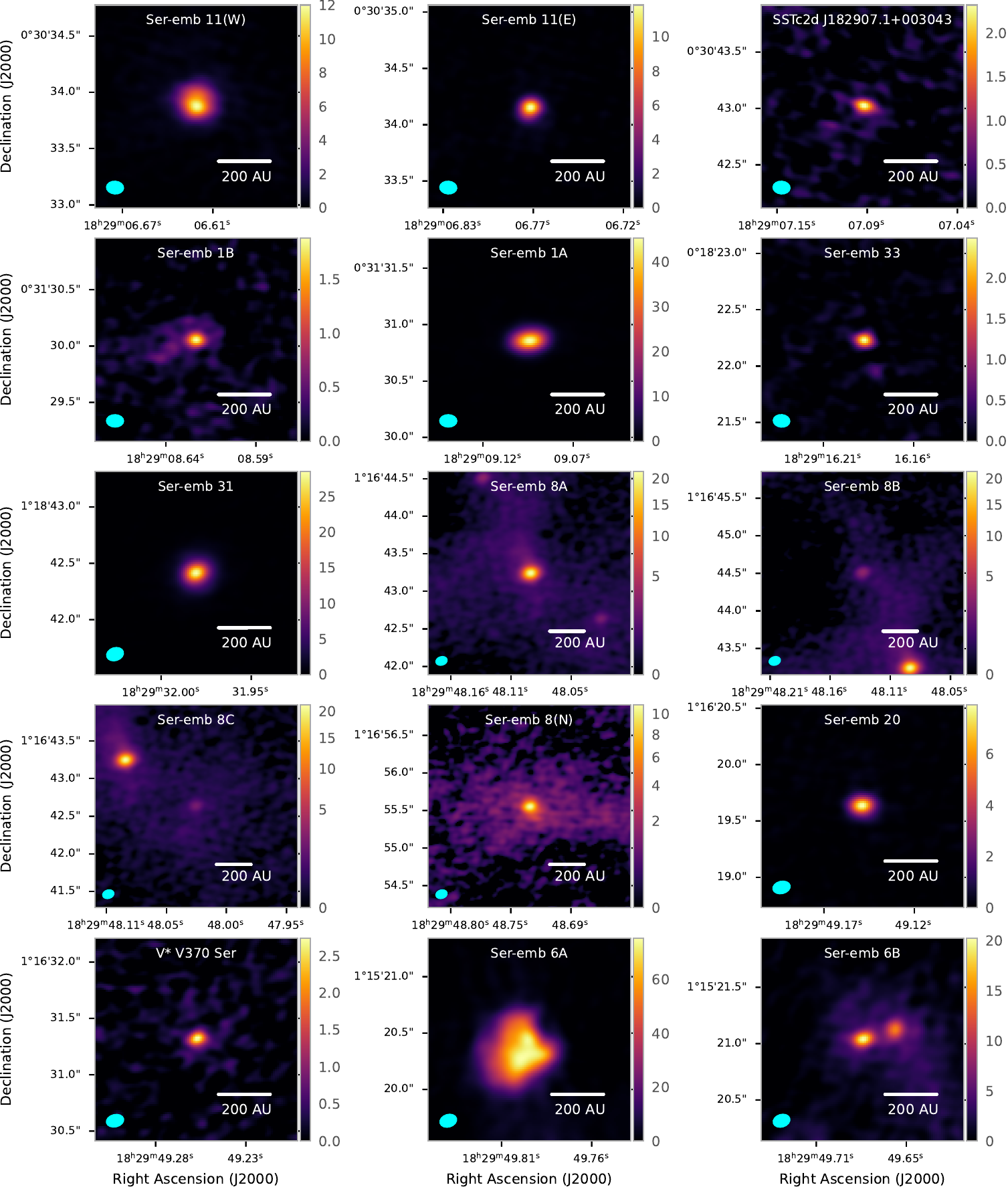}
    \caption{Part 2 of the Serpens disks detected in our ALMA Band 6 CAMPOS dust continuum survey. The cyan-filled ellipse represents the synthesized beam size. All the color scales are in units of mJy/beam. The white line marks a scale of 200\,au.}
\label{fig:Serp2}
\end{figure*}

\begin{figure*}[tbh!]
    \includegraphics[width=.99\textwidth]{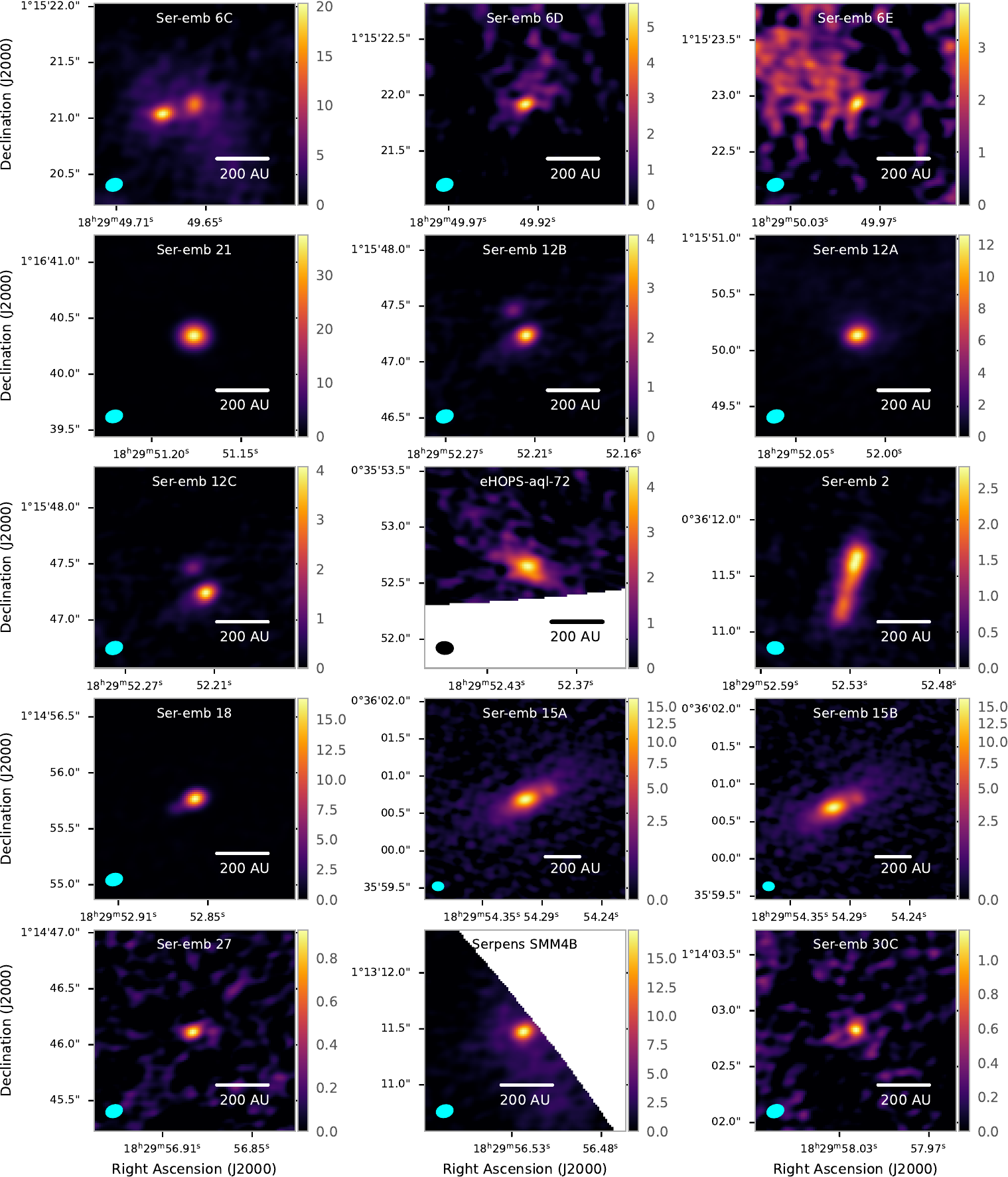}
    \caption{Part 3 of the Serpens disks detected in our ALMA Band 6 CAMPOS dust continuum survey. The cyan-filled ellipse represents the synthesized beam size. All the color scales are in units of mJy/beam. The white line marks a scale of 200\,au.}
\label{fig:Serp3}
\end{figure*}

\begin{figure*}[tbh!]
    \includegraphics[width=.99\textwidth]{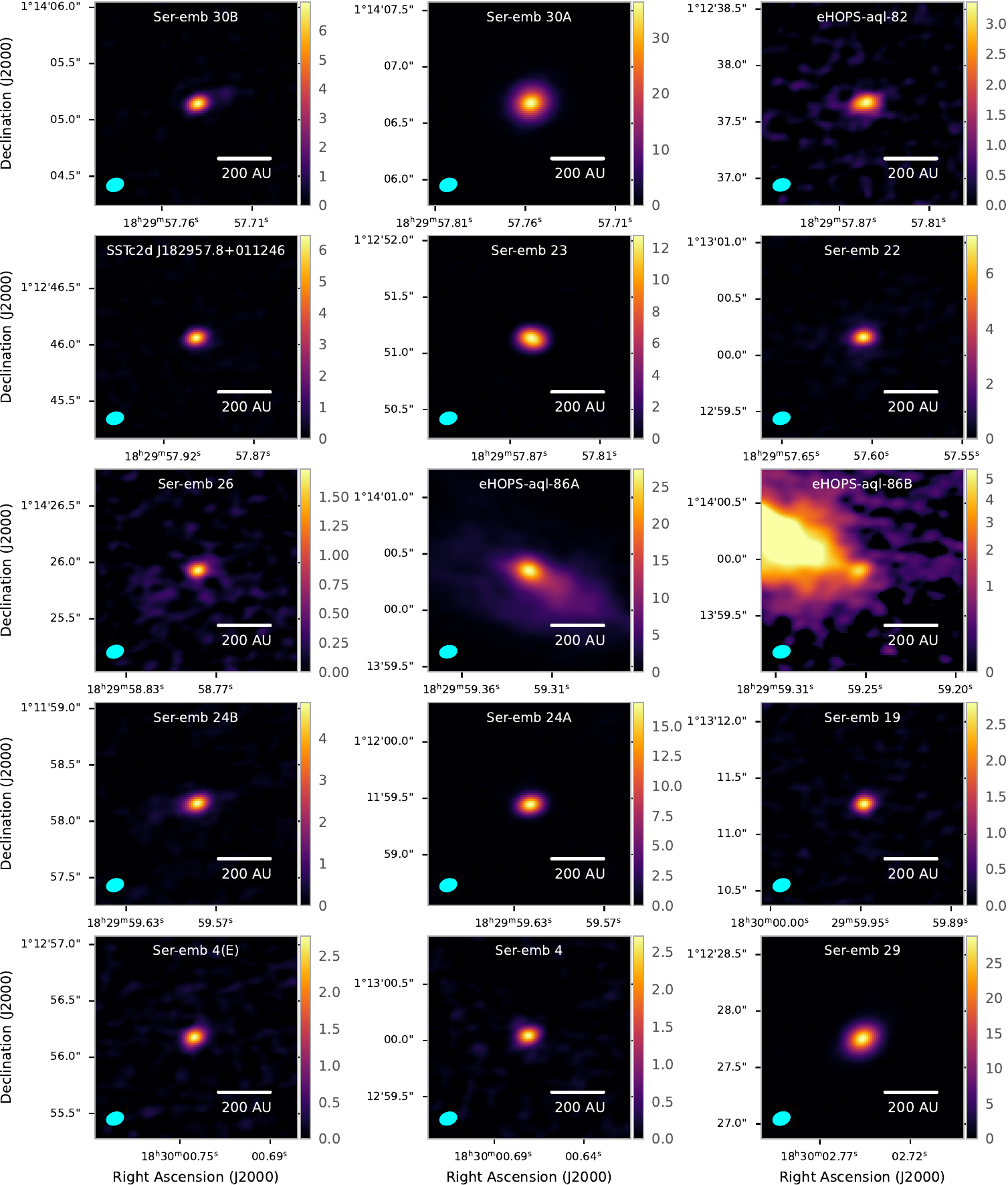}
    \caption{Part 4 of the Serpens disks detected in our ALMA Band 6 CAMPOS dust continuum survey. The cyan-filled ellipse represents the synthesized beam size. All the color scales are in units of mJy/beam. The white line marks a scale of 200\,au.}
\label{fig:Serp4}
\end{figure*}

\begin{figure*}[tbh!]
    \includegraphics[width=.66\textwidth]{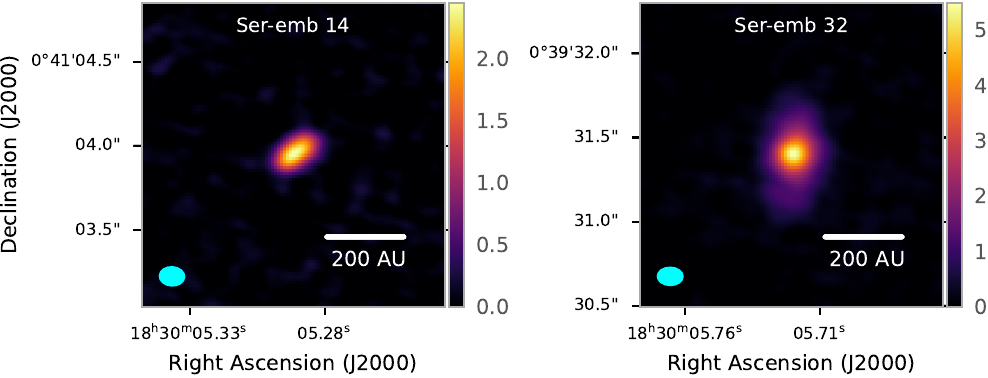}
    \caption{Part 5 of the Serpens disks detected in our ALMA Band 6 CAMPOS dust continuum survey. The cyan-filled ellipse represents the synthesized beam size. All the color scales are in units of mJy/beam. The white line marks a scale of 200\,au.}
\label{fig:Serp5}
\end{figure*}

\clearpage
\section{Image Gallery of candidate sources detected in the CAMPOS survey}
\label{Appendix:D}
\begin{figure*}[tbh!]
    \includegraphics[width=.99\textwidth]{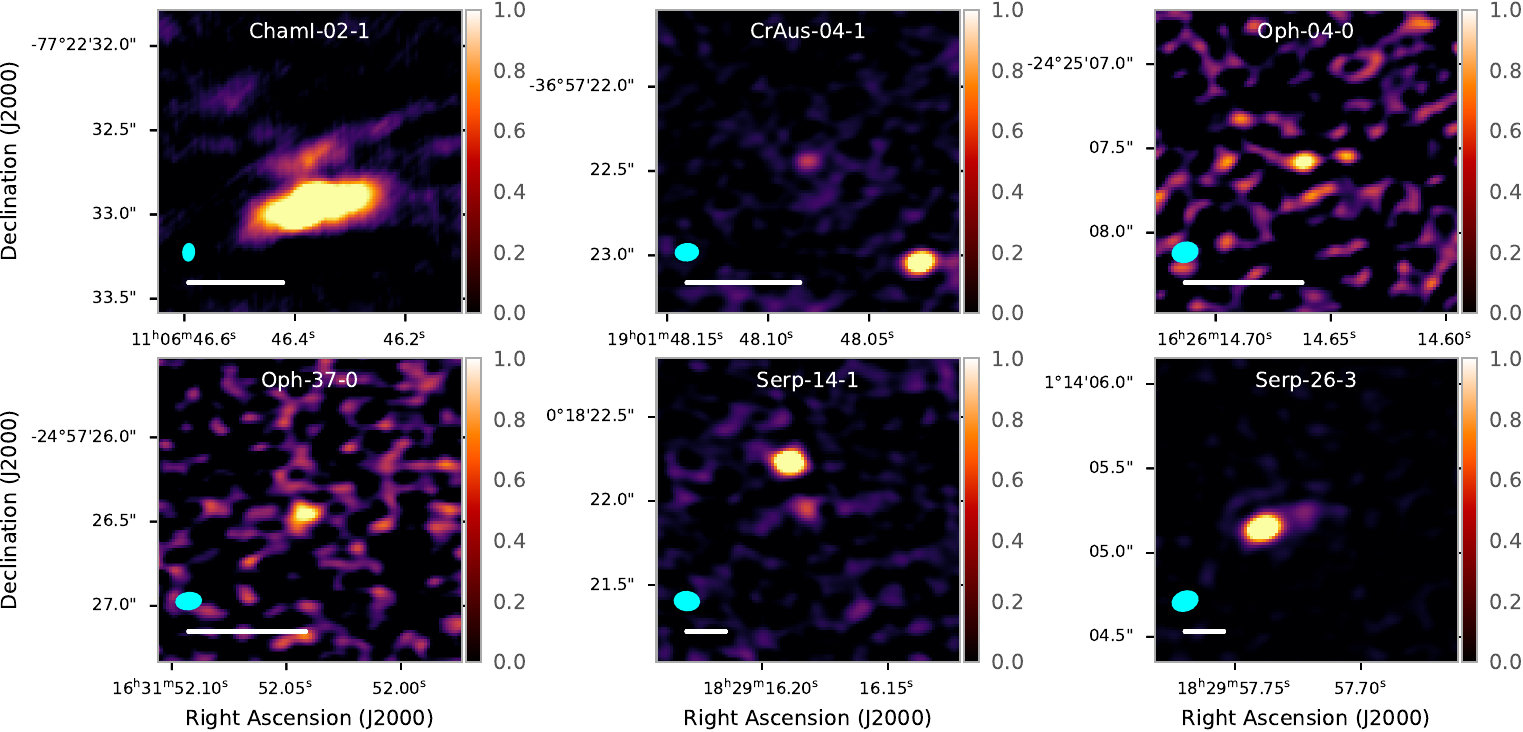}
    \caption{Candidate disks detected in our ALMA Band 6 CAMPOS dust continuum survey. The cyan-filled circle represents the synthesized beam size. All the color scales are in units of mJy/beam. The white line marks a scale of 100\,au.}
\label{fig:Candidate}
\end{figure*}

In this section, we present the candidate sources detected in our survey. The candidate sources are shown in \autoref{table:candidate} and the images are presented in \autoref{fig:Candidate}. Here we also include images of Class 0 protostar Oph-04-0 (J162614.6-242507) and Class I protostar Oph-37-0 (J163152.0-245726). Both of them exhibit faint emission ($\le 3\sigma$) at the known position of the protostar previously identified in observations conducted by the Spitzer Space Telescope.

\begin{table}[]
\centering
\caption{Summary of Candidate Sources}%
\begin{tabular}{ c c c c c c c}
\label{table:candidate}
\\
\hline \hline
  Name & Cloud & RA & DEC & Class & SN ratio & Note\\
 \hline 
CrAus-04-1 & Cor Australis  & 19:01:48.08 & -36:57:22.44 & 0 & 3.8 (natural) & Low S/N\\
ChamI-02-1 & Chamaeleon I &11:06:46.4 & -77:22:32.70 & I & 11.3 (uniform) & Dust clump or protostar?\\
Serp-14-1 & Serpens& 18:29:16.19 & +0:18:21.96 & II & 5.4 (briggs0.5) & No detection in uniform map (Dust clump?)\\
Serp-26-3 & Serpens & 18:29:57.73 & +1:14:05.25& II & 1.5 (briggs0.5) & Local peak near another source\\
\hline
\hline
\end{tabular}
\end{table}

\end{document}